\documentclass{tcibook}
\usepackage{fancyhea}
\usepackage{work}
\usepackage{bm}       %    enables bold math symbols  e.g.  \bm{\gamma}
\usepackage{graphicx}
\usepackage{multirow}
\usepackage{lineno}
\usepackage{threeparttable}
\usepackage[normalem]{ulem}
\usepackage{color}
\usepackage{textcomp}
\usepackage{arydshln} 
\usepackage[colorlinks = true,
            linkcolor = blue,
            urlcolor  = blue,
            citecolor = blue,
            anchorcolor = blue]{hyperref}

\def\beq{\begin{equation}}
\def\eeq#1{\label{#1}\end{equation}}
\def\eeqn{\end{equation}}

\newenvironment{Eqnarray}%
   {\arraycolsep 0.14em\begin{eqnarray}}{\end{eqnarray}}
\def\beqa{\begin{Eqnarray}}
\def\eeqa#1{\label{#1}\end{Eqnarray}}
\def\eeqan{\end{Eqnarray}}

\let\bar=\overbar

\def\lsim{\mathrel{\raise.3ex\hbox{$<$\kern-.75em\lower1ex\hbox{$\sim$}}}}
\def\gsim{\mathrel{\raise.3ex\hbox{$>$\kern-.75em\lower1ex\hbox{$\sim$}}}}

\def\del{\partial}
\def\Dslash{\not{\hbox{\kern-4pt $D$}}}
\def\dslash{\not{\hbox{\kern-2pt $\del$}}}
\def\pslash{\not{\hbox{\kern-2pt $p$}}}
\def\ETmiss{\not{\hbox{\kern-4pt $E$}}_T}

\def\Dlr{\mathrel{\raise1.5ex\hbox{$\leftrightarrow$\kern-1em\lower1.5ex\hbox{$D$}}}}

\def\MSB{{\bar{M \kern -2pt S}}}
\def\msb{{\bar{\scriptsize M \kern -1pt S}}}

\def\drb{{\bar{\scriptsize D \kern -1pt R}}}

\usepackage{appendix}
\usepackage{lscape} 

\def\authorlist#1#2{
    \vskip 0.4in
\begin{center}\begin{large} {\bf  #1 } \end{large}
    \vskip 0.2in
              #2
     \vskip 0.2in
   \end{center}
}

\begin{document}

\chapter{Accelerators for Electroweak Physics and Higgs Boson Studies}

\authorlist{A.~Faus-Golfe, G.H.~Hoffstaetter, Q.~Qin, F.~Zimmermann (editors)}{T.~Barklow,  E.~Barzi, S.~Belomestnykh, M.~Biagini, M.~Chamizo Llatas, J.~Gao, E.~Gianfelice,  
B.~List, V.~Litvinenko, E.~Nanni,  T.~Raubenheimer, 
T.~Roser, T.~Satogata,  V.~Shiltsev, 
S.~Stapnes, V.~Telnov  (contributors)
}

\paragraph{Abstract:} We discuss the goals, the designs, the
state of technical readiness, and the critical R$\&$D needs of the accelerators that are currently under discussion as Higgs and electroweak factories. 
We also address the respective staging options enabling 
future energy-frontier colliders. The accelerators covered are based on many different techniques and approaches. They include several circular colliders, various linear colliders, colliders based on energy recovery linacs (ERLs),  ERL-ring combinations, as well as $\gamma \gamma$ colliders. 
The linear colliders proposed consist of options for the International Linear Collider (ILC), for the Compact Linear Collider (CLIC), for the Cold Copper Collider (C$^3$), and for the more recent Higgs-Energy Lepton Collider (HELEN). ERLs are key components of the Recycling Linear e$^+$e$^-$ Collider (ReLiC),of  the Energy Recovery Linear Collider  (ERLC), and of the Circular Energy Recovery Collider (CERC). Among the more conventional ring colliders  
the following proposals are featured: the Future Circular Collider (FCC-ee), the Circular Electron Positron Collider (CEPC),  the Electron Positron Circular Collider at Fermilab (EPCCF), and the Large Electron Positron collider $\#$3 (LEP-3). In addition, we consider the X-ray FEL based gamma-gamma Collider Higgs Factory (XCC) and the High-Energy High-Luminosity $\gamma$-$\gamma$ collider (HE$\&$HL $\gamma\gamma$). Finally, a Higgs factory based on a circular muon collider is mentioned for completeness.

%%%%%%%%%%%%%%%%%%%%%%%%%%%%%%%%%%%%%%%%%%%%%%%%%%%%%%%

\newpage
\section{Executive summary}

The field of proposed Electroweak $\&$ Higgs Factories is broad and possible accelerators have not been strongly narrowed down in the accelerator and Particle Physics community. On the contrary, several additional new options have been put forward during the Snowmass'21 process. 
For the sake of comparison, the White Papers have been reclassified in three groups: linear colliders, circular colliders and $\gamma\gamma$ colliders. 
In the following, we assemble and compare the main parameters for the various proposals, summarized in Tables \ref{all-linear} and \ref{all-circular}. For all but two of these proposed colliders, White Papers were submitted to the Snowmass'21 process. These remaining two proposals are briefly mentioned, for the sake of completeness, with references to the corresponding literature.

It is our hope that this comparison will help the community to narrow down and focus our field on the most promising proposals, and to focus ongoing research on topics of largest impact. We, therefore, describe the critical R$\&$D items for each proposed collider and sketch consolidated R$\&$D efforts that would benefit jointly several of the projects.

By e$^+$e$^-$ Higgs factories, we refer to colliders that operate at 240/250 GeV (or 380 GeV in the case of CLIC) and, alternatively, also colliders with a staged program, including a stage above the top quark threshold (which for CLIC would be the initial stage).  
The latter colliders deliver superior physics performance for three reasons: (1)  The top quark is another important object that needs precision study. (2)  Above 350 GeV, the primary Higgs boson production mechanism changes from e$^+$e$^-$ $\rightarrow$ ZH to WW fusion production of the Higgs. This capability is essential to prove the influence of Beyond Standard Model (BSM) physics on the Higgs boson. (3)  Some measurements  require data at two different, sufficiently well separated CM energies.  The most important of these is the determination of the Higgs self-coupling from single Higgs production.

Some of the proposed Higgs Factories also propose a significant run at and around the Z pole (TeraZ) and a shorter run at the WW threshold. 
This opens a new capacity of discovering new physics through QCD, flavour and electroweak precision measurements (EWPO) and through Beyond-Standard-Model (BSM) searches, in particular for feebly coupled, long-lived particles and for axions and axion-like particles.

\begin{landscape}
\begin{table}[htbp]
\centering
\caption{Key parameters for the proposed Linear Higgs and Electroweak factory colliders. }
\begin{tabular}{l| l | c c c l l }
\hline 
\bf{Name} & \bf{Concept} &  \\
& 
& \bf{$\sqrt{s}$} & {\bf{\it L}/IP} & no.~IPs & \bf{Technology} & \bf{Comment}   \\
 &  &  [TeV]      & [nb$^{-1}$s$^{-1}$] &  &  \\ 
 \hline \hline \hline
\bf{ILC-250} & linear & 0.25 & 13.5 & 1 & 1.3 GHz SRF Nb (31.5 MV/m), 2 K  &   \\ 
ILC-250-HL & & 0.25 &  27 & 1 & 1.3 GHz SRF Nb (31.5 MV/m), 2 K &  2$\times n_{bunch}$\\ 
ILC-91 & & 0.09 &  0.21 & 1 & 1.3 GHz SRF Nb(31.5 MV/m), 2 K & $\downarrow f_{rep}$ \\ 
ILC-91-HL & & 0.09 &  0.41 & 1 & 1.3 GHz SRF Nb (31.5 MV/m), 2 K & 2$\times n_{bunch}$, $\downarrow f_{rep}$ \\ 
ILC-500 & &  0.5 & 18 & 1  & 1.3 GHz SRF Nb (31.5 MV/m), 2 K & \\ 
ILC-500-HL & & 0.5 & 36 & 1 & 1.3 GHz SRF Nb (31.5 MV/m), 2 K &  2$\times n_{bunch}$, $\downarrow f_{rep}$ \\ 
ILC-250-VHL & &  0.25 & 54 & 1 & 1.3 GHz SRF Nb (31.5 MV/m), 2 K &  2$\times n_{bunch}$, 2 $\times f_{rep}$ \\ ILC-1000 & & 1000 & 51 & 1 & 1.3 GHz SRF Nb (45 MV/m), 2 K &  $\downarrow n_{bunch}$, $\downarrow f_{rep}$  \\  
\hline \hline
\bf{CLIC-380} & linear & 0.38 & 23 & 1 & NCRF X-band (72 MV/m)  & two-beams/klystrons \\ 
CLIC-1500 &  &  1.5 & 37 & 1 & NCRF X-band (72-100 MV/m) & two-beams \\ 
CLIC-3000 &  &  3.0 & 59 & 1 & NCRF X-band (72-100 MV/m) &  two-beams \\ 
 \hline \hline 
{\bf{C$^{3}$-250}} & linear & 0.25  & 13 & 1 & NCRF C-band (70 MV/m) Cu, 80 K  &  
distr.~coupl.~RF \\ 
C$^{3}$-550 &  & 0.55  & 24  &1  & NCRF C-band (120 MV/m) Cu, 80 K &  distr.~coupl. RF, $\uparrow $RF power    \\ \hline
\bf{C$^{3}$}-Nb$_3$Sn & linear &  0.25  & $\ge 13$  & 1  & 1.3 GHz SRF Nb$_3$Sn (100 MV/m), 4.5 K  &   distr.~coupl.~RF
\\ \hline \hline
\bf{HELEN} & linear &   0.25 & 13.5 & 1  & TW SRF (70 MV/m) & \\ 
\hline \hline\hline
\bf{ReLiC} & linear ERL&  0.25 & 215 & 2 & SRF w.~separators & \\ \hline
\bf{ERLC-250}$^{*}$ & linear ERL &  0.25 & 390/1600 & 1 & 1.3/0.65 GHz SRF Nb/Nb$_3$Sn (20 MV/m) & w. 2-axis cavities \\
ERLC-500$^{*}$ & linear ERL &  0.5 & 175/780 & 1 & 1.3/0.65 GHz SRF Nb/Nb$_3$Sn (20 MV/m) & w. 2-axis cavities 
\\ 
\hline \hline \hline
\bf{XCC}  & $\gamma \gamma$  & 0.125--0.28  &  100$^{**}$ & 1 & based on C$^3$ &  with FEL \\ 
\hline \hline
\bf{HE$\&$HL $\gamma \gamma$} & $\gamma \gamma$  &  0.5--10 & 10\%LC$^{\ddagger}$  & 1 & based on any LC$^{\ddagger}$  &  with FEL \\ 
\hline \hline \hline
\multicolumn{6}{l}{$^{\ast}${\small {No white paper but listed for completeness.}} } \\
\multicolumn{6}{l}{$^{\ast \ast}${\small {(e$^-$e$^-$ geom.)}} } \\
\multicolumn{6}{l}{$^{\ddagger}${\small {For second interaction region with respect to base linear collider (LC)}} } \\
\end{tabular}
\label{all-linear}
\end{table}
\end{landscape}

\begin{landscape}
\begin{table}[htbp]
\centering
\caption{Key parameters for the proposed Circular Higgs and Electroweak factory colliders.}
\begin{tabular}{l| l | c c c l l }
\hline 
\bf{Name} & \bf{Concept} &  \\
&  % & species 
& \bf{$\sqrt{s}$} & {\bf{\it L}/IP} & no.~IPs & \bf{Technology} & \bf{Comment}   \\
 &  &  [TeV]      & [nb$^{-1}$s$^{-1}$] &  &  \\ 
 \hline \hline \hline
\bf{FCC-ee} & circular &  0.09 & 1810 &   4 &  400 MHz 1-cell Nb/Cu, 4.5 K & Z pole \\
    &    &  0.16 & 173 & 4  &  400 MHz 2-/4-cell Nb/Cu, 4.5 K & WW threshold \\ 
     &    & 0.24 & 72 &  4 &  400 MHz 2-/4-cell Nb/Cu, 4.5 K & ZH \\ 
      &   & 0.365 & 12.5 &  4 &  800 MHz 5-cell Nb, 2 K & ${\rm t}\bar{\rm t}$ \\ \hline
FCC-ee-H & circular &  125 & 230 &  4 & 400 MHz 2-/4-cell Nb/Cu, 4.5 K  & $e^+e^-\rightarrow {\rm H}$,  monochr.  \\
        \hline\hline 
\bf{CEPC} & circular &  0.24 & 83 & 2 & 2-cell 650 MHz, Nb, 2 K &  \\ 
 & & 0.09 & 1917 & 2  & 1-cell 650 MHz, Nb, 2 K &  \\ 
 & & 0.16 & 2660 & 2 & 2-cell 650 MHz, Nb, 2 K &  \\ 
 & & 0.36 & 8 & 2  & 5-cell 650 MHz, Nb, 2 K &  \\ 
\hline \hline
\bf{EPCCF} & circular & 0.24  & 10 & 1  & 650 MHz, bulk Nb &  on FNAL site \\ 
\hline \hline
\bf{LEP-3}$^{\ast}$ & circular & 0.24 & 11 & 2 &  & in LHC tunnel, not pursued
\\ 
\hline \hline \hline
\bf{CERC} & circular ERL & 0.16 & 870 & 1 & 100 km ring with two SRF linacs & \\  
& & 0.25 & 780 & 1 &  & \\ 
& & 0.365 & 280 & 1 & &  \\
& & 0.5 & 130 & 1 &  &  \\
& & 0.6 & 90 & 1 &  &  \\
\hline \hline\hline 
\bf{MC-HF}$^{\ast}$ & $\mu$ & 0.125 & 0.1 & 1 & ionization cooling &  4 MW p driver\\
\hline \hline \hline
\multicolumn{6}{l}{ $^{\ast}${\small {No white paper but listed for completeness}} } \\
\end{tabular}
\label{all-circular}
\end{table}
\end{landscape}

\begin{figure}[htbp]
\centering
\includegraphics[width =0.5\textwidth]{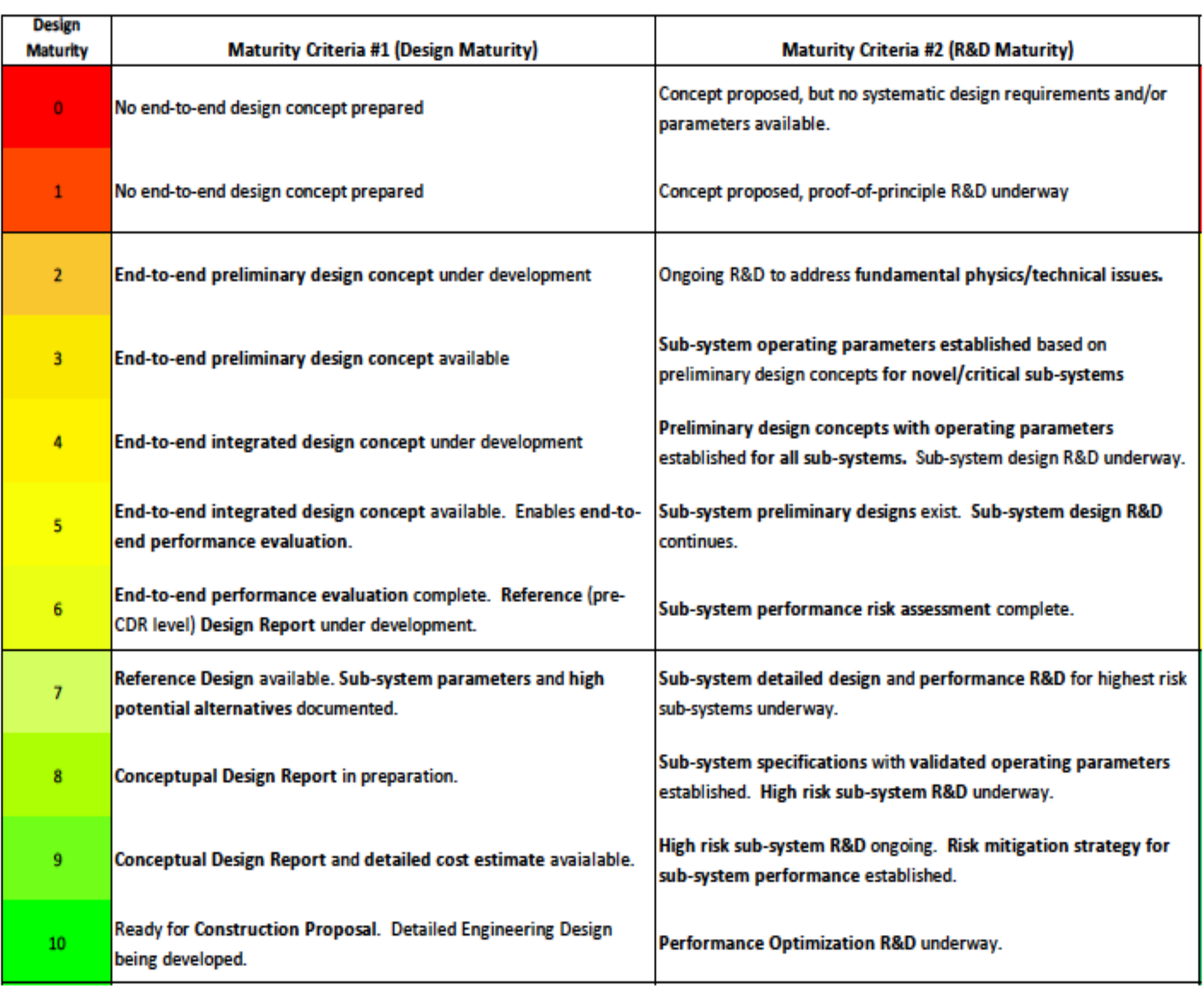}
\includegraphics[width =0.65\textwidth]{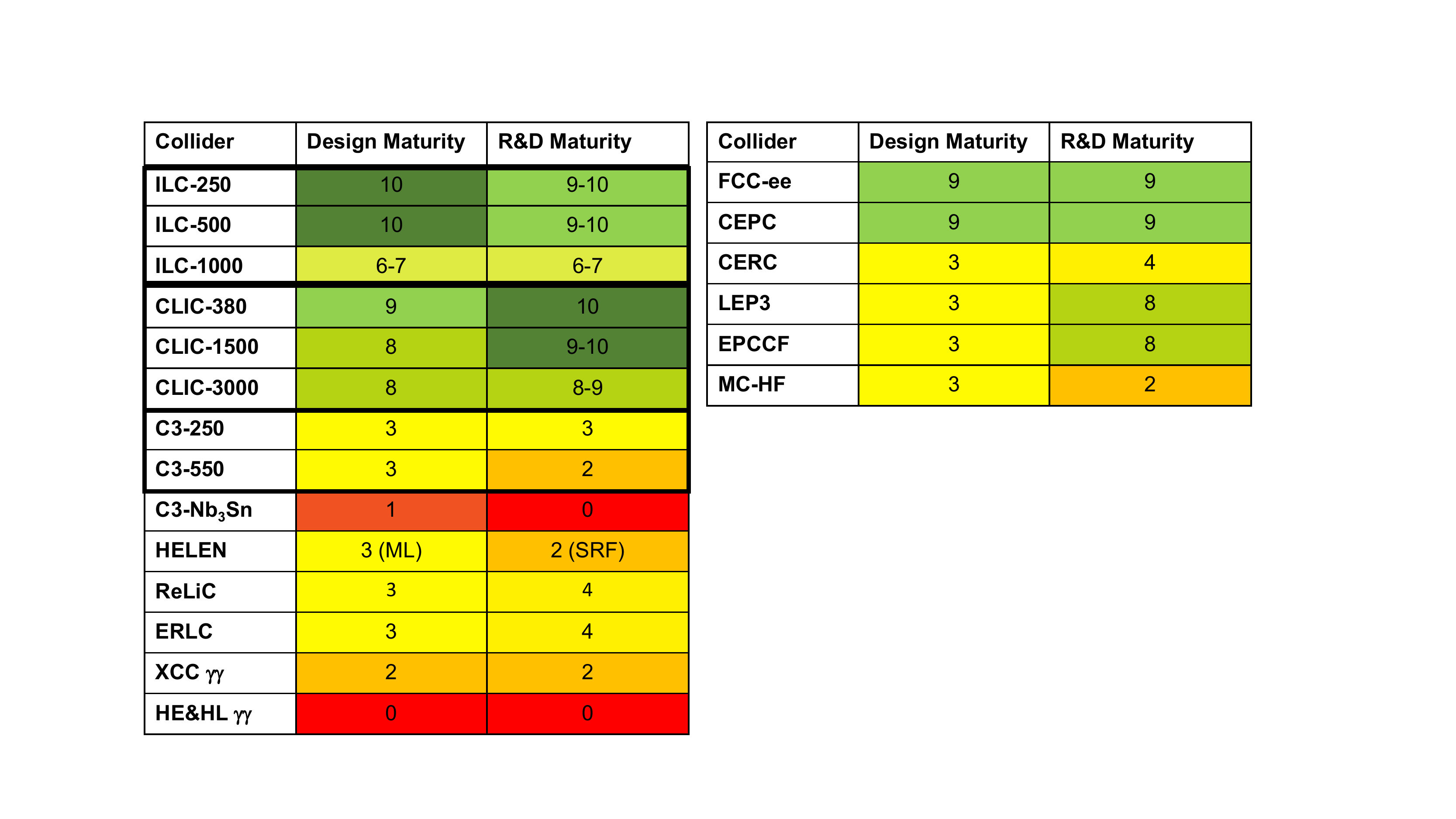}
\caption{Concept Maturity Evaluation: Design and R$\&$D for EW$\&$Higgs factories.}
\label{maturity-color}
\end{figure}

Between all the EW $\&$ Higgs factory proposals considered, only a small number are ``shovel-ready'' or close to a construction phase, while many of the proposals are still at an early design phase, i.e., in a  
(pre-)conceptional design stage.  The latter 
proposals should focus on their main R$\&$D tasks in order to move 
forward towards a technical design. 
A detailed estimation of the Technical Readiness Level (TRL), risk factors, technology validation, cost reduction impact, performance reach, and 
time scales for the various proposals   
are presented in detail in the Collider Implementation Task Force document \cite{ITFdocument}. As a complement, we have evaluated the proposal maturity regarding two aspects, namely design and R$\&$D, based on the criteria listed in the top part of Fig.~\ref{maturity-color}. The resulting evaluation is summarized in the bottom part of Fig.~\ref{maturity-color}.

The main R$\&$D technical topics to be further pursued 
for the different proposals are:
\begin{itemize}
    \item ILC: 
     (polarized) e$^{+}$ production,  FFS tunability and long term stability,  FFS doublet vibration issues, Injection/extraction devices.
    For upgrades: SRF with higher Q, higher gradient, Traveling Wave SRF cavities, and  Nb$_{3}$Sn.

    \item CLIC: mechanical vibration mitigation and stability, cost-efficient X-band components and RF sources.
    
    \item C$^{3}$: RF optimized structure including cost and industrialisation, cryomodule R$\&$D, RF source optimization and cost reduction, ultra-low emittance e$^{-}$ source. Overall accelerator layout.
    
    \item C$^3$-Nb$_3$Sn: R\&D on producing Nb$_3$Sn on inexpensive and thermally efficient metals such as Cu or bronze, and scale-up of current methods of Nb$_3$Sn coatings on Cu or bronze, geared towards distributed coupling structures and standard RF cavity cells.
    
    \item HELEN: SRF with higher Q, higher gradient, Traveling Wave SRF cavities, and  Nb$_{3}$Sn. Cryogenic optimization. Overall accelerator layout. 
    
    \item ReLiC: SRF with higher Q, higher gradient, Traveling Wave SRF cavities, and  Nb$_{3}$Sn. Cryogenics optimization. Test of high current, low loss energy recovery.  Self-consistent and coherent parameter table and overall accelerator layout.
    
    \item ERLC: Dual-axis SRF cavities, higher Q, higher gradient. Cryogenics optimization. Test of high-current, low loss energy recovery.  Self-consistent and coherent parameter table and overall accelerator layout.
    
    \item XCC: X-ray transport and focusing. Interaction-region layout with Compton collision point. FEL design. Overall accelerator layout and integration issues.
    
    \item HE$\&$HL $\gamma\gamma$: Development of a specific collider concept, self-consistent and coherent parameter table, overall accelerator layout and integration issues.
    
    \item FCC-ee: High-$Q_0$ SC cavities for the 400--800 MHz range; efficient RF power sources; 
    cryomodules and cryogenic system; ``low-field'' HTS magnets for final focus, arcs, and e$^+$ source; booster magnets. High-field magnets based on Nb$_3$Sn and HTS in preparation for FCC-hh.
    
    \item CEPC: two-cell 650 MHz bulk Nb cavities; efficient RF power sources at 650 MHz and 1.3 GHz; cryogenic system; booster magnets. High-field magnet systems using iron based HTS for SPPC.
    
    \item LEP-3: SRF cavities, RF power sources, cryogenic system, and booster magnet system. Self-consistent and coherent parameter table and overall accelerator layout.
    
    \item EPCCF: SRF cavities, RF power sources, cryogenic system, and booster magnet system. Self-consistent and coherent parameter table and overall accelerator layout.
        
    \item CERC: SRF with higher Q, ultra-small emittance preservation, damping rings with very flat beams and large energy acceptance, use of small gap magnets for power and cost reduction, High repetition rate extraction and injection kickers. Self-consistent and coherent parameter table and overall accelerator layout.
    
%This list makes it apparent that the research item that will benefit most projects, in a nutshell, are the following: SRF, RF sources, and cryomodules
    
\end{itemize}

EW $\&$ Higgs factories operating in the 240--250 GeV energy range are not very challenging from the point of view of beam energy; the target energy is less than 20\% higher than what had already 
been achieved at the former LEP collider.   
Indeed, a large part of the ongoing R$\&$D is focused on making these colliders more energy efficient and on reducing their construction cost. On the other hand, EW $\&$ Higgs factories are high-precision machines and luminosity will be their main figure of merit.
The joint technology R$\&$D  topics identified that will be most beneficial overall are:

\begin{itemize}

\item Energy

\begin{itemize}
\item SCRF: TW structures and Nb$_{3}$Sn (70 MV/m HELEN). Special attention has to be paid to the SCRF for ERLs.
\item NCRF: Cryo-cooled Copper structures (120 MV/m C$^3$), HTS coatings.
\item Cryogenics: massive production, plug compatibility, transport issues, gas-pressure regulations, more efficient gas coolers.
\item Efficient RF power sources. 
\end{itemize}

\item Luminosity

\begin{itemize}
\item Positioning, Monitoring, Alignment and Stabilisation: global strategies, instrumented girders, radiation-hard ground motion sensors.
\item e$^+$ production optimization: flux concentrators, pulsed or dc solenoids, capture linacs, targetry issues ... 
\item Nanobeam collision techniques: concepts and feedback
\item Damping Rings, booster and collider rings: low-emittance lattices 
\item Magnets: Interaction Region FFS and Injection/Extraction devices
\end{itemize}

\item Sustainability

\begin{itemize}
\item Energy consumption, efficiency, sustainability, carbon footprint; 
\item High-Efficiency RF power sources: Klystrons, Solid State Amplifiers and IOTs; 
\item Permanent magnets and HTS magnet systems.
\end{itemize}

\item Others

\begin{itemize}
\item Manufacturing techniques including additive, cost reduction and massive production; 
\item High power Beam Dumps (multi-MW); 
\item Machine protection and collimation; 
\item Polarized beams and polarimetry; 
\item Beam instrumentation;  
\item Robotics and automation. 
\end{itemize}

\end{itemize}

%%%%%%%%%%%%%%%%%%%%%%%%%%%%%%%%%%%%%%%%%%%%%%%%%%%%%%%

\newpage
\section{Accelerators with Higgs-factory potential \cite{Higgs}}

An e$^{+}$e$^{-}$ Higgs factory has been identified by the International Committee for Future Accelerators (ICFA) and the 2020 update of the European Strategy of Particle Physics (EPPSU) as  the highest-priority next collider after the LHC. 

During the Snowmass'21 process a set of potential lepton colliders operation in the energy range from the Z boson mass to the TeV scale was considered. The different projects considered include circular colliders, linear colliders, and Energy-Recovery-Linac-based colliders, ERL-ring combinations and as well as $\gamma \gamma$ colliders. The linear colliders covered are various options for the international linear collider (ILC), for the Compact Linear Collider (CLIC), for the Cold Copper Collider (C$^3$), and for the more recent Higgs-Energy Lepton Collider (HELEN). ERLs are components of the Recycling Linear e$^+$e$^-$ Collider (ReLiC), the Energy Recovery Linear Collider  (ERLC), and the Circular Energy Recovery Collider (CERC). And we include the following ring colliders: the Future Circular Collider (FCC-ee), the Circular Electron Positron Collider (CEPC),  the Electron Positron Circular Collider at Fermilab (EPCCF), and the Large Electron Positron collider $\#$3 (LEP-3). Further more, we consider the X-ray FEL based gamma-gamma Collider Higgs Factory (XCC) and the High-Energy High-Luminosity $\gamma$-$\gamma$ collider (HE$\&$HL $\gamma\gamma$).

The scientific questions to be addressed at a lepton collider are quite well defined; and given the recent discoveries in the hadrons colliders the operation energies are well known. The Physics questions that must be addressed with highest-priority are:  the precision measurement of Higgs couplings to SM fermions and gauge bosons, the measurement of Higgs self-couplings and the sensitivity to rare or non-SM Higgs decays. Most of the leptons colliders cited above have the capability to address these questions, but differ in their capabilities for exploring other fundamental questions, such as the search for new physics with SM couplings or feebly coupled that could provide explanations for the baryon asymmetry of the Universe, neutrino masses, dark matter, etc. 

The baseline energy of the e$^+$e$^-$ Higgs factories, is centered around 240/250 GeV, but operation below and or above the energy Higgs production is considered in most of the projects. 

Most of the proposed Higgs and Electroweak Factories can also not be seen in isolation of other accelerators that are being  proposed for higher energies and that would either greatly benefit from the infrastructure and accelerator developments described here, or that would be a direct upgrade or extension of the Higgs and Electroweak Factory. The upgrade or extension possibility is important for three reasons. The first is that the top quark is another important object that needs precision study.  This includes the measurement of the top quark mass but, more importantly, the measurement of the top quark electroweak couplings, which give some of the best tests of models with Higgs boson compositeness or new strong interactions.  This is an essential part of the program of precision Standard Model tests. The second is that above 350 GeV, the primary Higgs boson production mechanism changes from e$^+$e$^-$ $\rightarrow$ ZH to WW fusion production of the Higgs.  This means that, in the second stage, one can acquire  a new Higgs boson data set, comparable to the earlier one, with different characteristics.   Thus it is possible to discover an effect in the 240--250 GeV stage and confirm it using measurements with different initial conditions at the higher-energy stage. It is easy to suggest anomalies from precision measurements, but it is difficult to prove that these anomalies are real.  Thus, this capability is essential to prove the influence of beyond Standard Model physics on the Higgs boson. Finally, there are some measurements that require data at two different, sufficiently well separated CM energies.  The most important of these is the determination of the Higgs self-coupling from single Higgs production \cite{higgssc1,higgssc2}.  At any single energy, the effect of changing the Higgs self-coupling is degenerate with effects of changing other Higgs parameters.  However, by making measurements at two different CM energies, these effects can be separated.  A similar statement applies for measurement of beyond Standard Model effects in e$^+$e$^-$ $\rightarrow$ 2 fermion processes, for example, fermion compositeness or Z' searches. For these reasons, we also discuss the extendability of each proposed accelerator. This provides a natural connection to the accelerator energy frontier, e.g.~to the report of working group AF4 of the Snowmass'21 process.

Some of the proposed Higgs Factories also propose a significant run at and around the Z pole (TeraZ). This opens a new capacity of discovering new physics through QCD \cite{fcceeqcd,fcceeflavor1}, flavour \cite{fcceeflavor1,fcceeflavor2} and electroweak precision measurements (EWPO) \cite{fcceeaqed,fcceeewpo} 
and through Beyond-Standard-Model (BSM) searches, in particular for feebly coupled, long-lived particles and for axions and axion-like particles. A competitive flavour physics as well as  a serious impact on QCD  requires at least $10^{11}$~Z bosons. With the production of $10^{12}$ Z's the new collider's flavor physics programme will greatly outshine Belle II.  
A breakthrough benchmark for Electroweak Physics comes with a direct measurement of $\alpha_{\rm QED}$ at the Z pole \cite{fcceeaqed}, and a breakthrough for feebly coupled particle search arises when the search for Heavy Neutrinos reaches the seesaw limit
\cite{fcceeflavor1}. Both call for $5\times 10^{12}$ Z's or more. Circular colliders have luminosities that increase sharply at lower CM energy, making it possible to collect samples of $5\times 10^{12}$ Z events (``TeraZ'').  At linear colliders, the luminosity decreases at lower CM energies, but still it is possible to collect samples of $5\times 10^9$ Z events (``GigaZ''),  
already two orders of magnitude more than in the LEP program.  The Higgs factory detectors are expected to be greatly superior to those at LEP in tracking and hadron calorimetry and especially in flavor tagging.  The linear collider programs will be done with beam polarization, and key electroweak parameters can be measured most sensitively through polarization asymmetries,   
improving significantly (by factors up to 10-fold) the precision on parity violating Z couplings. The circular colliders benefit from exquisite beam energy calibration (with $<$100 keV accuracy for $\sqrt{s}$) based on resonant depolarization, allowing a highly precise  determination of Z and W masses and widths. 

So, it is possible, in either case, to revisit the electroweak measurements of LEP, increasing the precision by more than one or two orders of magnitude for linear or circular colliders, respectively.   
The huge event yields at circular colliders also enable a unique program on the physics of $\tau$, $c$, and $b$, including $b$ baryons. Circular colliders also envision a program at the WW threshold, to push the precision on the W mass below 0.4 MeV.  Both types of colliders will also measure the W mass from WW and single W production at 240/250 GeV, to a precision of about 2.5 MeV. Increased precision in electroweak observables is valuable for two reasons.  First, reaching a relative precision on electroweak observables of 10$^{-3}$ enables a new set of tests, significantly beyond LEP and LHC, for the influence of new physics 
either through loop corrections or by the 
mixing of new particles with the known ones.  
Second, since Higgs couplings are measured through a global fit that also includes electroweak observables, improvement of our knowledge of precision electroweak also improves our knowledge of the Higgs boson.  For this second purpose, a sample of 10$^8$ Z events with beam polarization, available at 250 GeV using radiative return, is already sufficient.  

The precision measurements are controlled by the statistical uncertainty, the integrated luminosity is of prime importance and the upgradability to higher luminosity is another characteristic considered in many of the projects.

The extent to which a proposed facility is ready for construction depends on the  R$\&$D issues still remaining, as well as --- and in some cases more importantly ---  
on the readiness of a laboratory, or country,  
to include it in its budget and schedule.   
In other cases the R$\&$D may be absolutely critical to confirm basic feasibility. For all the projects considered in the following the main R$\&$D topics and plans are identified.

Environmental sustainability, personnel availability for operation, and carbon footprint are major considerations for the acceptance of a new collider. This need was highlighted also by the EPPSU and will also be taken in consideration.

The power consumption estimates, including the underlying assumptions and level of completeness and maturity, differ significantly between proposals. 
At the ICFA workshop eeFACT'22 \cite{eefact22}, organized at Frascati in September 2022,
a special session was devoted to this theme, with pertinent brief presentations from all e$+$e$^-$ Higgs and Electroweak Factory proposals.
The eeFact'22 discussion and presentations resulted in the power budgets  compiled in Table \ref{tab:eefact}.

\begin{landscape}
\begin{table}[htbp]
\centering
\caption{Electrical Power Budgets for the proposed
Higgs and Electroweak factory colliders, and, for comparison the EIC,
based on invited contributions to the special session at eeFACT'22
\protect\cite{eefact22}. 
NI: Not Included; NE: Not Existing ; n/a: not available.}
\begin{tabular}{l| c c |c c | c c | c | c  | c| c| c c | c}
\hline\hline
Proposal & 	\multicolumn{2}{c|}{CEPC} & \multicolumn{2}{c|}{FCC-ee}
& \multicolumn{2}{c|}{CERC} & C3	& HELEN & CLIC	& ILC$^{*}$
& \multicolumn{2}{c|}{RELIC} & EIC \\
\hline
Beam energy [GeV] & 120	& 180 & 120 & 	182.5 & 	120	& 182.5& 	125 & 	125& 	 190	& 125	& 120& 	182.5 & 10 or 18 \\
Average beam current [mA] & 16.7 & 	5.5	& 26.7	& 5	& 2.47 &	0.9 & 	0.016	& 0.021 & 0.015	& 0.04	& 38 & 39
& 0.23--2.5 \\
Total SR power [MW] & 	60	& 100	& 100	& 100 & 	30& 	30 & 	0& 	0	& 2.87& 	3.6	& 0	& 0
&  9 \\ 
Collider cryo  [MW] & 	12.74& 	20.5 & 	17 & 	50& 	18.8& 	28.8& 	60	& 14.43	& 0	& 18.7 & 	28	& 43 & 12 \\
Collider RF  [MW] 
& 103.8 &	173.0 &	146 & 	146	& 57.8	& 61.8	& 20 & 24.80 &  26.2 	& 42.8	& 57.8	& 61.8  & 13 \\
Collider magnets [MW]	& 52.58	& 119.1 & 	39	& 89	& 13.9 & 	32& 	20	& 10.40& 	 19.5& 	9.5	& 2	& 3  & 25 \\
Cooling \& ventil.~ [MW]	& 39.13& 	60.3 & 
36	& 40 & -- & --	& 		15& 	10.50& 		18.5 & 	15.7	& NE & 	NE & 5  \\
General services  [MW]	& 19.84	& 19.8 & 	36& 	36& -- & -- & 	20& 	6.00	&  5.3& 	8.6& 	NE	& NE & 4 \\
Injector cryo  [MW]	& 0.64 & 	0.6 & 	1	& 1 & -- & -- & 6	& 1.96	&  -- & 	2.8 & 	NE	& NE & 0 \\
Injector RF  [MW]	 & 1.44 & 1.4 & 	2	& 2& -- & 	--	& 5& 	0.00 	& 14.5 &	17.1 &	192& 	196 & 5 \\
Injector magnets [MW]	& 7.45& 	16.8 &  2 &	4  & -- & -- & 4	& 13.07	& 6.2 & 10.1  & -- & -- & 5  
\\		
Pre-injector  [MW] & 	17.685 & 	17.7&  	10& 	10	& --& --& 	n/a	& 13.37	&  NE	& -- & 
NE& NE   & 10  \\
Detector  [MW]	& 4	& 4.0 &	8 &	8 & -- & --	& n/a	& 15.97 & 		2	& 5.7 &	NE&	NE & --   \\
Data center  [MW]	& -	& - & 	4	& 4 & 	-- & -- & 	n/a		& --& 	NI	& 2.7	& NE	& NE & --  \\
% Margin 										3.3	
Total power [MW] & 	259.3 & 	433.3 &	301	& 390 & 	89	& 122 & 	150& 	110.5	& 107& 	138	& 315 & 	341 & 79  \\

Luminosity [$10^{34}$~cm$^{-2}$s$^{-1}$]  &	10.0 & 	1.7 & 	7.7 &	1.3 &	
78 &	28 & 	1.3	& 1.35	& 2.3  & 2.7 & 
398& 	395 & 1 \\
Tot.~integr.~L/yr [1/fb/yr] & 	1300	& 217.1	& 1900& 	330	& 8800	& 4200	& 210 & 	390.7	& 
276 & 430 & 
39800 & 	39500 & 145 \\
Eff.~physics time / yr % assumed/needed  to achieve integrated annual luminosity 
[$10^7$~s] 	& 1.3	& 1.3	& 1.2	& 1.2 & 	1	& 1& 	1.6&	2.89 & 	 1.2 & 1.6 & 		2	& 2 & 1.45\\
Energy cons./yr [TWh] &	0.9 &	1.6& 	1.51	& 1.95&	0.34 &	0.47&	0.67&	0.89	&	
0.6 & 0.82 & 2 &	2.2 & 0.32\\
\hline\hline
\multicolumn{14}{l}{ $^{\ast}${\small {${\mathcal{L}}$ Upgrade}} } \\
\end{tabular}
\label{tab:eefact}
\end{table}
\end{landscape} 

Although fundamental scientiﬁc questions motivate the new facilities, broader impact on society through technological innovations are also expected. These aspects will also be addressed.

%%%%%%%%%%%%%%%%%%%%%%%%%%%%%%%%%%%%%%%%%%%%%%%%%%%%%%%

\newpage
\subsection{International Linear Collider (ILC) \cite{ILC} } 

The ILC is a proposed next generation linear e$^{+}$e$^{-}$ collider, under development by an international collaboration and to be hosted in Japan. The main objective is the Higgs precision study at $\sqrt{s}=$ 250 GeV.  The linear design, based on 1.3 GHz superconducting radio-frequency (SCRF) allows further phases at lower and higher energies: 91.2 GeV for the $Z$ resonance, 500-550 GeV for $t\bar{t}$ and 1 TeV for for the Higgs self-coupling measurement and many new physics searches. Further improvements on SCRF cavity technology will allow 3-4 TeV. Future technologies as plasma wakefield or dielectric laser accelerators will allow tens of TeV energy ranges. Furthermore, at the ILC beam polarization for either e$^{+}$ and e$^{-}$ should be provided. Ancillary experiments with beam dump and/or near IP detectors can be also hosted, making available the most intense highest e$^{+}$ and e$^{-}$ beams for beam dump and fixed target experiments to search the light weakly interacting particles.

\subsubsection{Design outline}
\begin{figure}[h]
\centering
\includegraphics[width = 0.75\textwidth]{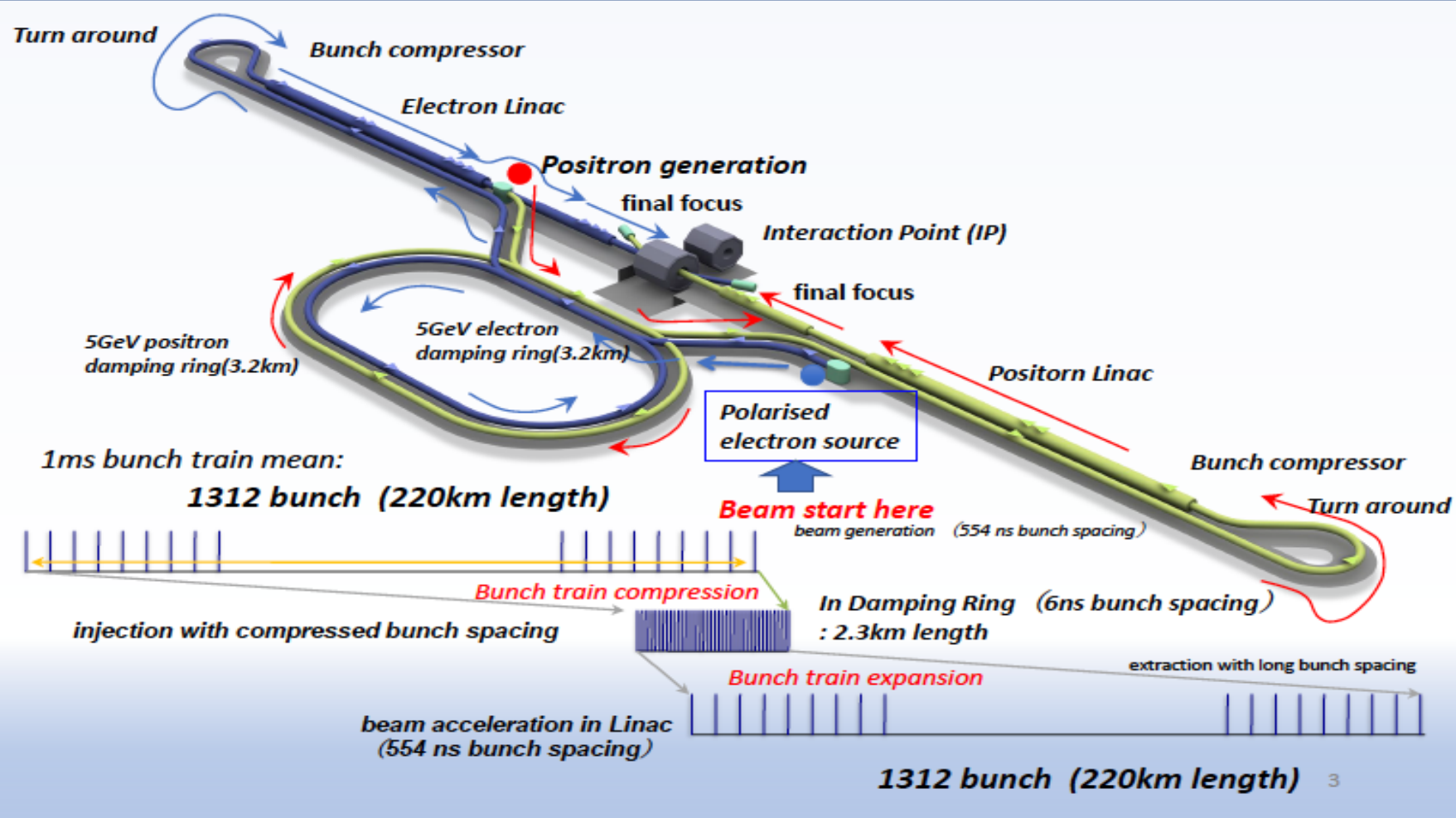}
\caption{Baseline ILC beam accelerator sequence.}
\label{ILC-sequence}
\end{figure}

 The ILC is a 250 GeV e$^{+}$e$^{-}$ linear collider (extendable up to 1 TeV), based on 1.3 GHz SCRF technology, designed to achieve 1.35  10$^{34}$cm$^{-2}$s$^{-1}$ (400 fb$^{-1}$ 4 years running). e$^{-}$ beams will be polarized to 80 $\%$ and e$^{+}$ to 30 $\%$ if undulator based positron source concept is used. 
 The design is governed by the goal of high power-efficiency. The overall power consumption is 111 MW for 250 GeV (limited to 300 MW at 1 TeV), by using SCRF cavities at 1.3 GHz powered by commercial klystrons. The accelerating gradients are in the range of 31.5 to 35 MV/m at 2 K. These values are the result of an optimization of overall efficiency and reasonable investments costs. The SRF technology is mature with a broad industrial base throughout the world. In recent years big improvements in material preparation, high-gradients and high-quality factors have been made. The main parameters for all the options, including the upgrades are summarized in Table \ref{ILC-tab}. The base beam accelerator sequence is shown in Figure \ref{ILC-sequence}.
 
\begin{landscape}
\centering
\begin{table}[tbhp] 
\caption{Summary table of the ILC accelerator parameters in the initial 250 GeV staged configuration  and possible upgrades. A 500 GeV machine could also be operated at 250 GeV with 10 Hz repetition rate, bringing the maximum luminosity to 5.4 10$^{34}$cm$^{-2}$s$^{-1}$. For operation at the $Z$-pole additional beam power  of 1.94/3.88 MW is necessary for positron production.}

\begin{tabular}{lcccccccc}
& & & & & & & &  \\
Quantity & Symbol & Unit & Initial & ${\mathcal{L}}$ Upgrade & Z pole & \multicolumn{3}{c}{${\mathrm{E}}$ / ${\mathcal{L}}$ Upgrades} \\
\hline
Centre of mass energy & $\sqrt{s}$ & ${\mathrm{GeV}}$ & $250$ & $250$ & $91.2$ & $500$ & $250$ & $1000$ \\
Luminosity & ${\mathcal{L}}$ & $10^{34}{\mathrm{cm^{-2}s^{-1}}}$ & $1.35$ & $2.7$ &  $0.21/0.41$ & $1.8 / 3.6$ & $5.4$ & $5.1$ \\
Polarization for $e^-/e^+$ & $P_{-}(P_{+})$ & $\%$ & $80(30)$ &  $80(30)$ & $80(30)$ & $80(30)$ & $80(30)$ & $80(20)$ \\ 
\hline
Repetition frequency &$f_{rep}$ & ${\mathrm{Hz}}$  & $5$ & $5$ &  $3.7$ & $5$ & $10$ & $4$ \\
Bunches per pulse  &$n_{bunch}$ & $1$  & $1312$ & $2625$ &  $1312/2625$  & $1312 / 2625$ & $2625$ & $2450$ \\
Bunch population  &$N_{e}$ & $10^{10}$ & $2$ &  $2$ &  $2$ & $2$ & $2$ & $1.74$ \\
Linac bunch interval & $\Delta t_{b}$ & ${\mathrm{ns}}$ & $554$ & $366$ &  $554/366$ & $554 / 366$ &  $366$ &$366$ \\
Beam current in pulse & $I_{pulse}$ & ${\mathrm{mA}}$& $5.8$ & $8.8$ &   $5.8/8.8$ & $5.8/8.8$ & $8.8$ & $7.6$  \\
Beam pulse duration  & $t_{pulse}$ & ${\mathrm{\mu s}}$ & $727$ & $961$ & $727/961$ & $727 / 961$ & $961$ & $897$ \\
Accelerating gradient & $G$ & ${\mathrm{MV/m}}$ & $31.5$ & $31.5$ & $31.5$ & $31.5$ & $31.5$ & $45$ \\
Average beam power  & $P_{ave}$   & ${\mathrm{MW}}$ & $5.3$ & $10.5$ &   $1.42/2.84^{*)}$ &  $10.5 / 21$  & $21$ & $27.2$ \\ 
\hline
RMS bunch length  & $\sigma^*_{z}$  & ${\mathrm{mm}}$ & $0.3$ & $0.3$ & $0.41$ & $0.3$ & $0.3$ &  $0.225$ \\
Norm. hor. emitt. at IP & $\gamma\epsilon_{x}$ & ${\mathrm{\mu m}}$& $5$ & $5$ & $5$ &  $5$ & $5$ & $5$  \\
Norm. vert. emitt. at IP & $\gamma\epsilon_{y}$ & ${\mathrm{nm}}$ & $35$ & $35$ &  $35$ &  $35$ & $35$ & $30$ \\
RMS hor. beam size at IP  & $\sigma^*_{x}$ & ${\mathrm{nm}}$  & $516$ & $516$ &  $1120$ & $474$ & $516$ &  $335$ \\
RMS vert. beam size at IP &$\sigma^*_{y}$ & ${\mathrm{nm}}$ & $7.7$  & $7.7$  &   $14.6$ & $5.9$ & $7.7$  & $2.7$ \\
Luminosity in top $1\,\%$ & ${\mathcal{L}}_{0.01} / {\mathcal{L}}$ &  & $73\,\%$  &  $73\,\%$ &  $99\,\%$ & $58.3\,\%$ & $73\,\%$ & $44.5\,\%$\\
Beamstrahlung energy loss & $\delta_{BS}$ &  & $2.6\,\%$  & $2.6\,\%$  &   $0.16\,\%$ & $4.5\,\%$ &$2.6\,\%$  & $10.5\,\%$ \\
\hline
Site AC power  & $P_{site}$ &  ${\mathrm{MW}}$ & $111$ & $138$ &   $94/115$ & $173 / 215$ & $198$ & $300$ \\
Site length & $L_{site}$ &  ${\mathrm{km}}$ & $20.5$ & $20.5$  &  $20.5$ & $31$ & $31$ & $40$ \\ \hline
\end{tabular}
\label{ILC-tab}
\end{table}
\end{landscape}

 \paragraph{SRF Technology}
 The ILC linacs are based on TESLA technology: 1.3 GHz nine-cell SC cavities made of niobium and operated at 2K. Pulsed klystrons supply the necessary RF power to the cavities by means of one input coupler per cavity and the corresponding waveguide power distribution.
 \begin{itemize}
 \item The TDR average gradient is 31.5 MV/m with 20 $\%$ spread between individual cavities and with a quality factor of 1.0 10$^{10}$. Recent progress in high-gradient R$\&$D raises the gradient to 35 MV/m with 1.6 10$^{10}$ quality factor. It should be noted that operating cost rise when the gradient increases. 
 
 \item The choice of the frequency is the result of a compromise between the higher cost of larger, lower-frequency cavities and the increased cost at higher frequencies associated with the lower sustainable gradient from increased surface resistivity. 1.3 GHz was chosen due to the commercial availability of high-power klystrons.
 
\item SRF ILC cavities are nine-cell structures (1.25 m) made of high-purity niobium. Cavities are produced from niobium ingots, the cavity cells are fabricated by deep-drawing the sheets into half cells joined by electron beam welding. After welding the inner part is prepared and treated by electropolishing or buffered chemical polishing. Being one of the major cost drivers, this process  is being optimized since the TDR. New treatment methods as the nitrogen infusion or two step baking  are being implemented and gradients near of 50 - 60 MV/m could be envisaged for future upgrades. Other more aggressive approaches as the use of Travelling Wave (TW) Structures with the possibility of achieving 70 MV/m are also under study.

\item Power coupler design is the result of an optimization of the Tesla Test Facility (TTF) and applied massively at the European XFEL. A lot of experience has been gained in this production.

\item Cryomodules accommodate 8/9 cavities (12.6 m), thermally insulated, and contain all the necessary tubes and supply for liquid helium 2-80 K. Cryomodules operate at 2 K and are cooled by superfluid helium. Nine cryomodules are connected to form a cryostring with a common helium supply. The liquid helium is supplied by several cryogenic plants. The RF power for each of these strings is provided by two klystrons (three for the luminosity upgrade) klystron. A "plug-compatible" design ensure the components and interchanges between the modules from different companies.

\item ILC design foresees the use of novel solid state Marx Modulators, being used successfully at KEK. The RF power is provided by 10 MW L-band multi-beam klystrons with a 65 $\%$ efficiency. Recently the new developments in High-Efficiency klystrons promise increased efficiencies. In the baseline design a single RF station (modulator + klystron) supplies 4.5 cryomodules (39 cavities). All cavities from the 9-cavity module and half of the 8-cavity module are connected to one Local Power Distribution System (LPDS), and three LPDS units to one klystron. This design is a cost-effective solution with minimal losses and enough flexibility to optimise the power during operation or to refurbish for luminosity upgrades.

\end{itemize}

\paragraph{Accelerator Design}

\begin{itemize}

\item The e$^{+}$e$^{-}$ sources are designed to produce 5 GeV beam pulses with a bunch charge of 50 $\%$ higher than the nominal bunch charge 3.2 nC (2 10$^{10}$). The e$^{-}$ source is based on SLC polarized e$^{-}$ source. The long bunch ILC train require a newly developed laser and powerful pre-accelerator structures, for which preliminary designs are available. 85 $\%$ polarization is expected at the source, enough for 80 $\%$ at the IP.
The baseline e$^{+}$ source is based on hard, circularly polarized photons produced by a helical undulator driven by the main e$^{-}$ beam at 126.5 GeV and with a positron yield of 1.5 e$^{+}$ / e$^{-}$. Positrons inherit a longitudinal polarization of 30 $\%$ from the circularly polarized photons. An intensive R$\&$D program has undergone various modifications and improvements on: the target heat load, radiation load in the flux concentrator, photon dump and specially the e$^{+}$ / e$^{-}$ yield. These studies have demonstrated the required performances and, in particular, the e$^{+}$ / e$^{-}$ yield is technically feasible. The level of maturity of the positron source for the ILC is fairly advanced and the R$\&$D is ongoing to optimize this option. As an alternative a conventional e$^{-}$-driven source (3 GeV e$^{-}$) has been developed. This alternative low-risk option offers some advantages from the flexibility operation point of view, but no polarization is provided.

\item The e$^{+}$e$^{-}$ polarisation vector  is rotated into transverse plane (perpendicular to the damping ring (DR) plane) before entering the DR, and rotated back to the longitudinal at the end of the RTML, to avoid depolarisation in the damping rings and beam transport. The possibility to flip spins for each beam independently on a pulse to pulse basis ensures an effective control of systematic effects.

\item Damping rings for e$^{+}$e$^{-}$  share a common tunnel of 3.2 km circumference with a normalized horizontal / vertical/horizontal emittance of 4 $\mu$ / 20 nm in 100 ms. These values are  rather conservative compared to the values for the 4th generation Synchrotron Radiation sources. 54 wigglers in each ring provided the needed small damping times. 650 MHz SCRF cavities provide the RF. Fast Injection and extraction bunch by bunch devices with a rise/fall time of around 3 ns will be used. Although, the values have been demonstrated, ongoing R$\&$D is being made to  improve its performances.

\item Ring-to-Main Linac transport consists in two arms of 14 km each for e$^{+}$e$^{-}$ consisting in DR extraction line, long low-emittance transfer (including collimators), turn around section, spin rotation and diagnostic section. The system has been optimized from the point of view of cost-effective and emittance preservation. 

\item Two main linacs (ML) accelerate the beams from 5 to 125 GeV. The first part of the ML is a two-stage bunch compressor to reduce the bunch length from 6 to 0.3 mm, the main linac continues with 6 km of cryomodules.  Cryomodules comprise nine cavities or eight cavities plus a quadrupole/corrector/beam position monitor unit, and all necessary cryogenic supply lines.

\item Beam Delivery System (BDS) is 2.254 km long and comprises: diagnostic and collimation section, followed by the Final Focus System (FFS). The FFS demagnifies the beam sizes to 516 /7.7 nm horizontal/vertical respectively by means of SC final focus doublet of quadrupoles. To bring the beams to the collision with  nanometer accuracy requires a feedback system to compensate drifts and vibrations effects. The BDS system is designed such that it can be upgraded to 1 TeV. The FFS design, in particular the Local Chromaticity Correction (LOC) has been validated in ATF2 in KEK. ATF2 operates at 1.3 GeV and its design goal is to achieve 37 nm vertical beam size with nm stability, this value corresponds to the 7.7 nm vertical beam size at 125 GeV for ILC.  A vertical beam size of 41 nm, which essentially satisfies the ATF2 design goal, has been produced at ATF2, with a bunch population of approximately 10$\%$ of the nominal value of 10$^{10}$ e$^{-}$ and with a reduced aberration optics. Recent studies indicate that the vertical beam size growth with the beam intensity owing the effects of wakeﬁelds
is acceptable at the real ILC.  Concerning the feedback, the 5th generation of the FONT5 feedback system has been tested successfully at ATF2, where a beam stabilisation of 41 nm has been demonstrated in excellent agreement with the predicted one given bunch jitter and bunch-to-bunch correlation. R$\&$D on tunability and long-term stability is going on in ATF2 to overcome these apprehensions and to maximize the luminosity potential of ILC.

\item 3.9 GHz crab cavities to rotate the bunches to compensate the 14 mrad beam crossing angle are envisaged. An intense R\&D is going on to optimize the design and performances of these devices.

\item Machine Detector Interface (MDI), two detectors are sharing the Interaction Point (IP) in push-pull configuration. The innermost FF quadrupole combined with a LOC sextupole is installed inside the detectors at 4.1 m from the IP. In contrast to TDR a large vertical access shaft (CMS style) is foreseen. 

\item Main beam dumps are designed to stand a maximum power of 17~MW enough for 1 TeV upgrade of the accelerator. The design is based on the SLAC 2.2 MW beam dump. A photon beam dump for the e$^{+}$ undulator scheme is under design.

 \end{itemize} 
 
 \paragraph{Civil Engineering and site}

\begin{itemize}

\item The Kitakami site (Tohoku region) was largely selected because of its excellent geological conditions (homogeneous granite formation). The site provides up to 50 km space, enough for a possible  1TeV upgrade or more. Proven technologies will be used to cope with seismic events, including magnitude 9 earthquakes. 

\end{itemize}

\paragraph{Sustainability}
ILC is based on SCRF which is more efficient than the NCRF in terms of energy consumption. ILC consumption is 111 MW at 250 GeV, 163 MW at 500 GeV and 300 MW at 1 TeV. Table \ref{ILC-tab} summarize the energy consumption for the  ILC baseline and the upgrades. In order to reduce the energy consumption the Advance Accelerator Association (AAA) has created the "Green-ILC working group" (WG) that collaborates with the ILC to study the efficient design of ILC components, accelerator sub-systems, overall system design, city hosting and laboratory campus.

\begin{itemize}

\item  Various proposals for green ILC components have been made. Some examples are: efficient refrigerators, recuperating the waste heat from the refrigerators to be used in the heat circuit (7 $\%$ reduction); efficient power sources as Solid State Amplifiers (SSA) (now available at lower prices) or high-efficiency klystrons (85 $\%$); high-Q and high-gradient cavities types; new dumps design using wake-field deceleration.

\item The energy consumption of ILC depends on the operation mode, switching between full beam, reduced beam, standby and stop. The various modes could be scheduled according to the available regenerative resources, electricity cost and demand for electric power in the ILC region site. Use of pre-chilled water and or liquid helium could help in the modulation.

\item A complete study has been made for the a "green ILC city and campus" in the framework of the AAA Green ILC WG. Some of the concepts are: smart power grid, including solar power farms and biomass power network making use of waste heat from ILC tunnel.

\item ILC is expected to emit 320 kilotons of CO$\_{2}$ per year (871 kilotons for Ichinoseki city close to ILC site), giving the fact that forests in the area can absorb 300 kilotons/year ILC lab in collaboration with local authorities should work on a mitigation plan.

\end{itemize}

Studies in the sustainability area will be synergetic with any future accelerator collider. In particular, there are clear plans for future work with CLIC regarding sustainability and power/energy optimisation.

\subsubsection{Proposals for upgrades or extensions}
 Two types of upgrades/extensions are considered: luminosity upgrades and energy upgrades.
 
 \paragraph{Luminosity upgrades}
 Luminosity upgrades are based on baseline existing technology. ILC Luminosity could be improved by increasing the charge per bunch or by increasing the number of bunches per second. Increasing the brightness per bunch will require smaller vertical beam sizes (tighter focusing) or lower emittances, given the high-risk of this option this is not considered at this stage.
 
 \begin{itemize}
 
\item Doubling the number of bunches per pulse from 1312 to 2625 will require to decrease the bunch separation from 554 ns to 366 ns, leading to an increase of the beam current per pulse from 5.8 mA to 8.8 mA, which will require the installation of 50$\%$ more klystrons and modulators. The cryogenic load is unchanged since RF pulse duration is not changed. Beam pulse duration will increase  from 714 $\mu$s to 961 $\mu$s. Doubling the number of bunches would double the beam current in DRs, e$^{+}$ rings could suffer from electron cloud instabilities. Tunnel is large enough to accommodate a third ring if needed.
All details are found in Table \ref{ILC-tab}.
 
\item In the 500 GeV configuration (longer ML) if operated at 250 GeV, the pulse repetition rate could be increased from 5 Hz to 10 Hz, further doubling the luminosity. But some upgrades are also needed on different sub-systems. DR RF and wigglers magnets have to be reinforced. Klystrons, bunch compressors and the ML could operate at 10 Hz. The ML power and the cryogenics can accept this mode because accelerating gradient is low. The positron source must be improved for higher heat load.
 
\end{itemize}
  
\paragraph{Energy extension and upgrades}
The energy upgrade is one of the big advantages of the linear colliders. In principle the ML could be extended in length, turnaround and compressors have to be moved. ILC BDS and dump have been designed to be easily-modify to operate at 1 TeV. Any expansion can be accomplished by adding new cryomodules at the low-energy ends of the accelerators without need of moving already installed modules.

  \begin{itemize}
 \item Running at Z-pole CM energy of 91.2 GeV is possible, if the  e$^{+}$ production is adapted  to this energy. In the case of the undulator scheme a re-configuration of the operation mode is needed to have the an e$^{-}$ beam at 125 GeV available to produce the e$^{+}$. This could be achieved by operation the e$^{-}$ at twice the repetition frequency (3.7Hz as in this case as in Table \ref{tab:eefact} due to the total power limit), alternating for physics and for e$^{+}$ production. 
 Some other issues as : DRs shorter damping time,  MLs operated at low-gradients, collimation, wakefields and beam-beam, have also been studied. In the case of the e$^{-}$-driven scheme simple configuration is possible, but polarization is not possible.
      All performances are summarized in Table \ref{ILC-tab}.
       
\item Increasing the beam energy from 250 GeV to 500 GeV   will require the extension of the SCRF linear accelerators resulting in a 31 km overall site length. By using similar SCRF technology as for 250 GeV (31.5 MV/m, $Q_{0}$ = 1.0 10$^{10}$). Given the fact that, the TDR baseline energy was 500 GeV, no major issues are identified. Some concerns are:  polarization, relocation of RTML elements and BDS  are no major issues.    
      
\item Increasing the beam energy  from 500 GeV to 1 Tev will require a further extension of the SCRF linear accelerators. Given the ongoing R$\&$D on high-gradient / high-Q, it is envisaged to achieve at that time 45 MV/m accelerating gradient and $Q_{0}$ = 2 10$^{10}$, which results in  a 40 km linac length. Taking the 500 GeV values a comprehensive scaling has been made and the performances are summarized in Table \ref{ILC-tab}. Some considerations are: wall plug-power limit (300 MW), beam current compatible with injectors, DRs and MLs, acceptable beamstrahlung losses (10 $\%$) and positron source shorter undulator (longer period, smaller field).
\end{itemize}

\subsubsection{Stageability to future experiments}
An extension of the Physics program has been made recently, for instance: using the main dump, the extracted beam or the far detector between others. All details in the Physics sections of \cite{ILC}.

\subsubsection{State of Technical Design Report (TDR)}
The TDR has been published in 2013 \cite{ILC}. Since then an intense R$\&$D program is ongoing. A detailed description of this program is given in the section: State of Proposal and R$\&$D plans.

\subsubsection{State of Proposal and R$\&$D plans}
\begin{figure}[h]
\centering
\includegraphics[width = 0.75\textwidth]{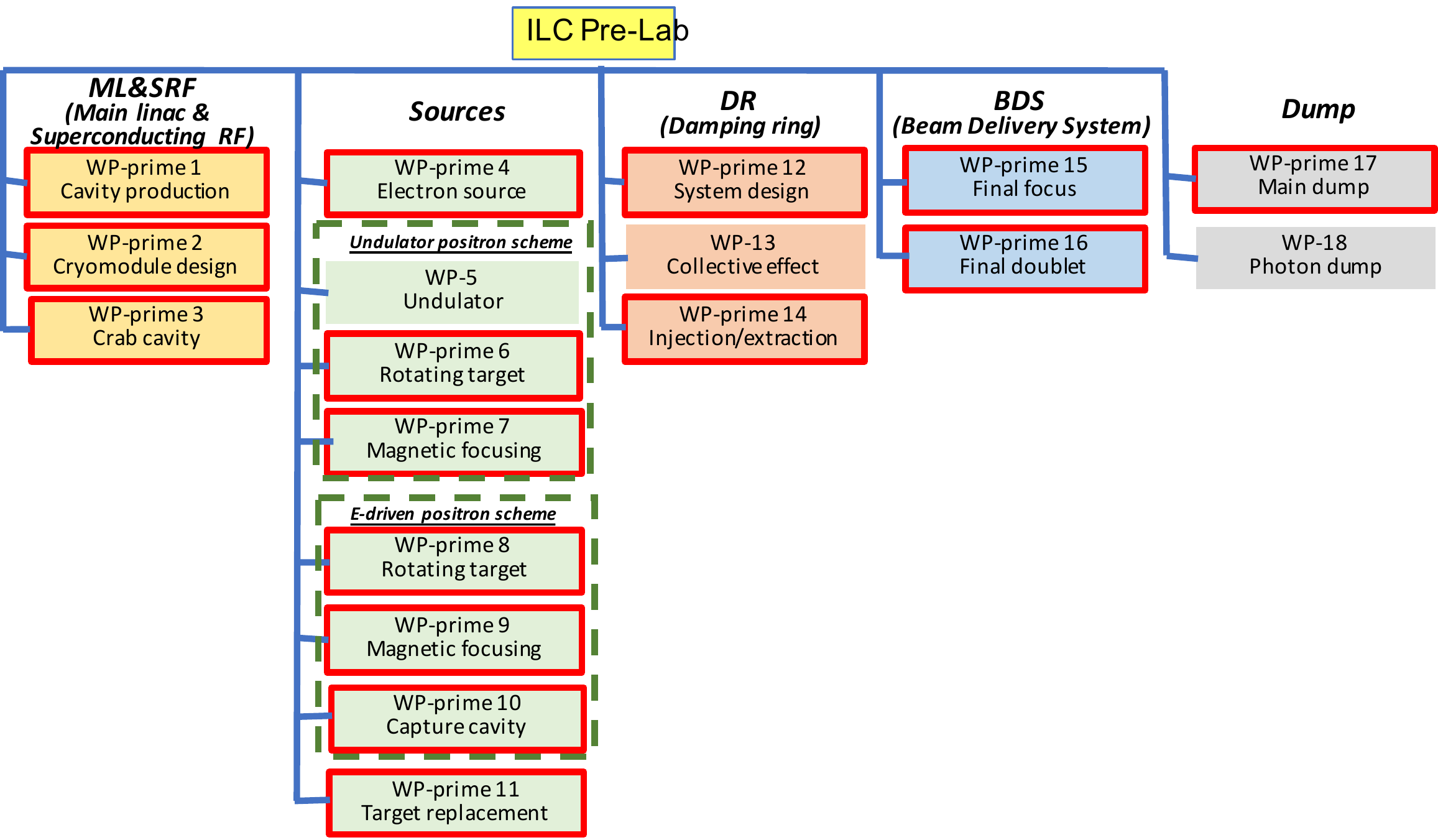}
\caption{Summary of the WPs for technical preparation of ILC Pre-lab.}
\label{ILC-IDT-WG2}
\end{figure}

The technical basis of the ILC was fully documented in the TDR. Still three issues need to be studied: revisiting/understanding/updating the recent SRF recent R$\&$D results, including cost; issues with the specific Tohoku site; and finally remaining technical issues. In order to resolve these issues the International Development Team (IDT) has been created by International Committee for Future Accelerators (ICFA) in August 2020. In particular, the WG2 is identifying the accelerator related activities for the ILC Pre-lab necessary to before starting the the ILC construction. The ILC Pre-lab activities are expected to continue about 4 years. A summary of the Working Packages (WPs) is illustrated in Figure \ref{ILC-IDT-WG2} and described in detail in \cite{ILC}. The technical preparation document was reviewed by the international review committee. The total global cost of the Pre-Lab project is about 60 MILCU (1MILCU=1M$\$$ in 2012) and about 360 FTE-year (cost of the infrastructure for the WPs not included). The cost will be shared internationally as in-kind contribution.
 
The explicit tasks of the WPs are:
\begin{itemize}
\item WP1 ML$\&$SRF: Cavity Industrial Production Readiness.
\item WP2 ML$\&$SRF: Cryomodule Assembly, Global Transfer and Performance Assurance.
\item WP3 ML$\&$SRF: Crab Cavity for BDS including cryomodule.
 
\item WP4 Sources: e$^{-}$ Sources driven lasers and GaAs/GaAsP photocathodes.
\item WP5 Sources: Undulator e$^{+}$ Source simulations.
\item WP6 Sources: Undulator e$^{+}$ Source rotating target (magnetic bearing, radiative cooling).
\item WP7 Sources: Undulator e$^{+}$ Source magnetic focusing (pulsed solenoid, yield calculations).
\item WP8 Sources: e$^{-}$-driven e$^{+}$ Source rotating target (water cooling, vacuum seal).
\item WP9 Sources: e$^{-}$-driven e$^{+}$ Source rotating target flux concentrator.
\item WP10 Sources: e$^{-}$-driven e$^{+}$ Source rotating capture system (Alternative Periodic Structure APS, capture linac, solenoid).
\item WP11 Sources: Target maintenance (undulator and e$^{-}$-driven).
\item WP12 Damping Rings: optics and permanent magnets.
\item WP13 Damping Rings: collective effects (e$^{-}$-cloud, ion-trapping, Fast Ion stability and fast feedback).
\item WP14 Damping Rings: Injection/Extraction devices (fast kicker and e$^{-}$ driven kicker).
\item WP15 BDS: FFS system design ( ATF3: Long-term stability, High-order aberrations) 
\item WP16 BDS: FFS doublet design optimization
\item WP17 Beam Dump: Main beam dump system design (water flow optimization, window sealing, failure and safety system)
\item WP18 Beam Dump: photon dump for undulator system (window water dump, graphite dump)
 \end{itemize}

A prioritization panel has identified the  most time-critical and essential work packages for ILC construction, compiled in order to address key technology issues for a next electron positron collider in the most efficient manner, with a concerted international effort.

%%%%%%%%%%%%%%%%%%%%%%%%%%%%%%%%%%%%%%%%%%%%%%%%%%%%%%%
\newpage

\subsection{Compact Linear Collider (CLIC)  \cite{CLIC}}

The CLIC is a proposed multi-TeV high-luminosity e$^{+}$e$^{-}$ collider under development by an international collaboration and to be hosted at CERN. CLIC has been optimized for three energy stages at $\sqrt{s}=$ 380, 1500 and 3000 GeV. The linear design based on 12 GHz (X-band) high-gradient normal conducting RF (NCRF), uses a novel two-beam acceleration technique as power supply. An alternative design based on X-band klystrons for the 380 GeV has also been studied.
The baseline 380 GeV CM energy gives access to SM Higgs and top quark physics, providing direct and indirect sensitivity to BSM effects. The second stage at 1.5 TeV open more Higgs production channels and rare decays and allows further sensitivity to BSM effects. Finally the ultimate stage at 3 TeV gives the best sensitivity to the Higgs self-coupling and new physics scenarios. Furthermore, polarization for e$^{-}$ should be provided.

\subsubsection{Design outline}

\begin{figure}[h]
\centering
\includegraphics[width = 0.75\textwidth]{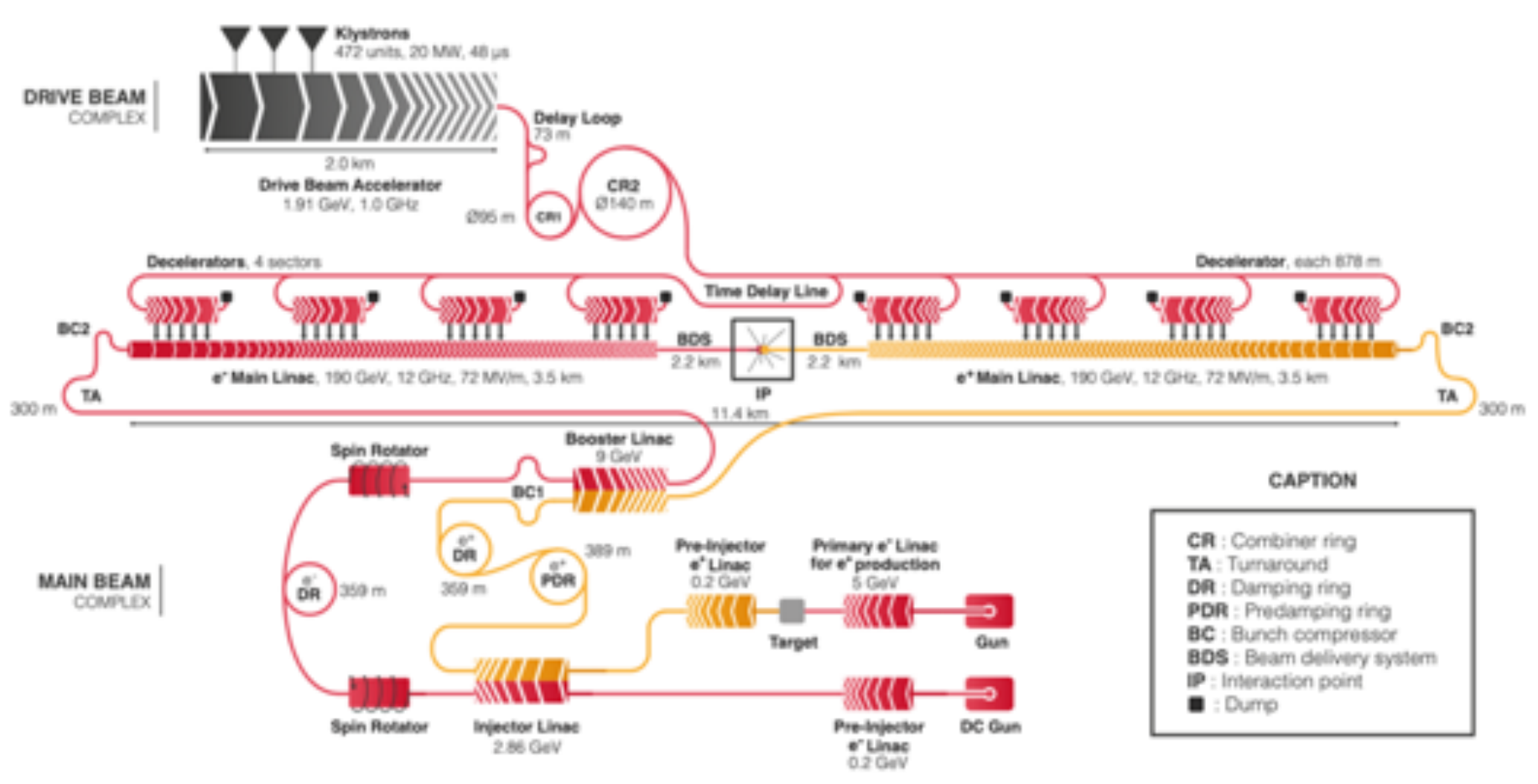}
\includegraphics[width = 0.5\textwidth]{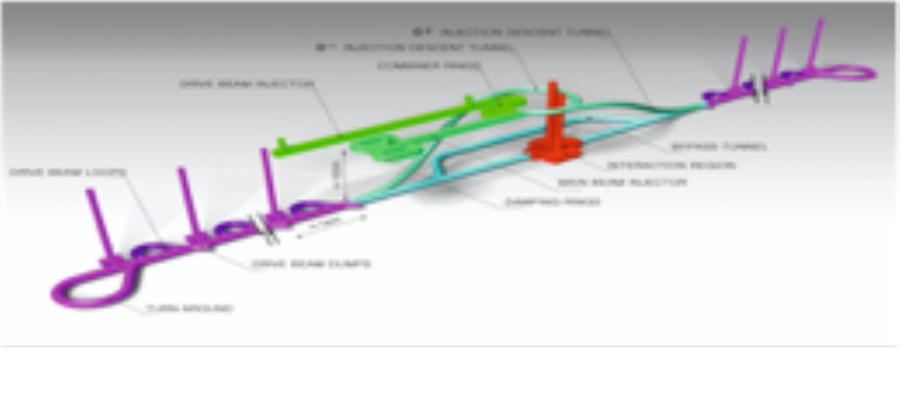}
\caption{Baseline CLIC beam accelerator sequence and layout.}
\label{CLIC-sequence}
\end{figure}

The CLIC is a 380 GeV e$^{+}$e$^{-}$ linear collider (extendable up to 3 TeV), based on 12 GHz (X-band) NCRF, designed to achieve 2.3  10$^{34}$cm$^{-2}$s$^{-1}$ (1.5 ab$^{-1}$ 8 years running). e$^{-}$ beams will be polarized to 80 $\%$. 
 The design is governed by the goal of acceptable length and affordable cost by using high accelerating gradients of 72 MV/m and larger beams currents for 380 GeV CM energy (100 MV/m for 1.5 and 3 TeV). The overall power consumption is 110 MW for 380 GeV in the two powering options (two-beams or high-efficient klystrons). The NCRF X-band technology is mature and an industrial base is being developed. The main parameters for the three energy stages, are summarized in Table \ref{CLIC-tab}. The base beam accelerator sequence and layout are shown in Figure \ref{CLIC-sequence}.

\begin{table}[tbhp] 
\caption{Summary table of the CLIC accelerator parameter for the different energy stages. The estimated power for 1500 and 3000 GeV has not optimized. }
\begin{center}
\begin{tabular}{lccccc}
Quantity & Symbol & Unit & Stage 1 & Stage 2 & Stage 3  \\
\hline
Centre of mass energy & $\sqrt{s}$ & ${\mathrm{GeV}}$ & $380$ & $1500$ & $3000$ \\
Luminosity & ${\mathcal{L}}$ & $10^{34}{\mathrm{cm^{-2}s^{-1}}}$ & $2.3$ & $3.7$ &  $5.9$ \\
Polarization for $e^-$ & $P_{-}$ & $\%$ & $80$ &  $80$ & $80$  \\
\hline
Repetition frequency &$f_{rep}$ & ${\mathrm{Hz}}$  & $50$ & $50$ &  $50$  \\
Bunches per train  &$n_{bunch}$ & $1$  & $352$ & $312$ &  $312$  \\
Bunch separation  & $\Delta t_{b}$ & ${\mathrm{ns}}$ & $0.5$ & $0.5$ &  $0.5$  \\
Bunch population  &$N_{e}$ & $10^{9}$ & $5.2$ &  $3.7$ &  $3.7$ \\
Accelerating gradient & $G$ & ${\mathrm{MV/m}}$ & $72$ & $72/100$ & $72/100$  \\
\hline
RMS bunch length  & $\sigma^*_{z}$  & $\mu{\mathrm{m}}$ & $70$ & $44$ & $44$  \\
RMS hor. beam size at IP  & $\sigma^*_{x}$ & ${\mathrm{nm}}$  & $149$ & $60$ &  $40$ \\
RMS vert. beam size at IP &$\sigma^*_{y}$ & ${\mathrm{nm}}$ & $2.0$  & $1.5$  &   $1.0$ \\
RMS energy spread at IP &$\sigma^*_{\epsilon}$ & $\%$ & $0.35$  & $0.35$  &   $0.35$ \\
Crossing angle at IP & $\theta_{c}$ &  ${\mathrm{mrad}}$ & $16.5$  & $20$  &   $20$ \\
\hline
Site AC power  & $P_{site}$ &  ${\mathrm{MW}}$ & $110$ & $364$ &   $589$  \\
Site length & $L_{site}$ &  ${\mathrm{km}}$ & $11.4$ & $29.0$  &  $50.1$ \\ \hline
\end{tabular}
\label{CLIC-tab}
\end{center}
\end{table}

\paragraph{NCRF Technology}
The CLIC main linacs are based on X-band copper cavities, using a novel two-beam accelerating technique to supply the necessary RF power to the main cavities by means of a waveguide power distribution. An alternative option with conventional pulsed klystrons have also been studied.

\begin{itemize}
 \item The main linac accelerating structures for 380 GeV, have to accelerate a train of bunches with a gradient 72 MV/m with a breakdown rate (BDR) less than 3 10$^{-7}$ m$^{-1}$.  The structures are Travelling Wave built in micron precision tolerances with a tapered inner aperture diameter ranging from 8.2 to 5.2 mm and approximately 25 cm length. They include damping features to suppress the transverse wakefields and they are equipped with special BPMs so-called wake-field monitors, in order to measure and correct micron-level misalignments. Being an important contribution to the overall cost a cost reduction effort is on-going in collaboration with industrial partners.
 
 \item The RF power for the main cavities is provided in the baseline option  by a high-current low-energy drive beam that runs parallel to the main beam through a sequence of power extraction and transfer structures (PETS). The drive beam generates the RF power in the PETS that is ten transferred to the  accelerating structures using a waveguide network. The drive beam is generated by a central  complex at fundamental frequency of 1 GHz. The drive beam is accelerated in the drive beam linac to 1.91 GeV. A complicated gymnastic in the delay loop and a series of combiner rings fabricate the appropriate bunching to feed the drive beam decelerator system, which run in parallel to the main colliding beam. This technique has been successfully tested experimentally in the CLIC Test Facility (CTF3) at CERN. This technique strongly reduce the cost and power consumption for 1.5 and 3 TeV energy stages in comparison to the conventional klystron option.
 
 \item An alternative design for the 380 GeV based on X-band klystrons to produce the RF power for the main linac has been studied as back-up solution. This solution increase the cost of the main linac, but avoids the construction cost of the drive-beam complex and makes the linac more modular.Furthermore a larger tunnel to accommodate the klystrons and modulators is needed.  The cost is competitive only at low energies.  An optimized layout and beam dynamics studies have been studied in detail, showing similar performances as the drive-beam case.
 
 \end{itemize}

 \paragraph{Alignment and Stabilisation}
 
 \begin{itemize}
     \item In the case of CLIC special attention has to be paid to the alignment and stability issues. In order to preserve the luminosity the total error budget allocated to the absolute positioning of the major accelerator components is 10-20 $\mu$m (compared to 100-500 $\mu$m at the LHC or HL-LHC). A detailed simulation study and strategy including ground motion and vibration operation effects have been performed. The strategy has been validated experimentally element by element (PACMAN) and in a CLIC module test-bench. Recent studies indicate that the shape of the transmissibility function is more important for luminosity than the integrated RMS displacement. This understanding has triggered the development of adapted ground motion sensors (hard radiation) for the stabilisation. The R$\&$D is ongoing.
 
 \end{itemize}
 
 \paragraph{Accelerator Design}
 
 \begin{itemize}
 
 \item The e$^{+}$e$^{-}$ sources are designed to produce 2.86 GeV beam pulses with a bunch charge  of 5.2 10$^{9}$. The main e$^{-}$ beam is able to provide a polarization of 80 $\%$ at the IP. The e$^{+}$ are produced by channeling in a crystal with a 5 GeV e$^{-}$ beam. The photons heat a second target and produced e$^{-}$e$^{-}$ pairs. The e$^{+}$ are captured and accelerated to 2.86 GeV. The e$^{+}$ are not polarized.
 
\item e$^{-}$ longitudinal beam polarization is rotated into transverse plane (perpendicular to the damping ring (DR) plane) before entering the DR, and rotated back to the longitudinal before entering in the booster linac, to be preserved. 

\item Damping Rings for e$^{+}$e$^{-}$ are 360 m circumference with a normalized horizontal / vertical emittance of  0.5 $\mu$m / 5 nm in 1.2 ms. In the case of the e$^{+}$ a pre-damping ring is used before entering in the DR. The DR contain SC wigglers (Nb-Ti) in each straight section to increase the radiation damping and to reduce the Intra Beam Scattering (IBS) effect. The  resulting vertical emittance is similar to the order of magnitude of the current 4th generation Synchrotron Radiation values. R$\&$D is ongoing  to achieve its challenging performances, in particular the prototyping of longitudinal variable field magnets (tunable permanent combined dipole-quadrupole). The most challenging pulsed magnets are the ones of the DRs. Strip-line kickers specially designed for achieving high-field uniformity and time stability has been specially designed. Inductive adders based on solid-state modulators are used to power the strip-line kickers. A full system has been successfully tested in the ALBA SR. Stability of combined flat-top ripple and drop of the field of $\pm$2 10$^{-4}$ has been achieved.

\item The Ring-to-Main Linac  accelerates the beams until 9 GeV and compresses the bunch length  from 1.8 mm until 70 $\mu$m.

\item Two main linacs (ML) accelerate the beams from 9 to 190 GeV in a 3.5 km sequence of 1456 identical RF modules. Each module supports four pairs of accelerating structures with an active length of 0.46 m and a gradient of 75 MV/m. Quadrupoles are interleaved to form a FODO lattice. In the alternative klystron based option, a two-pack solid-state modulator equipped with two 50 MW X-band commercial klystrons feed each RF module. Two pulse compressors systems with linearising cavities complete the system.

\item The Beam Delivery System (BDS) is 2.2 km and comprises: diagnostic and collimation section, followed by the Final Focus System (FFS). The FFS demagnifies the beam sizes to 149/2.0 nm horizontal/vertical respectively by means of hybrid technology (permanent and electro magnets) final focus doublet of quadrupoles. To bring the beams to the collision with  nanometer accuracy requires a feedback system to compensate drifts and vibrations effects. The BDS system is designed such that it can be upgraded to 1.5 and 3 TeV. As stated before on the ILC section, the FFS system designed is being validated in ATF2.

\item Crab cavities to rotate the bunches to compensate the 16.5 mrad beam crossing angle are envisaged. An intense R\&D is going on to optimize the design and performances of these devices.

\item Machine Detector Interface (MDI), the main change since the CDR in this aspect is the increased distance between the last quadrupole (QD0) and the IP, allowing the magnet to be installed in the tunnel and outside of the detector.

\item Main beam dumps are designed to stand a maximum power of 2.9, 7 and 14  MW enough for the 380, 1500 and 3000 GeV baseline and upgrades of the accelerator respectively.
\end{itemize}

\paragraph{Civil Engineering and site}
\begin{itemize}
 \item The civil engineering design has been optimized for the 380 GeV. The study include: tunnel length and layout, optimized injector complex, access shafts and related structures. A detailed study has also been made for the Klystron option giving the fact, a larger diameter tunnel (10 m instead of 5.6m) is needed. Using the Tunnel Optimization Tool (TOT), developed for CLIC,  a 380 GeV solution extensible for 1.5 and 3 TeV has been found with good geological conditions (molasse). This solution has the advantage to have the injection complex and the experimental located entirely on CERN site.
 
 \end{itemize}

 \paragraph{Sustainability}
 
 Since the CDR the CLIC energy consumption has been significantly reduced compared to earlier estimates. A detailed study has been made for 380 GeV. CLIC consumption is 110 MW at 380 GeV (similar for the klystron option), 364 MW at 1.5 TeV and 589 at 3 TeV. Table \ref{CLIC-tab} summarize the energy consumption for the  CLIC baseline and the energy stages. 
 
\begin{itemize}

\item  A reduction of  35$\%$ for the 380 GeV due to the optimisation of the injectors, the introduction of optimized accelerator structures, the improvement of the RF efficiency, the reduction of the number of klystrons with high-efficiency (70$\%$), not yet included in power estimates and the use of permanent magnets, has been obtained. A further reduction (30$\%$) has been obtained more recently by re-design of the DR RF systems and the prospects for higher-deficiency L-band klystrons for the drive beam. The estimates for higher-energy stages do not yet include these savings and will be revised in future.

\item The energy consumption of CLIC depends on the operation mode, switching between full beam, reduced beam, standby and stop. The various modes could be scheduled according to the available regenerative resources, electricity cost and demand for electric power in the CERN region site. 

\item Studies of "green CLIC" as: smart power grid, including solar power farms and biomass power network making use of waste heat from CLIC tunnel, should be completed.

\item CLIC Carbon footprint analysis has not been made. Studies will be made in near future. 
 
 \end{itemize}

Studies in the sustainability area will be synergetic with any future accelerator collider. In particular, there are clear plans for future work with ILC regarding sustainability and power/energy optimisation. The C$^{3}$ concept obviously has many commonalities with the CLIC klystron driven version also.

\subsubsection{Proposals for upgrades and extensions}

\paragraph{Luminosity upgrades}
\begin{itemize}
 \item At 380 GeV the luminosity could be doubled by doubling the repetition rate from 50 Hz to 100 Hz without major changes but with the increase of the overall power consumption and cost (at 55$\%$ and 5$\%$ respectively).
 The CLIC luminosity is largely determined by the vertical emittance  growth at IP, this value is dominated by the impact of static (misalignment) and dynamic (ground motion) imperfections. Simulation studies of the beam-based tuning including between others tuning bumps as an additional method to reduce the emittance growth, are ongoing in order to review the emittance budgets and hence to increase the luminosity. 
 
 \end{itemize}
 
\paragraph{Energy extension and upgrades}

\begin{itemize}
 \item Operating the 380 GeV complex at Z-pole (91.2 GeV) is possible by reducing the main linac gradient and the bunch charge by a factor 4 but keeping the emittances and bunch length, the expected luminosity is 2.3 10$^{32}$cm$^{-2}$s$^{-1}$. Alternatively, an initial installation of just the linac for the Z-pole with an appropriate optimization of the BDS, would result in a luminosity of 0.36 10$^{34}$cm$^{-2}$s$^{-1}$.
 
 \item CLIC 380 GeV can be easily upgraded to higher energies like 1.5 TeV and 3 TeV stages. Flexibility has been an integral part of the design choices for the first energy stage and the 3 TeV has been the baseline for the  CLIC CDR. In order to minimize the modifi cation to the drive beam complex, the drive beam current is the same at all energy stages. For the upgrade from 380 GeV to 1.5 TeV, only minor modifications of the drive beam complex are needed. The energy increase is achieved by adding more drive-beam modules. The upgrade from 1.5 to 3 TeV requires the construction of a second drive-beam generation complex. The preservation of the beam quality in the main linac is slightly more challenging at higher energies. Collimation system are longer to cope with the higher energies and the FFS longer to limit the SR and emittance degradation in the bending systems. Extraction lines and dumps have to be equipped with new magnets. The impact of the energy upgrades on main-beam injector and DRs is quite small, given the fact that charges for higher energies are smaller.
 In the case of the klystron option higher energies are possible by re-using the klystron-driven accelerating structures and the klystrons and by adding new drive-beam powered structures. Bunch charge will be slightly reduced. An important difference in this case is the placement of the modules given the fact the larger radius of the main linac, instead to be placed at the beginning of the new tunnel, as in the drive-beam case, will remain at the end of the new tunnel close to the BDS.
 
 \end{itemize}

 \subsubsection{Stageability to future experiments}
 
 \begin{itemize}
 \item Studies for allocating a 2nd experimental region/ detector by means of a dual Interaction Region has been performed in detail. 
 
 \item $\gamma \gamma$ collisions up to $\sim$315 GeV are also possible with a luminosity spectrum interesting for physics. Studies are ongoing to have a consolidated physics scenario.
 
 \end{itemize}
 
\subsubsection{State of Technical Design Report}
TDR is not existing, but after the CDR publication on 2012, having as baseline the 3 TeV, and with the discovery of the Higgs-boson, the initial stage was changed to 380 GeV, and a comprehensive technical prototyping program was carried on 2013-2019. This new baseline, the detailed engineering design and, in particular, the pre-series in industry of assembled units for the complete modules, have been reported in the Project Implementation Plan (PIP 2018), together with a a report submitted to the European Strategy Update in 2018-2019. The quality of the PIP approaches the TDR level.

\subsubsection{State of Proposal and R$\&$D plans}
The CLIC study is mature and the basic R\&D challenges has been addressed, however a preparation period period is needed for final engineering design of some aspects.

\begin{itemize}

 \item Performance validation of a X-band linac in a small scale facility (FEL or other applications) will be important to demonstrate the reliability, the technical parameters, the simulation and the modelling tools.
 
 \item Comprehensive luminosity upgrade study for each energy stage.
 
 \item X-band components, cost-effective, industrial fabrication and massive production issues, including conditioning strategy and module integration.
 
 \item Investigations on cost-effective X-band power sources klystrons and  modulators. Including high-efficiency klystrons for X-band and L-band.
 
 \item Thermo-Mechanical studies of main linacs.
 
 \item Hard-radiation ground motion  and vibration sensors.
 
 \item  Complete the optimization of the power consumption at 1500 and 3000 GeV, the "green" CLIC and realize the Carbon footprint study.

 \end{itemize}

%%%%%%%%%%%%%%%%%%%%%%%%%%%%%%%%%%%%%%%%%%%%%%%%%%%%%%%
\newpage
\subsection{Cool Copper Collider (C$^{3}$) \protect\cite{C3,C3_physics} }

 The C$^{3}$ is a recent new proposed linear e$^{+}$e$^{-}$ collide, under development at SLAC, UCLA, INFN, LANL and Radiabeam. The main objective is to carry out precision studies of the Higgs boson at $\sqrt{s}=$ 250 GeV. This new proposed linear collider is based on cold copper with distributed coupling of the RF power. Relatively inexpensive upgrade to 550 GeV  are achievable on the same footprint. This energy upgrade will open the possibility to top quark measurement and will provide a basis for the extension into multi-TeV energy ranges. 3 TeV could be achieved  with the same technology and extending the machine. Polarization for e$^{-}$ should be provided, e$^{+}$ polarization using a similar production as for ILC should be possible as an upgrade.
 
\subsubsection{Design outline}

\begin{figure}[h]
\centering
\includegraphics[width = 0.75\textwidth]{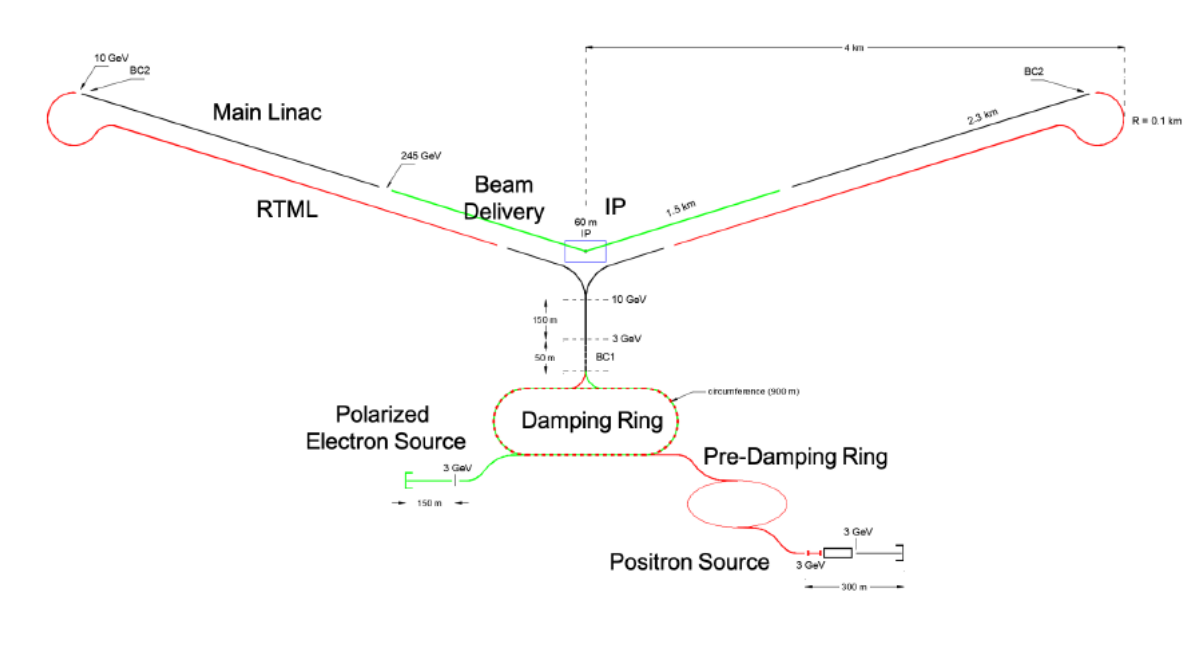}
\caption{Conceptual layout of the C$^{3}$ accelerator complex at 250 - 550 GeV.}
\label{C3-sequence}
\end{figure}

The C$^{3}$ is a 250 GeV e$^{+}$e$^{-}$ linear collider (extendable up to 3 TeV), based on  5.712 GHz (C-band) NCRF, designed to achieve 1.3  10$^{34}$cm$^{-2}$s$^{-1}$ (2 ab$^{-1}$ 10 years running). e$^{-}$ beams will be polarized to 80 $\%$. 
 The design is governed by two aspects, the first is the use of cryogenic copper (77 K) to get high-gradients (70-120 MV/m) and to low breakdown rates and the second is the individual feed to each cavity from a common RF manifold, all in the same copper block. The total length is the 8 km for 250 and 550 GeV (RF power upgrade). The overall power consumption is around 150 and 175 MW respectively. The NCRF C-band cryo-cooled is well grounded but many technical aspects related to cryomodules need a dedicated R$\%$D. The main parameters for the two energy stages, are summarized in Table \ref{C3-tab}. The base beam accelerator sequence and layout are shown in Figure \ref{C3-sequence}.
 
\begin{table}[tbhp] 
\caption{Summary table of the C$^{3}$ accelerator parameter for the different energy stages. Final focus parameters are preliminary.}
\begin{center}
\begin{tabular}{lcccc}
Quantity & Symbol & Unit & Baseline &  ${\mathrm{E}}$  Upgrade \\
\hline
Centre of mass energy & $\sqrt{s}$ & ${\mathrm{GeV}}$ & $250$ & $550$  \\
Luminosity & ${\mathcal{L}}$ & $10^{34}{\mathrm{cm^{-2}s^{-1}}}$ & $1.3$ & $2.4$  \\
Polarization for $e^-$ & $P_{-}$ & $\%$ & $80$ &  $80$   \\
\hline
Repetition frequency &$f_{rep}$ & ${\mathrm{Hz}}$  & $120$ & $120$   \\
Bunches per train  &$n_{bunch}$ & $1$  & $133$ & $75$  \\
Bunch separation  & $\Delta t_{b}$ & ${\mathrm{ns}}$ & $5.26$ & $3.5$   \\
Bunch charge  & $N_b$ & $nC$ & $1$ &  $1$  \\
Accelerating gradient & $G$ & ${\mathrm{MV/m}}$ & $70$ & $120$  \\
\hline
RMS bunch length  & $\sigma^*_{z}$  & $\mu{\mathrm{m}}$ & $100$ & $100$  \\
Norm. hor. emitt. at IP & $\gamma\epsilon_{x}$ & ${\mathrm{nm}}$& $900$ & $900$   \\
Norm. vert. emitt. at IP & $\gamma\epsilon_{y}$ & ${\mathrm{nm}}$ & $20$ & $20$  \\
hor. beta function at IP  & $\beta^*_{x}$ & ${\mathrm{mm}}$  & $12$ & $12$  \\
vert. beta function at IP &$\beta^*_{y}$ & ${\mathrm{mm}}$ & $0.12$  & $0.12$   \\

Crossing angle at IP & $\theta_{c}$ &  ${\mathrm{mrad}}$ & $14$  & $14$   \\
\hline
Site AC power  & $P_{site}$ &  ${\mathrm{MW}}$ & $\sim 150$ & $\sim 175$   \\
Site length & $L_{site}$ &  ${\mathrm{km}}$ & $8$ & $8$   \\ \hline
\end{tabular}
\label{C3-tab}
\end{center}
\end{table}

\paragraph{NCRF cryo-cooled Technology}
The  C$^{3}$ main linacs are based on C-band cryo-cooled copper cavities, feed individually by a distributed coupling waveguide. 

\begin{itemize}

 \item The main linac accelerating structures for 250 GeV, have to accelerate a train of bunches with a gradient 70 MV/m with a low breakdown rate (BDR) thanks to the operation at 77 K (liquid Ni). Operating copper structures at cryogenic temperatures allows to reduce the RF power requirements while increasing the beam loading and accelerating gradient. Structures are 40 cells 1 m long operating at $\pi$ mode.
 
 \item The RF power is distributed to the cells individually through a distributed coupling waveguide. The geometry of the cavities have been optimized to used efficiently the power. 
 
 \item A detailed study of cavity geometry for 1 nC bunches, showed that the C-band (5.712 GHz) is the optimal frequency. At this frequency we could keep highly efficient structure, with high gradients and excellent beam dynamics with proper damping and detuning of the cavities.
 
 \item Cavities are fabricated in split-blocks, with a length of $\sim$1 m, with all cavities machined in two/four copper slabs. Cavities are CNC machined to reduce the cost and the two/four copper slabs are bonded. High-order detuning is incorporated by adjusting the cavity geometry of each cell during fabrication. Slot damping with lossy materials will suppress the long-range wakefields. 
 
 \item The main linac is composed of 9 m cryomodules units. Each cryomodule houses eight 1 m distributed coupling accelerating structures supported on four 2 m girders. Each girder also support permanent magnet quadrupoles, BPMs and alignment movers and positioning devices. Each cryomodule has 4 four RF power sources with two waveguides, each waveguide power one accelerating structure. Thermal load is removed with liquid Ni.
 
 \end{itemize}
 
 \paragraph{Accelerator Design}
 
 The Main linac has been studied with some detail, the others element needed: sources, DRs and BDS has been taken from ILC or CLIC designs
 
 \begin{itemize}
 \item The baseline  polarized are produced by standard DC guns, buncher and accelerator. A polarized RF gun is also being investigated. The unpolarized e$^{+}$ is based on the CLIC design. Polarized e$^{+}$ are possible by extracting 125 GeV e$^{-}$ of the main linac an using an undulator as in ILC.
 
 \item DRs are used for e$^{-}$ and e$^{+}$, pre-DR is used in the case of the e$^{+}$. If a polarized e$^{-}$ RF gun is used the e$^{-}$ DR could be eliminated.
 
 \item The BDS system is 1.5 km length and follow the ILC-CLIC design.
 
 \end{itemize}

\paragraph{Civil Engineering and site}

No specific site has been chosen, a C$^{3}$ linear e$^{+}$e$^{-}$ could be sited anywhere in the world. Specific studies for Fermilab supposes 7 km footprint with 155 MV/m Gradients. The C$^{3}$ linac could also be an upgrade of the ILC.

 \paragraph{Sustainability}
 
The preliminary values for the power consumption for the 250 GeV and 550 Gev are 150 and 175 MW (100 and 118 for both linacs with 65$\%$ RF source effciency).
A dedicated study will be necessary including a real cryomodule as well as the others systems of the LC.

\subsubsection{Proposals for upgrades and extensions}

\paragraph{Luminosity upgrades}
Luminosity upgrades are not been studied at the current state of the project.
 
\paragraph{Energy extension and upgrades}

\begin{itemize}

 \item An intermediate stage in the construction at the Z-pole (91.2 GeV) could be possible, but this is not part of the baseline.
 
 \item The main strategy for the C$^{3}$ collider is to have the same length (8 km) for the two energy stages: 250 and 550 Gev but operating at 70 and 120 MV/m, by increasing the RF sources in the linac. The RF power per meter will increase from 30 to 80 MW/m. In the upgraded case, RF efficiency improving should be considered.
 Additional magnets will be required in the BDS.
 
 \item Starting from the design parameters of the C$^{3}$ 550 GeV collider is straight  forward to go at 3 TeV by a simple extension of the linac. In this regime the CLIC two-beam accelerator could be envisaged instead of klystrons.
 
 \end{itemize}

\subsubsection{Stageability to future experiments}
Not developed.

\subsubsection{State of Technical Design Report}
TDR is not existing.

\subsubsection{State of Proposal and R$\&$D plans}
To complete the TDR a full demonstration facility of the C$^{3}$ collider at GeV scale will be needed. The main R$\&$D developments are:

\begin{itemize}

 \item Fully engineered cryomodule including supports, alignment, permanent quadrupoles,  BPMs, girders and liquid and gaseous Ni devices.
 
 \item Operation of cryomodule: liquid and gaseous Ni use and vibrational issues.
 
 \item Single bunch studies for high-gradient (120 MV/m) including breakdown rates and multibunch for wakefield impact.
 
 \item Development of an ultra-low emittance polarized cryo-RF gun
 
 \item Development in partnership with the industries of a baseline RF source.
 
 \item Industrialisation and massive production.
 
 \end{itemize}

%%%%%%%%%%%%%%%%%%%%%%%%%%%%%%%%%%%%%%%%%%%%%%%%%%%%%%%
\newpage

\subsection{Cool Copper collider with Nb$_3$Sn Coating (C$^3$-Nb$_{3}$Sn), \cite{C3-Nb3Sn}}

Because of the considerable progress achieved in Normal-Conducting RF, particularly with the C$^3$ concept aimed at high gradient and high RF-to-beam efficiency, the latter would be an ideal candidate as a bulk structure geometry of an SRF cavity. The necessary structure is machined in two halves by low-cost numerically-controlled milling machines. This machining process produces ultra-high vacuum quality surfaces that need no further machining before a standard Cu surface etch. This manufacturing technique provides an ideal Cu surface to be coated with superconducting films, as it allows complete access to the inner cavity surface for the coating process. The system is then assembled simply by joining the two blocks. The geometry and coupling could be optimized further for SRF linac designs. 

A devoted global effort in developing Cu cavity structures coated with Nb$_3$Sn would make Electroweak and Higgs factories (as well as their energy upgrades) more affordable and more likely to be built. Using the next decade for R\&D on producing Nb3Sn on inexpensive and thermally efficient metals such as Cu or bronze, while pursuing in parallel the novel U.S.~concept of parallel-feed RF accelerator structures, would compound the best of both worlds. Not only do parallel-feed RF structures enable both higher accelerating gradients and higher efficiencies, but they would be applicable to both Cu and Nb$_3$Sn coated Cu cells. Increased effort on these two techniques would synergize expenditures towards 10-year progress, which will naturally converge to a clear decision by the community on which path to take for the RF of an ILC or any future accelerator.  If for any reason, the C3 structures were not ready in ten years, the current methods of Nb$_3$Sn coatings on Cu or bronze are geared towards standard cavity cells. Were one to succeed, it could still be implemented on conventional Cu RF cavity. In conclusion, the use of distributed coupling structure topology within improved performance parameters together with Nb$_3$Sn coating technology can lead to a paradigm shift for superconducting linacs, with higher gradient, higher temperature of operation, and reduced overall costs for any future collider. 

%%%%%%%%%%%%%%%%%%%%%%%%%%%%%%%%%%%%%%%%%%%%%%%%%%%%%%%
\newpage

\subsection{Higgs-Energy LEpton Collider (HELEN), \cite{HELEN}}

\subsubsection{Design outline}

The International Linear Collider (ILC) based on superconducting radio frequency (SRF) accelerator technology stores a standing wave with little loss in a superconducting cavity structure. Because the standing wave is comprised of a forward traveling and a backward traveling wave, of which only the forward traveling one contributes to acceleration, the ILC is not optimally using it's stored fields. HELEN, on the other hand, excites a circularly traveling wave which can achieve significantly stronger accelerating gradients. The accelerator consequently becomes much shorter than the ILC but is otherwise very similar. The peak accelerator gradients have progressed beyond the ILC specs of 31.5 MV/m, demonstrating already 50 MV/m. HELEN cavities could therefore likely achieve accelerator gradients of 70MV/m for average cavities with focused R$\&$D of only a couple of years. This could make this linear collider short enough to fit onto the Fermilab site. The reduced length also reduces the overall cost.

The HELEN collider can be upgraded to higher luminosities in the same way as was proposed for the ILC or to higher energies either by extending the linacs or with higher accelerating gradients as they become available, for example through Nb$_3$Sn.

\paragraph{SRF Traveling Wave Technology}

\begin{figure}[htbp]
\centering
\includegraphics[width =
0.75\textwidth]{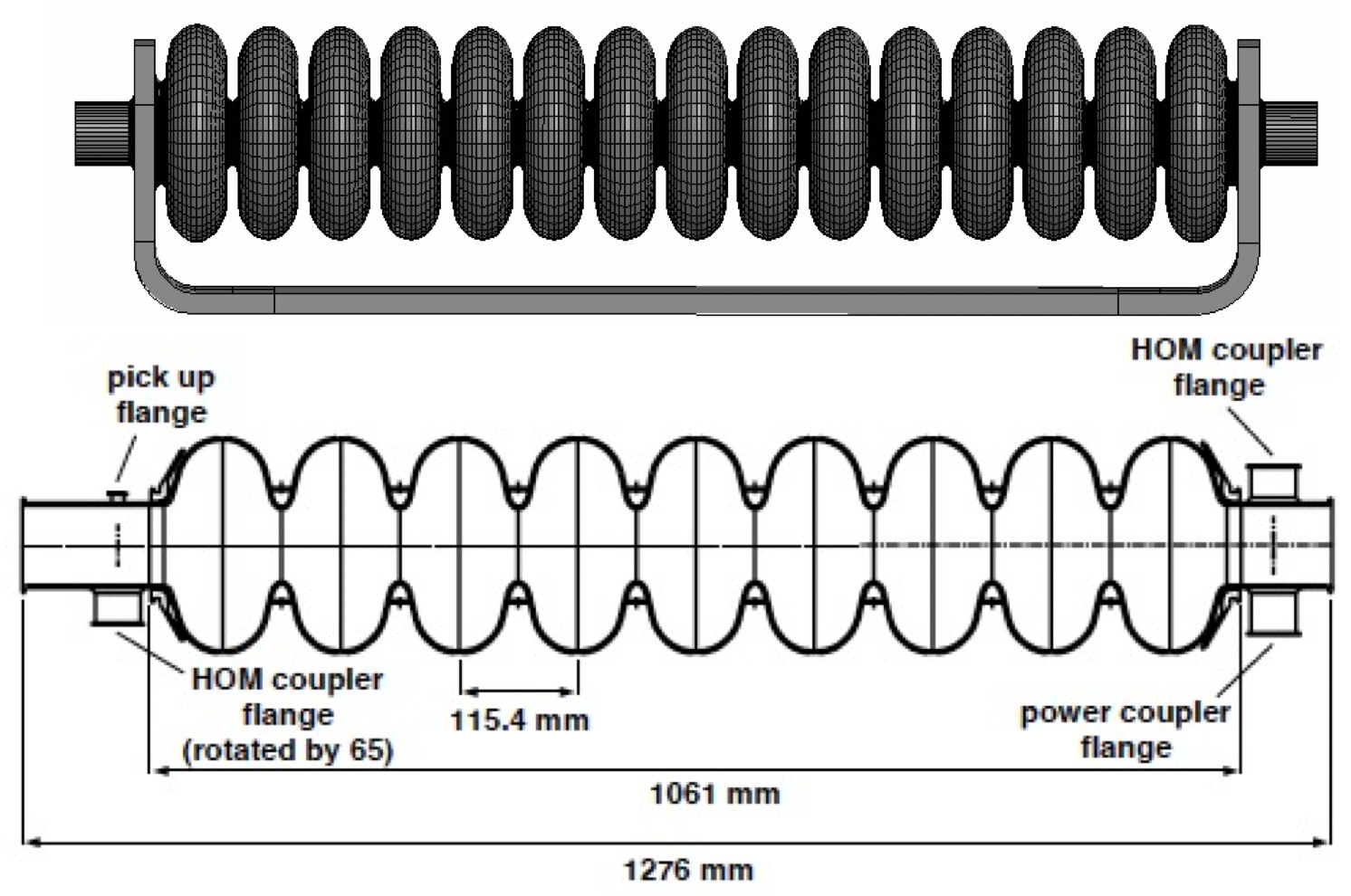}
\caption{Traveling wave SRF cavity, increasing the real estate gradient of the ILC. Top: TW with a 105◦ phase advance per cell. Bottom: the standing-wave TESLA structure.}
\label{fg:TravelingWaveSRF}
\end{figure}

The main technical detail to the HELEN is the use of a traveling wave SRF cavity. A resonant ring is constructed from an elliptical multi-cell cavity with a wave guide connecting the entrance to the exit, as shown in Fig.~\ref{fg:TravelingWaveSRF}. Instead of exciting a standing wave, a traveling wave is excited that circularly travels along the cavity and back thought the wave guide. Such resonant ring technology has regularly been applied for evaluating coupler pairs and is experimentally tested.
 
 \paragraph{Accelerator Design}
 
\begin{figure}[htbp]
\centering
\includegraphics[width = 0.75
\textwidth]{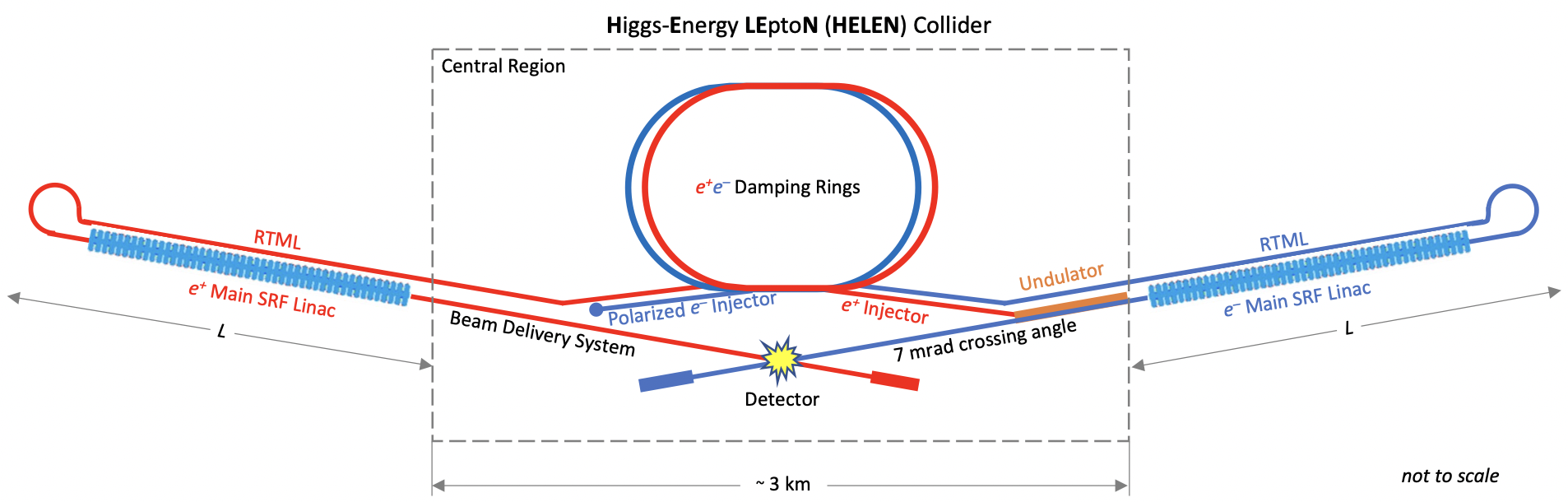}
\caption{Conceptual Layout of HELEN collider.}
\label{fg:HelenLayout}
\end{figure}

A conceptual layout of HELEN is shown in  Fig.\ref{fg:HelenLayout}. 

The total length depends on the real estate gradient of fill factor times accelerating gradient that can be achieved with SRF technology. The HELEN project evaluates three scenarios: (1) Improved ILC-type standing wave cavities at 55MV/m with 71\% fill factor. The layout would become 9.4 km long. (2) Traveling Wave cavities at 70MV/m with fill factor of 84\% would make the layout 7.5km long. (3) Ultimate Nb$_3$Sn standing wave gradients of 90MV/m with fill factor of 71\% would lead to a 6.5km long site. A traveling wave cavity based on Nb$_3$Sn technology has not been considered but would constitute the ultimate upgrade of this technology, with even shorter accelerator footprint.

Tables \ref{fg:HELENcavityTypes} and \ref{fg:HELENparameters}  show respectively the three cavity types  considered for HELEN and the main design parameters taking as baseline the 70 MV/m cavity option.

\begin{table}[htbp] 
\caption{The three cavity types evaluated by HELEN, with the 70MV/m version being the baseline.}
\begin{center}
\begin{tabular}{lcccc}
Parameter & Unit & Advanced SW & TW & Nb$_{3}$Sn \\
\hline
Accelerating Gradient & ${\mathrm{MV/m}}$ & 55 & 70 & 90  \\
Fill factor &  & 0.711 &  0.804 &  0.711 \\
Real state (effective) gradient & ${\mathrm{MV/m}}$ & 39.1 & 55.6 & 64.0 \\
Cavity $Q$ & $10^{10}$ & 1.0 (2 K) & 0.69 (2 K) & 1.0 (4.5 K) \\
Active cavity length & ${\mathrm{m}}$ & 1.038 & 2.37 & 1.038 \\
Cavity $R/Q$ & ${\mathrm{Ohm}}$ & 1158 & 4890 & 1158 \\
Geometry factor & ${\mathrm{Ohm}}$ & 279 & 186 & 279 \\
$B_{pk}/E_{acc}$ & ${\mathrm{mT / (MV/m)}}$ & 3.71 & 2.89 & 3.71 \\
$E_{pk}/E_{acc}$ &  & 1.98 & 1.73 & 1.98 \\
Number of cavities & & 4380 & 1527 & 2677 \\
Number of cryomodules & & 505 & 382 & 309 \\
Collider length & ${\mathrm{km}}$ & 9.4 & 7.5 & 6.9 \\
Main Linac AC power & ${\mathrm{MW}}$ & 49 &  39 &  58\\
Site AC power  &   ${\mathrm{MW}}$ & 121 & 110 & 129\\
\hline
\end{tabular}
\end{center}
\label{fg:HELENcavityTypes}
\end{table}

\paragraph{Civil Engineering and site}
In the rough layout estimate the beam delivery systems and IRs fit in a real estate of 3km, the turn around for the damping rings adds about 300m on each side so that the accelerating gradient and the fill fraction determine the length of the accelerator. The Fermilab site has a north-south extend of about 7.5 km and could make space available for HELEN with traveling wave cavities. 

The cost savings associated with the reduced length of the main linac is estimated to be about 25$\%$.
 
 \paragraph{Sustainability}
The power consumption of HELEN is little changed from the ILC, as the beam power remains the same even though the linac is shorter, and beam-related infrastructure remains largely unchanged. It is estimated that the site power would be 110 MW rather than 111 MW for the 250 GeV version of the ILC.

\begin{table}[htbp] 
\caption{Summary table of the HELEN accelerator parameters at 250 GeV.}
\begin{center}
\begin{tabular}{lccc}
Quantity & Symbol & Unit & Baseline  \\
\hline
Centre of mass energy & $\sqrt{s}$ & ${\mathrm{GeV}}$ & $250$  \\
Luminosity & ${\mathcal{L}}$ & $10^{34}{\mathrm{cm^{-2}s^{-1}}}$ & $1.35$  \\
Polarization for $e^-/e^+$ & $P_{-}(P_{+})$ & $\%$ & $80(30)$  \\ 
\hline
Repetition frequency &$f_{rep}$ & ${\mathrm{Hz}}$  & $5$  \\
Bunches per pulse  &$n_{bunch}$ & $1$  & $1312$  \\
Bunch population  &$N_{e}$ & $10^{10}$ & $2$  \\
Linac bunch interval & $\Delta t_{b}$ & ${\mathrm{ns}}$ & $554$  \\
Beam current in pulse & $I_{pulse}$ & ${\mathrm{mA}}$& $5.8$   \\
Beam pulse duration  & $t_{pulse}$ & ${\mathrm{\mu s}}$ & $727$  \\
Accelerating gradient & $G$ & ${\mathrm{MV/m}}$ & $70$ \\
Average beam power  & $P_{ave}$   & ${\mathrm{MW}}$ & $5.3$  \\ 
\hline
RMS bunch length  & $\sigma^*_{z}$  & ${\mathrm{mm}}$ & $0.3$  \\
Norm. hor. emitt. at IP & $\gamma\epsilon_{x}$ & ${\mathrm{\mu m}}$& $5$   \\
Norm. vert. emitt. at IP & $\gamma\epsilon_{y}$ & ${\mathrm{nm}}$ & $35$  \\
RMS hor. beam size at IP  & $\sigma^*_{x}$ & ${\mathrm{nm}}$  & $516$  \\
RMS vert. beam size at IP &$\sigma^*_{y}$ & ${\mathrm{nm}}$ & $7.7$  \\
Luminosity in top $1\,\%$ & ${\mathcal{L}}_{0.01} / {\mathcal{L}}$ &  & $73 \%$\\\
Beamstrahlung energy loss & $\delta_{BS}$ &  & $2.6\,\%$  \\
\hline
Site AC power  & $P_{site}$ &  ${\mathrm{MW}}$ & $110$  \\
Site length & $L_{site}$ &  ${\mathrm{km}}$ & $7.5$  \\ \hline
\end{tabular}
\end{center}
\label{fg:HELENparameters}
\end{table}

\subsubsection{Proposals for upgrades and extensions}

\paragraph{Luminosity upgrades}
While the traveling wave technology could be applied to any of the ILC's upgrade options, e.g. a 500 GeV version, the accelerator would become longer and would no longer fit onto the Fermilab site. A north-south orientation of the Higgs factory on the Fermilab site has been proposed where it is most likely that an extension off site of a longer accelerator would be possible.
 
\paragraph{Energy extension and upgrades}

An option beyond 250 GeV is not being proposed because of Fermilab's space limitations. But an extension to 500 GeV would be possible with similar implications as for the ILC.
 
 \subsubsection{Stageability to future experiments}
 
 HELEN is not proposed as a precursor of a future accelerator. But extensions to further interaction regions, e.g. for $\gamma\gamma$ collisions could be envisioned.
 
\subsubsection{State of Technical Design Report}
For most of the machine the Design and R\&D maturity are the same as for ILC. Only in three areas the state of the technology is different.
The SRF R\&D Maturity is 2, R\&D is ongoing to address fundamental physics and technology issues. The Main Linac design maturity is 3, its operating parameters are only established based on preliminary design concepts. The Site-specific Design maturity is also 2 as R\&D for fundamental questions is still ongoing.

\subsubsection{State of Proposal and R$\&$D plans}

Rapid developing, prototyping, and testing of new SRF cavities and cryomodules would be possible as significant work on traveling wave SRF cavities at high gradients has already been performed and Fermilab already has suitable R\&D infrastructure that supports the full cycle of R\&D, production, and verification (including testing cryomodules with beam) at the SRF accelerator test facilities and FAST linac. An R\&D period of about 5 years is estimated to produce workable and reliable cavities and associated crymodule prototypes.
If given high priority, the construction of the HELEN collider could start as early as 2031–2032 with first physics in 2040.
The HELEN collider can be upgraded to higher luminosities in the same way as was proposed for the ILC or to higher energies either by extending the linacs or with higher accelerating gradients as they become available.

%%%%%%%%%%%%%%%%%%%%%%%%%%%%%%%%%%%%%%%%%%%%%%%%%%%%%%%
\newpage

\subsection{Recycling Linear e$^{+}$e$^{-}$ Collider (ReLiC), \cite{ReLiC} }

\subsubsection{Design outline}

\paragraph{Critical Technology}

\begin{figure}[htbp]
\centering
\includegraphics[width =
\textwidth]{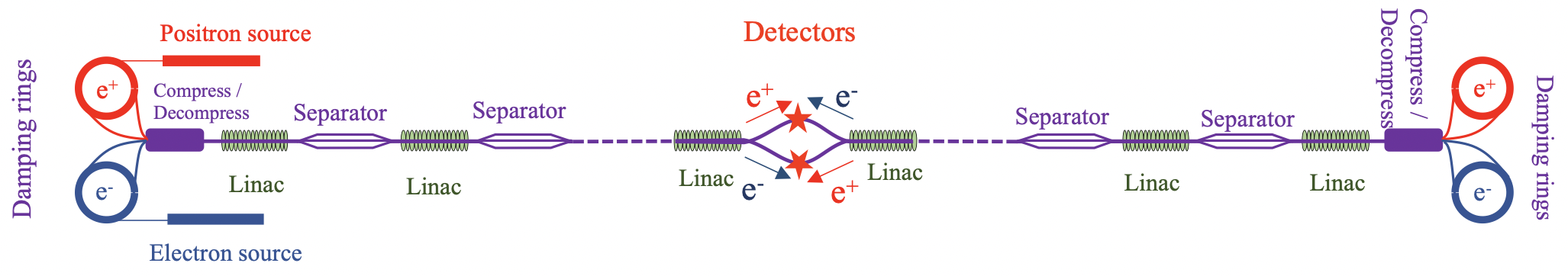}
\caption{Conceptual layout of ReLiC.}
\label{fg:ReLiClayout}
\end{figure}

ReLiC is an e$^+$/e$^-$ linear collider concept of recycling both the used particles and the used beam energy. While each linac accelerates particles for collisions, the spent beam after the collision is decelerated in the opposing linear accelerator respectively, recapturing its energy. A concept referred to as push-pull energy-recovery linac. After reaching low energy, the particles are not rejected but rather recycled in a damping ring where they regain their density and their polarization. Figure \ref{fg:ReLiClayout} shows it's outline. The SRF linacs are divided by separators, where used (decelerating) beams are separated from the accelerating beams to avoid damaging beam collisions. To avoid emittance growth, the separators provide for undisturbed straight trajectories of the accelerating beams, with steering only for the decelerating beams. All beams are on-axis in the linac’s SRF structures.

Many technical features are similar to other linear collider concepts:
\begin{itemize}
\item At the interaction regions, the beams are flat and have low emittances with an achievable vertical disruption.
\item Damping rings are used to achieve the required beam densities.
\end{itemize}

Main differences to other linear colliders are that:
\begin{itemize}
    \item The energy of the spent beam is recycled.
    \item The particles of the spent beam are recycled in the damping ring where they are
    \begin{itemize}
        \item replenished with an injection of only a few nano-Ampere.
        \item regain their density by the emittance damping that is also used in other colliders.
        \item regain their polarization via the Sokolov-Ternov process.
    \end{itemize}
    \item The beamstrahlung is limited by colliding mono-energetic beams
\end{itemize}

\begin{figure}[htbp]
\centering
\includegraphics[width =
0.8\textwidth]{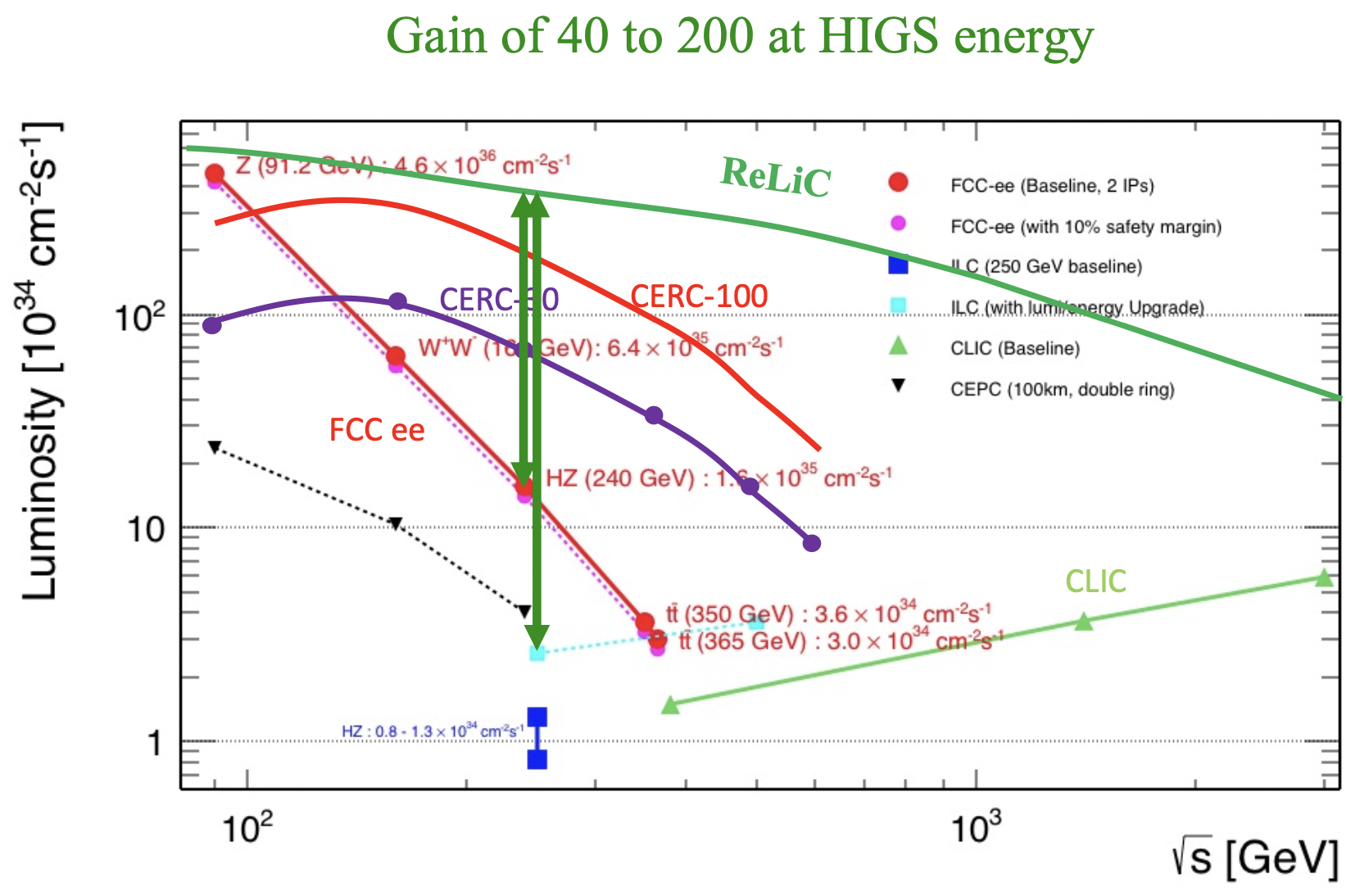}
\caption{The large luminosity potential of ReLiC compared with others Higgs/EW factories.}
\label{fg:ReLiCluminosity}
\end{figure}

These difference allow vastly higher luminosity (around two orders of magnitude) as shown in Fig. \ref{fg:ReLiCluminosity} and a vastly more efficient production of luminosity (also by two orders of magnitude).

This concept is extendable to higher energies as outlined in Fig. \ref{fg:ReLiCparameters}.
 
 \begin{table}[tbhp] 
\caption{Summary table of the ReLiC accelerator parameters for the different energy stages.}
\begin{center}
\begin{tabular}{lcccc}
Quantity & Symbol & Unit & Baseline &  ${\mathrm{E}}$  Upgrade \\
\hline
Centre of mass energy & $\sqrt{s}$ & ${\mathrm{GeV}}$ & $250$ & $3000$  \\
Luminosity & ${\mathcal{L}}$ & $10^{34}{\mathrm{cm^{-2}s^{-1}}}$ & $429$ & $40$  \\
Luminosity per IP & ${\mathcal{L}}/IP$ & $10^{34}{\mathrm{cm^{-2}s^{-1}}}$ & $215$ & $20$  \\
\hline
Collision frequency &  & ${\mathrm{MHz}}$  & $3$ & $18$   \\
Bunches per train  &$n_{bunch}$ &  & $5$ & $21$  \\
Bunch population  &$N_{e}$ & $10^{10}$ & $4$ & $1$  \\
Beam current in linac & $I_{pulse}$ & ${\mathrm{mA}}$& $18$ & $29$ \\
\hline
RMS bunch length  & $\sigma^*_{z}$  & ${\mathrm{mm}}$ & $1$ & $5$  \\
Norm. hor. emitt. at IP & $\gamma\epsilon_{x}$ & $\mu {\mathrm{m}}$& $4$ & $8$   \\
Norm. vert. emitt. at IP & $\gamma\epsilon_{y}$ & ${ \mathrm{nm}}$ & $1$ & $2$  \\
hor. beta function at IP  & $\beta^*_{x}$ & ${\mathrm{m}}$  & $5$ & $100$  \\
vert. beta function at IP &$\beta^*_{y}$ & ${\mathrm{mm}}$ & $0.34$  & $7$   \\
hor. disruption parameter  & $D_{x}$ &  & $0$ & $0$  \\
vert. disruption parameter &$D_{y}$ &  & $109$  & $3$   \\
\hline
Section length &  &  ${\mathrm{m}}$ & $500$ & $250$   \\ 
Site length & $L_{site}$ &  ${\mathrm{km}}$ & $21$ & $276$   \\ \hline
\end{tabular}
\label{fg:ReLiCparameters}
\end{center}
\end{table}
 
Key Technologies that need to be addressed by R$\&$D are:
\begin{itemize}
\item CW superconducting RF (SRF) linacs with high Q, similar to needs of the ILC.
\item 5-cell 1.5 GHz SRF cavities with effective HOM damping, as ReLiC is choosing a different frequency than the ILC.
\item Electro-magnetic separators for contra-propagating bunch-trains.
\item Low emittance damping rings with flat beams and large energy acceptance to capture the spent beam.
\item Bunch compressor for the accelerated beam, similar to other damping rings.
\item Bunch decompressor to capture the spent beam in the damping ring.
\item MHz rate injection/ejection kickers
\item nA-scale top-off e+e- injectors
\item Two collision areas (IPs) similar to other linear colliders.
\item Vertical beam stabilization at the IPs, also similar to other linear colliders.
\end{itemize}
 
 \paragraph{Accelerator Design}
 
 Several challenges are similar to other linear colliders, in particular to the ILC, as much of the technology in the linacs and the damping rings are related. The dominant design challenges are:
 \begin{itemize}
\item 1.5 GHz SRF cavities with quality factor Q $>$ 10$^{11}$ at 1.5 K
\item High-efficiency 1.5~K LiHe refrigerators
\item Reactive tuners to reduce power to suppressing microphonics
\item Damping rings with very flat beams ($\epsilon_h / \epsilon_v \approx 2,000 - 4,000$).
\item Damping rings with 10\% energy acceptance.
\item 10-fold bunch compressor/decompressor at 10 GeV.
\item MHz rate injection/ejection kickers
\item Vertical beam stabilization at the IPs
 \end{itemize}

\paragraph{Civil Engineering and site}
A site has not yet been proposed for ReLiC, and site considerations are similar to other linear colliders like the ILC.

\subsection{Sustainability}

With current SRF technology (LSLS HE) with $Q\approx 3\cdot 10^{10}$, ReLiC at 250 GeV c.m. energy will consume about 350 MW of AC power, which is about equally split between beam energy losses for radiation and cryogenics.

Increasing the energy to 3 TeV c.m. with current technology will result in an unsustainable AC power requirement exceeding 2 GW.

There is a potential of 5-fold in crease in Q, which would make ReLiC operation at all energy from Higgs to 3 TeV much more energy efficient. The Higgs factory ReLiC will then require about 200 MW of AC power, and the 3 TeV c.m. operation just under 1 GW.

There are potentials for reducing the energy consumption.
\begin{itemize}
\item The RF powers needed in damping rings is proportional luminosity. For the 250 GeV ReLiC, reducing the luminosity by a factor of 100 would reduce the power need by a factor of 4 to 50 MW.
\item Cryoplant power is proportional to the total collider energy. It can be  reduced by improving LiHe refrigerators from their current 19$\%$ the of theoretically possible Carnot efficiency. Investments in LiHe refrigerator R$\&$D is probably the best strategy of improving the Carbon footprint of SRF systems.
\end{itemize}

\subsubsection{Proposals for upgrades and extensions}

\paragraph{Luminosity upgrades}
\begin{itemize}
 \item Luminosity of ReLiC can be upgraded by increasing beam currents.
\item RF power required in damping rings will grow proportionally to the beam currents, e.g. proportionally to the luminosity.
\item This proportionally allows to stage luminosity upgrades by augmenting the ring’s RF system.
 \end{itemize}
 
\paragraph{Energy extension and upgrades}

An upgrade program from the Higgs energy to pup to 3TeV has been evaluated with the following aspects, and as illustrated by Fig. \ref{fg:ReLiCparameters}. 

\begin{itemize}
 \item Extending the c.m. energy in ReLiC to 3 TeV has been investigated
\item The main challenge is maintaining low energy of beamstrahlung photons.
\item This extension also requires increasing energy of damping ring.
\end{itemize}
 
\subsubsection{Stageability to future experiments}
 
ReLiC as a Higgs factory at 250 GeV is a natural precursor of a higher energy ERL-based linear collider, as much of the infrastructure can be reused.
 
\subsubsection{State of Technical Design Report}

The ReLiC concept has been published to the level summarized here, on the conceptual design stage. A small group of researchers continue to refine the concept, but a full Design Report is not currently being prepared.

\subsubsection{State of Proposal and R$\&$D plans}

The following items require R$\&$D:
\begin{itemize}
 \item The major processes (beam-beam collisions, beamstrahlung, effect of beam separators were either simulated or scaled from other projects such as CERC etc.) deeming the concept sound. A full design would require more detailed parameter studies.
 \item No realistic cost estimate has yet been generated.
\item The power estimations may still miss significant components, e.g., cooling of the tunnel and heat losses for LiHe transfer lines.
\item High efficiency LiHe refrigeration systems.
\item Very high-Q SRF cavities.
\item Reactive tuners in SRF systems.
\item Damping rings with large aperture.
\item MHz rated kickers.
\item Providing and controlling flat beams with $\epsilon_h/\epsilon_v \approx 2,000$.
\end{itemize}

Even a concentrated R\&D period would require at least 5 years for effective solutions to these important issues.

%%%%%%%%%%%%%%%%%%%%%%%%%%%%%%%%%%%%%%%%%%%%%%%%%%%%%%%
\newpage
\subsection{A high-luminosity superconducting twin e$^{+}$e$^{-}$ linear collider with energy recovery (ERLC), \cite{ERLC}}

\subsubsection{Design outline}

Even though no white paper has been submitted to this concept of an ERL-based linear collider at the Higgs energy, for completeness we include a brief reference to this concept. It is related ReLiC, in that it is a push-pull ERL, but it uses dual axis cavities rather than beam separators to avoid collisions between the accelerating and the decelerating beam.

\begin{figure}[htbp]
\centering
\includegraphics[width =
\textwidth]{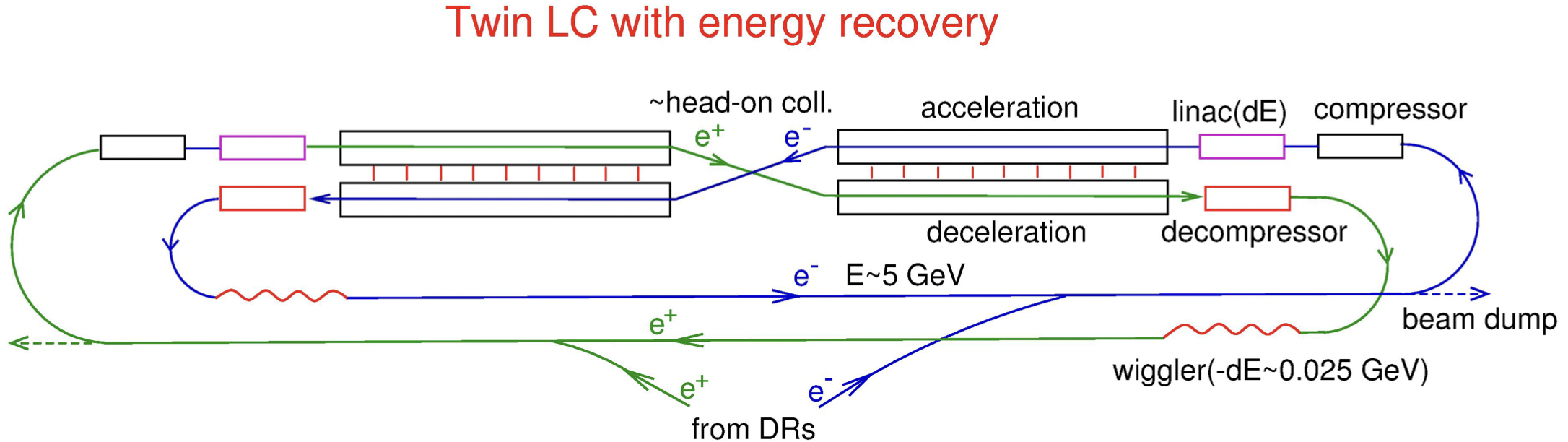}
\caption{Conceptual layout of the ERLC.}
\label{fg:ERLClayout}
\end{figure}

Superconducting technology and the continuous development of Energy Recovery Linacs has lead to superconducting linear collider designs that provide energy recovery. The ERLC, like the ReLiC, is one such design. To avoid parasitic collisions inside the linacs, a twin (dual) LC is proposed, as shown in  Fig.~\ref{fg:ERLClayout}. It uses single cavities with two axes, i.e. strongly coupled cavities, where a beam that is decelerated along one axis leaves a wake that can accelerate a beam along the other axis.  
Reference \ref{fg:ERLClayout} discusses the achievable luminosity in such a scheme and the power needed for such a collider. Vast luminosity gains are illustrated, similar to ReLic, of about two orders of magnitude. With current SRF technology of solid Nb cavities at 1.8 K and 1.3 GHz as for the ILC one expects to obtain a luminosity of $39\cdot 10^{34} \mathrm{cm^{-2}s^{-1}}$ for a 250 GeV Higgs factory. The power need for the accelerating SRF and for Higher Order Mode removal is around 200~MW with ILC parameters for the SRF (cavities
with $Q\approx 10^{10}$). With SRF advances towards cavities with $Q\approx 3\times 10^{10}$, power needs are of order 100 MW for this large a luminosity. Further SRF advances, using superconductors operating at 4.5 K with high $Q_0$ values, as envisioned for Nb$_3$Sn, at 0.65 GHz, the luminosity can be  $140\cdot 10^{34} \mathrm{cm^{-2}s^{-1}}$ for the same power needs. Energy upgrades are also analyzed, estimating for a 500~GeV c.m. collider with a  power limit of 150 MW a luminosity of $80\cdot 10^{34} \mathrm{cm^{-2}s^{-1}}$. These estimates are almost two orders of magnitude greater than at the ILC, where the beams are not energy recovered.

\paragraph{Critical Technology}

The most critical technology are dual axes, high Q SRF cavities with high gradients. All other R\&D items are similar to those outlined for ReLiC. The advantages in luminosity rely on ERL technology, the suitable cavities and their stabilization is therefore the dominant research item. Advances in high Q research and in more efficient refrigeration are the most important items to limit the power consumption.
 
 \paragraph{Accelerator Design}
 
The design parameters for the different ERLC options are shown in Fig. \ref{fg:ERLCparameters}.

\begin{landscape}
\centering
\begin{table}[tbhp] 
\caption{Summary table of the ERLC  accelerator parameters in the initial 250 GeV staged configuration energy upgrade.}
\begin{tabular}{lcccccccccc}
& & & & & & & & & & \\
Quantity & Symbol & Unit & \multicolumn{4}{c}{Baseline} & \multicolumn{4}{c}{${\mathrm{E}}$ Upgrades} \\ \hline
Technology &  &  & \multicolumn{2}{c}{Pulsed} & \multicolumn{2}{c}{Contin.} & \multicolumn{3}{c}{Pulsed} & Contin. \\
Operation mode &  &  & \multicolumn{2}{c}{Nb} & \multicolumn{2}{c}{Nb$^3$Sn} & \multicolumn{2}{c}{Nb} & \multicolumn{2}{c}{Nb$^3$Sn} \\
Cavity frequency &  & ${\mathrm{MHz}}$ & $1.3$ & $0.65$ & $1.3$ & $0.65$ & $1.3$ & $0.65$  & $1.3$ & $0.65$ \\
\hline
Centre of mass energy & $\sqrt{s}$ & ${\mathrm{GeV}}$ & $250$ & $250$ & $250$ & $250$ & $500$ & $500$ & $500$ & $500$ \\
Luminosity & ${\mathcal{L}}$ & $10^{34}{\mathrm{cm^{-2}s^{-1}}}$ & $39$ & $75$ &  $83$ & $160$ & $17.4$ & $34.2$  &  $41.2$ & $78$ \\
\hline
Accelerating gradient & $G$ & ${\mathrm{MV/m}}$ & $20$ & $20$ & $20$ & $20$ & $20$ & $20$ &  $20$ & $20$ \\
Duty Cycle & $DC$ &   & $0.19$ & $0.37$ &  $1$ & $1$ & $0.121$ & $0.237$ &  $0.47$ &  $1$ \\
Cavity Quality  & $Q$ & $10^{10}$ & $3$ & $12$ &  $3$ & $12$ & $3$ & $12$ &  $3$ &  $12$ \\
Active cavity length & &  ${\mathrm{km}}$ & $12.5$ & $12.5$ &  $12.5$ & $12.5$ & $25$ & $25$ &  $25$ &  $25$ \\
Bunch population  &$N_{e}$ & $10^{10}$ & $1.13$ &  $2.26$ &  $0.46$ & $1.77$ & $1.13$ & $2.26$ & $0.685$ & $1.23$ \\
Linac bunch interval & $\Delta t_{b}$ & ${\mathrm{m}}$ & $0.23$ & $0.46$ &  $0.23$ & $0.46$ &  $0.23$ & $0.46$ &  $0.23$ & $0.46$  \\
Repetition frequency &$f_{rep}$ & $10^{8} {\mathrm{Hz}}$ & $2.47$ & $2.37$ &  $13$ & $6.5$ & $1.57$ & $1.54$ & $6.1$ & $6.5$ \\
\hline
RMS bunch length  & $\sigma^*_{z}$  & ${\mathrm{mm}}$ & $0.3$ & $0.3$ & $0.3$ & $0.3$ & $0.89$ & $0.89$ &  $0.89$ &  $0.89$ \\
Norm. hor. emitt. at IP & $\gamma\epsilon_{x}$ & ${\mathrm{\mu m}}$& $10$ & $10$ & $10$ &  $10$ & $10$ & $10$ & $10$ &  $10$ \\
Norm. vert. emitt. at IP & $\gamma\epsilon_{y}$ & ${\mathrm{nm}}$ & $35$ & $35$ &  $35$ &  $35$ & $35$ & $35$ & $35$  &  $35$ \\
RMS hor. beam size at IP  & $\sigma^*_{x}$ & ${\mathrm{nm}}$  & $1050$ & $2100$ &  $430$ & $1660$ & $1260$ &  $2500$ & $760$ & $1380$ \\
RMS vert. beam size at IP &$\sigma^*_{y}$ & ${\mathrm{nm}}$ & $6.2$  & $6.2$  &   $6.2$ & $6.2$ & $7.4$  & $7.4$ & $7.4$  & $7.4$ \\
Beamstrahlung energy loss & $\delta_{BS}$ &  & $0.2\,\%$  & $0.2\,\%$ & $0.2\,\%$ & $0.2\,\%$ &$0.1\,\%$  & $0.1\,\%$  & $0.1\,\%$  & $0.1\,\%$ \\
\hline
Site AC power  & $P_{site}$ &  ${\mathrm{MW}}$ & $120$ & $120$ &   $120$ & $120$ & $150$ & $150$ & 150 & 150 \\
Site length & $L_{site}$ &  ${\mathrm{km}}$ & $30$ & $30$  &  $30$ & $30$ & $50$ & $50$ & $50$ & $50$\\ 
\hline
\end{tabular}
\label{fg:ERLCparameters}
\end{table}
\end{landscape}

\paragraph{Civil Engineering and site}

The design is on the conceptional analysis stage. There is no design with fully simulated parameters for which a site or civil engineering considerations could be started. The merit of the study is that it illustrates the ERL path to large luminosity increases, compatible with arguments for ReLIC. Evaluations of luminosity and power needs come to similar conclusions, strengthening both accelerator analyzes.
 
 \paragraph{Sustainability}

With a limit to the 100 to 200MW power limit, the luminosity efficiency is larger by two order of magnitudes than the ILC, leading to large energy savings per physics impact. But this relies on significant advances in high Q cavities and in refrigerator efficiencies. A breakdown power for two different types of cavities is shown in Table \ref{fg:ERLCpower}.

\begin{table}[tbhp] 
\caption{Breakdown of power needs for  two ERLC options, after significant SRF improvements beyond ILC parameters.}
\begin{center}
\begin{tabular}{lcccc}
Quantity & Symbol & Unit & \multicolumn{2}{c}{Baseline}  \\ \hline
Technology &  &  & Pulsed & Contin.  \\
Operation mode &  &  & Nb & Nb$^3$Sn \\
Cavity frequency &  & ${\mathrm{MHz}}$ & $1.3$ & $0.65$  \\
\hline
Centre of mass energy & $\sqrt{s}$ & ${\mathrm{GeV}}$ & $250$ & $250$  \\
Luminosity & ${\mathcal{L}}$ & $10^{34}{\mathrm{cm^{-2}s^{-1}}}$ & $39$  & $160$ \\
\hline

Duty Cycle & $DC$ &   & $0.19$ & $1$  \\
Bunch population  &$N_{e}$ & $10^{10}$ & $1.13$ &   $1.77$ \\
\hline

Beam generation power  &  &  ${\mathrm{MW}}$ & $small$ & $small$ \\
Radiation in wigglers power  &  &  ${\mathrm{MW}}$ & $4.45$ & $18.4$  \\
HOMs beam energy power  &  &  ${\mathrm{MW}}$ & $5.5$ & $9$  \\
HOMs cool.  1.8 (4.5) K power  &  &  ${\mathrm{MW}}$ & $24.8$ & $10$  \\
HOM cool.  77 K power  &  &  ${\mathrm{MW}}$ & $27.6$ & $44.7$  \\
RF diss. cool.  1.8 (4.5) K power  &  &  ${\mathrm{MW}}$ & $57.6$ & $38$  \\
\hline 
Site AC power  & $P_{site}$ &  ${\mathrm{MW}}$ & $120$ & $120$  \\
\hline

\end{tabular}
\end{center}
\label{fg:ERLCpower}
\end{table}

\subsubsection{Proposals for upgrades and extensions}

As shown in Table \ref{fg:ERLCparameters}, a 500 GeV options is analyzed as extension beyond a Higgs factory.
 
\subsubsection{State of Proposal and R$\&$D plans}

R\&D should focus on dual axes SRF cavities, including peak field studies, high Q studies, HOM damping analysis, and microphonics control. Design of a cryomodule and test with beam will also be essential.

%%%%%%%%%%%%%%%%%%%%%%%%%%%%%%%%%%%%%%%%%%%%%%%%%%%%%%%
\newpage

\subsection{X-ray FEL based  $\gamma \gamma$ Collider Higgs Factory (XCC) \cite{FEL_gg} and High-Energy High-Luminosity  $\gamma \gamma$ Colliders  (HE$\&$HL $\gamma \gamma$ ) \cite{gg}}

The concept of $\gamma\gamma$ collider was first proposed by Ginzburg, Telnov, et al.~\cite{ginzburg} 
around 1980, further developed by Telnov, e.g.~\cite{telnov}, 
and
by Kim, Zholents et al.~\cite{kkim}. A $\gamma\gamma$ collider 
Higgs factory only requires  
two electron beams, no positrons, no damping rings, 
and can operate at a lower centre-of-mass energy than an e$^+$e$^-$ Higgs factory, namely at an e$^-$e$^-$ energy of 125(--140) GeV 
instead of 240 GeV, thanks to the direct
production $\gamma\gamma \rightarrow $H, implying a 
shorter linac. 
The electron beams and the photon beam should ideally both be polarised. 

While the previously proposed $\gamma\gamma$ colliders considered
the creation of the gamma rays by backscattering optical laser pulses off high-energy electron beams, two proposals at Snowmass 2021 consider replacing the optical laser by photons from an FEL \cite{FEL_gg,gg}.
The new proposals do not limit operation to Compton $x$ parameters below 4.8 (which, in earlier proposals, was deemed to be necessary to avoid pair creation from gamma-photon scattering), but consider 
$x$ parameters larger \cite{gg} or even 
much larger than 4.8 \cite{FEL_gg}.

A concrete proposal XCC was worked out for a $\gamma\gamma$ collider based on the 
C$^3$ cool copper linac. There are evident synergies with ray FEL developments, e.g.~LCLS-II.

\subsubsection{Design outline}
\begin{figure}[h]
\centering
\includegraphics[width =
0.75\textwidth]{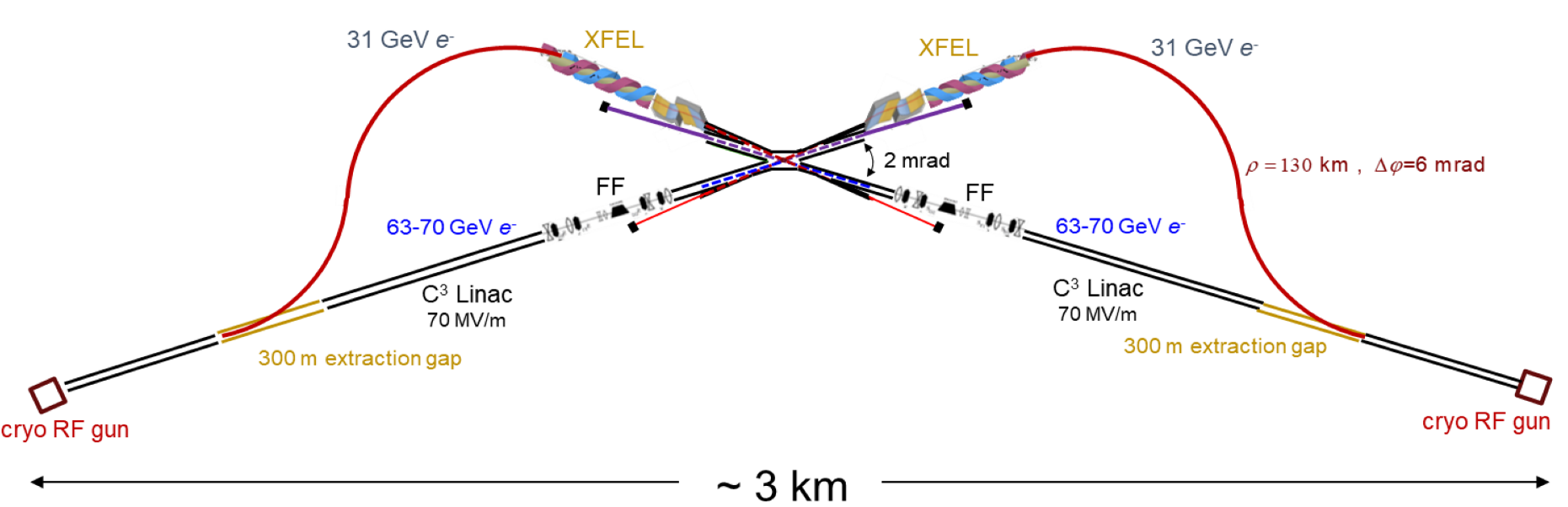}
\caption{Concept of XCC $\gamma\gamma$ collider Higgs factory based on 
C$^3$ technology \protect\cite{FEL_gg}.}
\label{XCCconc}
\end{figure}

\begin{figure}[h]
\centering
\includegraphics[width =
0.95\textwidth]{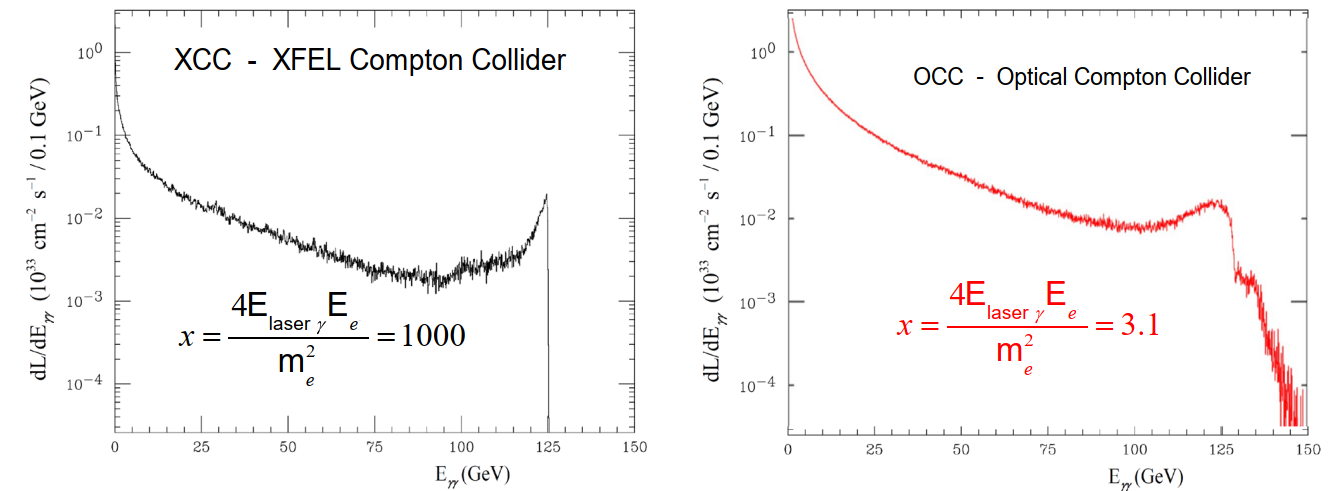}
\caption{Luminosity energy spectrum for XCC 
compared with the one obtained from optical laser approach,
for the parameters of Table \ref{tabxcc2}, 
illustrating the merit of operating at $x=1000$ \protect\cite{FEL_gg}.}
\label{XCC2}
\end{figure}

XCC assumes a cryogenic Cu RF Gun, producing 
76 electron bunches of 1 nC charge with 120 nm-rad normalized emittance,
with 90\% polarisation at 240 Hz.
The bunches are accelerated in a 
cryogenic Cu linac (C$^3$) with a gradient of 70 MV/m, to about 70 GeV, 
and then made to collider with 700 mJ/pulse 1 keV $\gamma$ pulses from an XFEL.
Electron beams of 31 GeV from the same
linacs are used to drive the two FELs.
The minimum bending radius for 
transporting and bending the e$^-$ beam to the respective undulator sections is chosen as 130 km, which avoid emittane growth due to coherent synchrotron radiation. 
The FEL X-rays are focused to 70 nm FWHM at the Compton collision points. 
The overall concept is illustrated in Fig.~\ref{XCCconc}.
Parameters are summarized in Table \ref{tabxcc1}. 
Figure~\ref{XCC2} compares luminosity spectra for XCC with a
classical optical laser approach, for the parameters of Table \ref{tabxcc2}.

\begin{table}[htbp]
\caption{ Parameters of XCC $\gamma\gamma$ collider Higgs factory}
\label{tabxcc1}
\begin{center}
\begin{tabular}{lc|lc}
\hline\hline
Final Focus parameters & Approx.~value & FEL parameters & Approx.~value \\
\hline
Electron energy & 62.8 GeV & Electron energy & 31 GeV \\
Electron beam power & 0.57 MW & Electron beam power & 0.28 MW \\
$\beta_{x,y}^{\ast}$ & 0.03 mm  & normalized emittance    &  120 nm \\
$\gamma \epsilon_{x,y}$ & 120 nm &  rms energy spread ($\Delta \gamma/gamma$) &  0.05\% \\
$\sigma_{x,y}^{\ast}$ at e$^-$e$^-$ IP & 5.4 nm & bunch charge & 1 nC \\
$\sigma_z$ & 20 $\mu$m &  linac-to-XFEL curvature radius & 133 km \\
bunch charge & 1 nC & undulator $B$ field & $\ge$1 T \\
rep.~rate at IP & 240$\times$38 Hz & undulator period & 9 cm\\
$\sigma_{x,y}$ at IPC & 12.1 nm & average $\beta$ function & 12 m\\
$L_{\rm geom}$ & $9.7\times 10{^34}$~cm$^{-2}$s$^{-1}$ & x-ray $\lambda$ (energy) &  1.2 nm (1 keV)\\
$(\Delta E/E)_{\rm rms}$ & 0.05\% &  x-ray pulse energy & 0.7 J \\
$l^{\ast}$ (QD0 exit to IP) & 1.5 m & pulse length & 40 $\mu$m \\
$d_{CP}$ (IPC to IP) & 60 $\mu$m & $a_{\gamma,x}$/$a_{\gamma,y}$ (x/y waist) & 21.2 nm / 21.2 nm \\
QD0 aperture & 9 cm diameter & non-linear QED $\xi^2$ & 0.10 \\
\hline\hline
Site Parameters & Approx. Value &  --- &  --- \\ \hline
crossing angle & 2 mrad &   & \\
site power & 85 MW &  &  \\
total length & 3.0 km & & \\
\hline\hline
\end{tabular} 
\end{center}
\end{table}

\begin{table}[htbp]
\caption{Parameters of XCC $\gamma\gamma$ collider Higgs factory compared with classical optical-laser 
$\gamma\gamma$ approach, and with ILC Higgs factory parameters for two directions of 
electron and positron polarization.}
\label{tabxcc2}
\begin{center}
\begin{tabular}{l|cccccc}
\hline\hline
Machine & $E_{e^-}$ [GeV] & $N_e^-$ [nC] & Polarization & $N_H$/yr & $N_H/N_{\rm hadr}$  & $N_{\rm minbias}$/BX \\
\hline
XCC & 62.8 & 1.0  & 90\% e$^-$ & 34,000 & 170  &  9.5 \\
OCC & 86.5 & 1.0 &  90\% e$^-$ & 30,000 & 540 &  50   \\
\hline
ILC & 125 & 3.2  & $-$80\% e$^-$, $+$30\% e$^+$ & 42,000  & 140  & 1.3\\
ILC & 125  & 3.2 & $+$80\% e$^-$, $-$30\% e$^+$ & 28,000 & 60  & 1.3 \\
\hline\hline
\end{tabular} 
\end{center}
\end{table}

\subsubsection{Sustainability}
The total electron beam power is close to 2 MW. The electrical power required for the 
linac RF systems is 23 MW, the power for the cryogenics system is estimated at 34 MW, and the remaining power for the accelerator complex is taken to be 31 MW, resulting in a total XCC site power of 88 MW.

\subsubsection{Upgrades and extensions}
The XCC scheme also allows for $\gamma$-e$^-$  
collisions with a luminosity of order
 $10^{35}$~cm$^{-2}$s$^{-1}$  \cite{FEL_gg}.

The XCC concept can be extended to higher collision energy.
An XCC energy upgrade to a center-of-mass energy $E_{\rm c.m.} \le 280$~GeV is proposed 
to study the Higgs self-coupling through
double Higgs production, $\gamma \gamma \rightarrow HH$  \cite{FEL_gg}. 
The double Higgs rate at the XCC with $E_{\rm c.m.}=280$~GeV 
is the same as the rate for $e^+e^- \rightarrow ZHH$  
at the ILC with $E_{\rm cm}=500$~GeV, 
and so this is an interesting XCC energy upgrade.

A concept to produce the required luminosity in $\gamma\gamma$ colliders of higher energy, i.e.~between 0.5 to 10 TeV c.m., 
was discussed in Ref.~\cite{gg}, considering 
the second interaction region of a baseline e$^+$e$^-$ linear collider.  
With the best of modern standard lasers, high-energy $\gamma\gamma$ colliders from electron beams of E$\ge$250 GeV are possible at the expense of photon luminosity, which could be as low as
1\% of the geometric e$^+$e$^-$ luminosity, i.e.~10 times lower than for photon colliders at c.m.~energies below 0.5 TeV. This is because the higher energy $\gamma$ rays annihilate in the Breit-Wheeler process, $\gamma \gamma_0\rightarrow e^+e^-$  and in the Bethe-Heitler disrupting process, 
$e \gamma \rightarrow  e e^+e^-$. These processes occur when the invariant quantity $x=12.3 E [{\rm TeV}]/\lambda_0 [\mu {\rm m}]$  exceeds 4.8, where $E$ is the energy of each e$^+$ or e$^-$ beam, and $\lambda_{0}$ is the wavelength of the incoming photon.  Instead, for $x<4.8$, the luminosity of the $\gamma\gamma$ collisions can be as high as 43\% of the geometric e$^+$e$^-$ luminosity, which translates to 10\% of the geometric luminosity at the desired $\gamma\gamma$ c.m.~ energy. 

In \cite{gg} a single Free Electron Laser (FEL) design meets the specifications to enable $\gamma\gamma$ colliders as second interaction regions of e$^+$e$^-$ colliders over the energy range of 0.5 TeV to 10 TeV c.m.
The same electron beams and accelerators of the original e$^+$e$^-$ collider are used for two identical high gain Self-Amplified Spontaneous Emission (SASE) FELs. At the appropriate energy required in the FEL design, i.e.~2.3 GeV, every other bunch from each beam is diverted to each FEL line where a helical undulator produces circularly polarized 0.5 eV light with 0.1--1 Joule per pulse.  The remaining bunches continue down the Linac until reaching their nominal  energy, and colliding with a geometric luminosity of 1--6$\times 10^{34}$~cm$^{-2}$s$^{-1}$. 
The central FEL wavelength of 2.4~$\mu$m, obtained with either standard warm magnet or superconducting technology for the undulator, and an $x$-factor in the range of 2 to 40, maximize the luminosity of the $\gamma\gamma$ collider as second interaction region of a 0.5--10 TeV c.m.~electron-positron collider. This means that, once installed, the FEL can be used even if the e$^+$e$^-$ collider itself undergoes energy upgrades. But more importantly, this FEL increases the expected $\gamma$ intensity by a factor of 10 for electron beams of up to 0.5 TeV, by a factor of 6 for electron beams of up to 1.5 TeV, and by a factor of 3 for electron beams of up to 5 TeV. This translates to a factor 10 increase in the luminosity of $\gamma\gamma$ colliders as second interaction regions of 0.5 TeV to 1 TeV c.m. e$^+$e$^-$ colliders, a factor of 6 for a 3 TeV c.m. e$^+$e$^-$ collider, and a factor of 3 for a 10 TeV c.m.~e$^+$e$^-$ collider. This FEL concept, therefore, paves the way for High Energy \& High Luminosity $\gamma\gamma$ colliders. The parameter $x$, the Compton backscattered maximum photon energy, and the $\gamma\gamma$ luminosity are shown in Table \ref{ggtab} as a function of the e$^+$e$^-$ collider energy in the c.m., 
up to 10 TeV energy \cite{gg}.

\begin{table}[tbhp] 
\caption{Possible parameters of FEL-based high energy $\gamma\gamma$ colliders \protect\cite{gg}.}
\begin{center}
\begin{tabular}{lcccccc}
Quantity & Symbol & Unit & \multicolumn{4}{c}{${\mathrm{E}}$ Upgrades} \\ \hline
Centre of mass energy & $\sqrt{s}$ & ${\mathrm{GeV}}$ & $500$ & $1000$ & $3000$ & $10000$ \\ \hline
x-factor & &  & $2(4)$ & $4$ & $12$ & $40$ \\
Maximum $\gamma$ energy &  & ${\mathrm{GeV}}$ & $170(200)$ & $400$ & $1380$ & $4880$ \\
 $L_{\gamma \gamma}/L_{ee}$& & $\%$  & $\leq 10$ & $\leq 10$ & $\leq 6$ & $\leq 3$ \\
\hline
\end{tabular}
\end{center}
\label{ggtab}
\end{table}

%%%%%%%%%%%%%%%%%%%%%%%%%%%%%%%%%%%%%%%%%%%%%%%%%%%%%

\newpage

\subsection{Future Circular Lepton Collider (FCC-ee) \cite{FCCee,FCC_summary}}

The Future Circular electron-positron Collider, FCC-ee, is a proposed 
new storage ring of 91 km circumference,  which has been  
designed to carry out a precision study 
of Z, W, H, and ${\rm t}\bar{\rm t}$ with an extremely high 
luminosity, ranging from 
$2\times 10^{36}$~cm$^{-2}$s$^{-1}$ per interaction point (IP),  
on the Z pole (91 GeV c.m.), 
$7\times 10^{34}$~cm$^{-2}$s$^{-1}$ per IP at the ZH production peak
and $1.3\times 10^{34}$~cm$^{-2}$s$^{-1}$ per IP at the ${\rm t}\bar{\rm t}$. 
In the case of four experiments,
the total luminosity on the Z pole will be
close to $10^{37}$~cm$^{-2}$s$^{-1}$. 
FCC-ee will also  offer unprecedented energy resolution, both
on the Z pole and at the WW threshold. 

The FCC-ee represents  a low-risk technical solution for an electroweak 
and Higgs factory, which is based on 60 years 
of experience with e$^{+}$e$^{-}$ circular colliders and particle detectors.
R$\&$D is being carried out on components for improved performance, but 
there is no need for ``demonstration'' facilities, as
LEP2, VEPP-4M, PEP-II, KEKB, DA$\Phi$NE, or SuperKEKB already 
used many of the key ingredients in routine operation.

The FCC shall be located in the Lake Geneva basin and be linked to the existing CERN facilities. 
The FCC utility requirements are similar to those in actual use at CERN.  
The FCC ``integrated programme'' consists
of the FCC-ee Higgs and electroweak factory as a first stage, succeeded by a 100 TeV hadron collider, FCC-hh, as the ultimate goal.
This sequence of FCC-ee and FCC-hh  
is inspired by the successful past Large Electron Positron collider (LEP) and Large Hadron Collider (LHC)  projects at CERN. 
The FCC-hh would re-use the tunnel and the technical infrastructure of the FCC-ee.   
% It represents a comprehensive long-term programme maximising physics opportunities. 
A similar two-stage project is under study in China, under the name CEPC/SPPC \cite{cepc}. 

The FCC complex will also allow for heavy-ion collisions,  
proton-lepton collisions (FCC-eh), high-energy high-luminosity 
electron-ion collisions and numerous other options, extending
to Gamma Factories and even muon colliders (FCC-$\mu\mu$). 

In summer 2021 a detailed FCC Feasibility Study focused on siting, 
tunnel construction, environmental impact,
financing, operational organisation, 
etc., was launched by the CERN Council.
This FCC Feasibility Study (FCC FS) 
should provide the necessary input to 
the next European Strategy Update expected in 2026/27.

The FCC technical schedule foresees the start of tunnel construction in the early 2030s, 
the first  $\rm e^{+}e^{-}$ collisions at the FCC-ee in the mid or late 2040s, and the first FCC-hh hadron collisions around the year 2070.

\subsubsection{Design outline}
The FCC-ee is conceived as a double ring e$^+$e$^-$  collider. 
It shares a common footprint with  the 100 TeV hadron collider, FCC-hh, 
that would be the second stage of the FCC integrated programme. 

The FCC-ee design features a novel asymmetric IR layout and optics to limit the 
synchrotron radiation emitted towards the detector (one of the lessons from LEP), 
and to generate the  large crossing angle 30 mrad, required for the 
virtual crab-waist collision scheme. 

The FCC-ee layout has a superperiodicity of four 
and can accommodate either two or four experiments.   
Key parameters are summarized in Table  \ref{tab:fccee}.
Thanks to resonant depolarization, at the two lower
energies, a precision energy calibration is possible, 
down to 100 keV accuracy for $m_Z$ and 300 keV for $m_W$. 

The crab waist collision scheme was first demonstrated at DA$\Phi$NE,
where it tripled the collider luminosity. 
More recently, in 2020, 
at SuperKEKB the ``virtual'' crab waist collision first developed for the  
FCC-ee \cite{koide}, 
was successfully implemented, and is now used in routine operation \cite{funakoshi:ipac22-moplxgd1}. 
SuperKEKB is also already operating with a vertical 
IP beta function $\beta_{y}^{\ast}$ of 1 mm when delivering luminosity to Belle II,  
and, during accelerator studies, 
has further squeezed $\beta_{y}^{\ast}$ down to 0.8 mm,
the smallest value 
considered for FCC-ee (see Table \ref{tab:fccee}). 

\begin{figure}[htb]
\begin{center}
\includegraphics[width=0.75\linewidth]{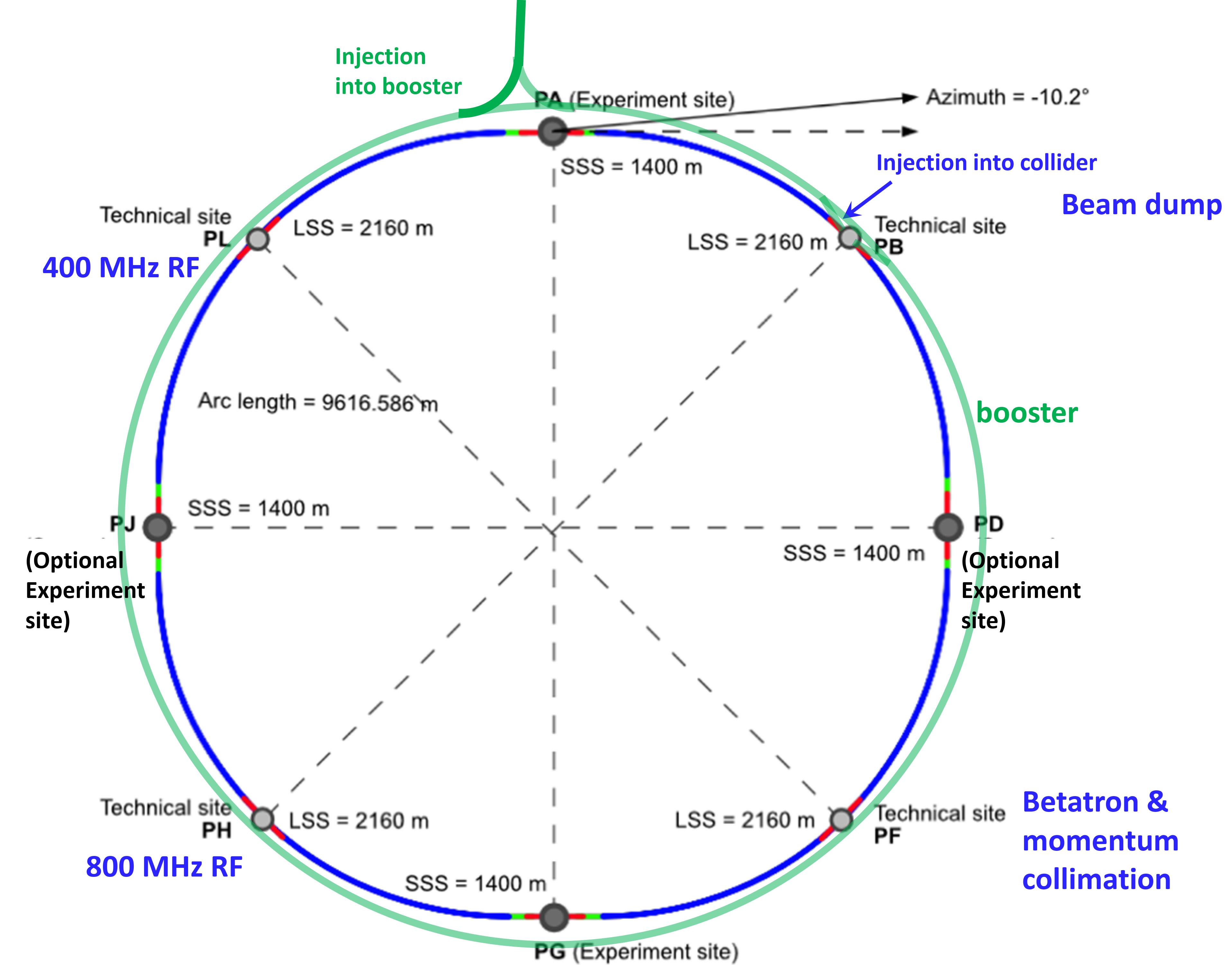}
\vspace*{-3 mm}
\end{center}
\caption{Schematic layout of FCC-ee and its booster 
with a circumference of 91.1 km and strict
four-fold superperiodicity. }  
\label{fcclayout}
\end{figure}

In Table \ref{tab:fccee} the synchrotron radiation power is assumed 
to be limited to 50 MW per beam.
As the centre-of-mass energy is increased, the synchrotron radiation 
power is kept constant,  
primarily by reducing the number of bunches.   
Top-up injection requires a full-energy 
booster synchrotron in the collider tunnel.
Figure ~\ref{fcclayout} sketches 
the layout and possible straight-section functions 
for the FCC-ee.

\begin{table}[htbp]
\caption{Preliminary key parameters of FCC-ee
(K. Oide, 2021), as evolved from the CDR parameters, 
now with a circumference of 91.1 km,
and a new arc optics for Z and W running. 
Luminosity values are given per 
interaction point (IP),
for a scenario with 4 IPs.  
Both the natural bunch lengths due to synchrotron
radiation (SR) and their values in 
collision including the effect of  
beamstrahlung (BS) are shown.
The FCC-ee considers a combination of 400 MHz radiofrequency systems  
(at the first three energies, up to $2\times$2 GV) 
and 800 MHz (additional cavities for ${\rm t}\bar{\rm t}$ operation),
with respective voltage strengths as indicated.
The beam lifetime shown represents the combined effect of
the luminosity-related 
radiative Bhabha scattering and 
beamstrahlung, the latter relevant
only for ZH and ${\rm t}\bar{\rm t}$ running
(beam energies of 120 and 182.5 GeV).}
\begin{center}
\begin{tabular}{lcccccc}
Quantity & Symbol & Unit & Z & W & ZH & $t \bar t$ \\
\hline
Centre of mass energy & $\sqrt{s}$ & ${\mathrm{GeV}}$ &45.6 & 80 & 120 & 183\\ 
Number of IPs & $N_{IP}$ & ~ & 4 & 4  & 4 & 4 \\
SR power per beam & $P_{SR}$ & MW & 100 &100&100&100 \\
Luminosity per IP & $L_{IP}$ & nb$^{-1}$s$^{-1}$ & 1810 & 173 & 72 & 12.5 \\
Energy loss per turn & $U_0$ &GeV & 0.04 & 0.37 & 1.87  & 10.0  \\
Circumference & C & km & 98&98&98&98 \\ \hline
Bunch number &$N_b$& ~ & 8800 & 1120 & 336 & 42 \\
Beam current & $I$ & mA & 1400  & 135 &  26.7 &  5.0 \\ \hline
Bunch length (wo/w BS) & $\sigma_z$ &mm& 4.4/14.5 &  3.6/8.0 &  3.3/6.0 & 2.0/2.8  \\
Beta function at IP &$\beta_y$& mm & 0.8 & 1.0 & 1.0 & 1.6 \\
Emittance &$\varepsilon_x/\varepsilon_y$&nm/pm & 0.71/1.41  & 2.17/4.34 & 0.64/1.29 & 1.49/2.98 \\ \hline
RF voltage 0.4/0.8GHz & $V_{RF}$& GV & 0.12/0 & 1.0/0 & 2.1/0 & 2.5/8.8 \\
Long.~damp.~time & & turns & 1170 & 216 & 64.5 & 18.5 \\
Beam lifetime & $\tau$ &min& 19 & 20 & 7 & 10 \\ \hline
\end{tabular}
\label{tab:fccee} 
\end{center}
\end{table}

\subsubsection{Project Cost and Profile}
The FCC CDR of 2019 
included a cost estimate for the first stage, the FCC-ee,
which is reproduced 
in Table \ref{tab:cost}.

\begin{table}[htbp]
\caption{Construction cost estimate (in 2018 CHF) for FCC-ee considering a 
machine configurations at the  Z, W, and  H working points. A baseline configuration with 2 detectors is assumed. The 
CERN contribution to 2 experiments is included.
\label{tab:cost} 
}
\begin{center}
\begin{tabular}{lcc}
\hline\hline
Cost category & MCHF & \% \\
\hline
Civil engineering & 5,400 & 50 \\
Technical infrastructure & 2,0009 & 18 \\
Accelerator & 3,300 &  30 \\
Detector & 200 & 2 \\
\hline total cost (2018 prices) & 10,900 & 100\\
\hline
\end{tabular}
\end{center}
\end{table}

\begin{figure}[htb]
\begin{center}
\includegraphics[width=0.98\linewidth]{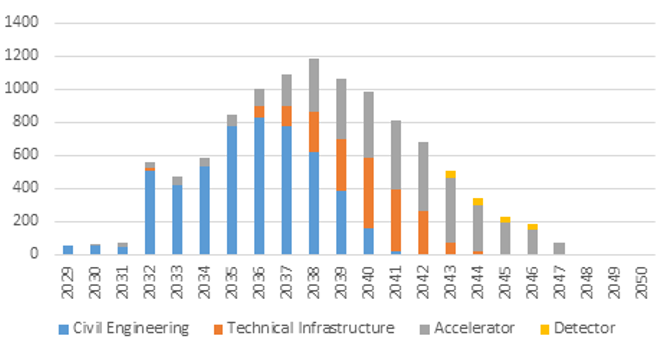}
\vspace*{-3 mm}
\end{center}
\caption{Example draft spending profile for 
FCC-ee, in units of MCHF versus the year
(M.~Benedikt). }  
\label{fig:spending}
\end{figure}

A draft spending profile for FCC-ee is displayed in Fig.~\ref{fig:spending}.
This figure assumes civil engineering construction from 2032 to 2040,
installation of technical infrastructure 
from 2037 to 2043,
construction of accelerator and experiments during the years 2032--2045, and, finally, commissioning and start of operation 
in the period 2045--2048.

\subsubsection{FCC-ee R\&D} 
Many of the technologies 
required for constructing an FCC-ee exist since several
decades \cite{richter}.  
Ongoing FCC-ee research and development (R\&D) efforts focus 
on further improving the overall energy efficiency, 
on obtaining the measurement precision required, and on achieving the 
target performance in terms of beam current and luminosity.  

Key FCC-ee R\&D items for improved energy efficiency  
include high-efficiency continuous wave (CW)
radiofrequency (RF) power sources 
(klystrons, IOTs and/or solid state), high-$Q_0$ SC cavities 
for the 400--800 MHz range, and possible applications of 
high-temperature superconductor (HTS) magnets. 
For ultra high precision centre-of-mass energy measurements, 
the R\&D should cover 
state-of-art and beyond in terms of spin-polarisation simulations and measurements (inv.~Compton, beamstrahlung, etc.). 
Finally, for high luminosity, high current operation,
FCC-ee requires a next generation beam stabilization/feedback 
system to suppress instabilities arising over a few turns, a robust
low-impedance collimation scheme, and a machine tuning system based on artificial 
intelligence. 

\paragraph{SRF Cavity Developments}
Since PETRA, TRISTAN and LEP-2,  superconducting RF systems are the underpinning technology for modern circular lepton colliders. 
The FCC-ee baseline foresees the use of 
single-cell 400 MHz Nb/Cu cavities for high-current low-voltage beam operation at the Z production energy,
four-cell 400 MHz Nb/Cu cavities at the W and H (ZH) energies, and 
a complement of five-cell bulk Nb 800 MHz cavities at 2 K  for low-current high-voltage
${\rm t\overline t}$ operation \cite{Abada:2019zxq}.
In the full-energy 
booster, only multi-cell 400 and 800 MHz cavities will be installed. 
For the FCC-ee collider,  also alternative RF scenarios, 
with possibly fewer changes between operating points, are being 
explored, such as novel 600 MHz slotted waveguide elliptical (SWELL) cavities \cite{swell}.

\paragraph{R\&D for the FCC-ee Arcs}
Aside from the various RF systems, another major
component of the FCC-ee is the regular arc, covering almost 80~km.  
The arc cells must be cost effective, 
reliable and easily maintainable. Therefore,  
as part of the FCC R\&D program it is 
planned to build a complete arc half-cell mock up including girder, vacuum system with antechamber and pumps, dipole, quadrupole and sextupole magnets, beam-position monitors, cooling and alignment systems, and technical infrastructure interfaces, by the year 2025. 
Constructing some of the magnets for the FCC-ee final focus or arcs with advanced  high-temperature superconductor (HTS) technology could lower energy consumption and increase operational flexibility. The focus  of this HTS R\&D will not 
be on reaching extremely high field, but on operating lower-field 
SC magnets at temperatures  between 40 and 77~K. 
Nevertheless, this development could also be a step towards 
higher field HTS magnets for the hadron collider FCC-hh, where
operation at 40 K instead of 2 K, would dramatically 
reduce the electric power consumption.

\paragraph{Polarimetry and Centre-of-Mass Energy Calibration}
Highly precise centre-of-mass energy calibration 
at c.m.~energies of 91 GeV (Z pole) and 
160 GeV (WW threshold), a cornerstone of the precision physics programme of the FCC-ee, relies on using resonant depolarisation 
of wiggler-pre-polarised pilot bunches \cite{blondelprec}.  The operation with polarised pilot bunches requires constant and high precision monitoring of the residual 3-D spin-polarization of the colliding bunches,  which --- if nonzero --- would affect the physics measurements.

\paragraph{FCC-ee Pre-Injector}
Concerning the FCC-ee  pre-injector, the CDR design foresaw 
a pre-booster synchrotron.  
At present, this choice is under scrutiny. 
As an alternative, and possibly new baseline,
it is proposed to extend the energy of the injection linac to 
10--20 GeV, for direct injection into the 
full-energy booster \cite{craievich:ipac22-wepopt063}. 
The higher-energy linac could be 
based on state-of-the-art S-band
technology as employed for the FERMI upgrade at the ELETTRA synchrotron radiation facility. Alternatively, a C-band linac could be considered, possibly
based on the C$^3$ technology \cite{bai2021c3}. 
The ongoing FCC-ee R\&D program 
includes the design, construction and, finally, testing   
with beam of a novel 
positron source \cite{craievich:ipac22-wepopt063,humann:ipac22-thpotk048}  
plus capture linac, 
at the PSI SwissFEL facility, with a primary electron 
energy that can be varied from 0.4 to 6 GeV. 
This will allow measuring the achievable positron yield and validating the 
innovative concept. 

% to measure the positron 
% yield and compare the performance with the simulations.
\paragraph{Full Energy Booster}
The injection energy for the full-energy booster is defined by 
the field quality of its low-field magnets.
Magnet development and prototyping of booster 
dipole magnets, along with field measurements (presently only available for the twin collider  CEPC \cite{Kang:2021mxp}),
should guide the choice of the injection energy.
Maintaining beam stability at injection into the booster 
may require the installation of wiggler magnets 
for increasing the beam energy spread. 
An alternative optics, which may both
increase the SR energy spread and avoid very low magnetic fields,
is based on alternating the polarity of arc  
dipole magnets at injection, reminiscent of 
what is being planned for the 
Electron Storage Ring (ESR) of the 
US Electron Ion Collider (EIC) \cite{marx:ipac22-wepopt042,osti_1765663},
although the FCC-ee booster is fast ramping, 
while the ESR will operate at different constant beam energies.

\paragraph{Role of SuperKEKB}
The SuperKEKB collider, presently being commissioned \cite{yukiyoshi}, 
features many of the key elements of FCC-ee:   
double ring, large crossing angle, low vertical IP beta function $\beta_{y}^{\ast}$ 
(design value $\sim$0.3~mm), 
short design beam lifetime of a few minutes, top-up injection, and a positron  production rate of up to several $ 10^{12}$/s.  
SuperKEKB has achieved, in both rings, the world's smallest ever 
$\beta_{y}^{\ast}$ of $0.8$~mm,
which also is the lowest value 
considered for FCC-ee. 
Profiting from a new 
``virtual'' crab-waist collision scheme,
first developed for FCC-ee \cite{koide},
in July 2022 SuperKEKB reached a world record
luminosity of  $4.71 \times 10^{34}$~cm$^{-2}$s$^{-1}$. 
However, several issues still need to be resolved, 
such as  a vertical emittance blow up,    
the transverse machine impedance, 
and single-bunch instability threshold, 
sudden beam losses, poor 
quality of the injected beam, etc.  
Training and learning at SuperKEKB is an excellent preparation for future Electroweak Higgs Factories, such as the FCC-ee. 

% SuperKEKB is pushing the frontiers of accelerator physics with a vertical rms beam spot size of about 300 nanometer, the lowest of any operating collider.  
% The future goal is pushing the luminosity to $6 \times 10^{35}$~cm$^{-2}$s$^{-1}$, and a beam spot size of 50 nm.
% SuperKEKB serves as an important test-bed for FCC-ee and other future electron-positron colliders and as a unique 
% facility for training the next generation of 
% accelerator physicists, who will be commissioning  
% future colliders like the FCC-ee.

\paragraph{Collaboration with the EIC}
The EIC ESR \cite{osti_1765663}  has almost identical beam parameters as FCC-ee, except that it foresees about twice the maximum electron beam current, or half the bunch spacing, 
and that it will operate at lower beam energy.
About ten domains of common interest have been identified by the FCC and EIC design teams, for each of which a joined EIC-FCC working group is being set up. 
The EIC will start beam operation about a decade prior to FCC-ee.
It would, thereby, provide another invaluable opportunity to train the next generation of accelerator physicist on an operating collider,
to test hardware prototypes, beam control schemes, etc.

\begin{figure}[htb]
\begin{center}
\includegraphics[width=0.7\linewidth]{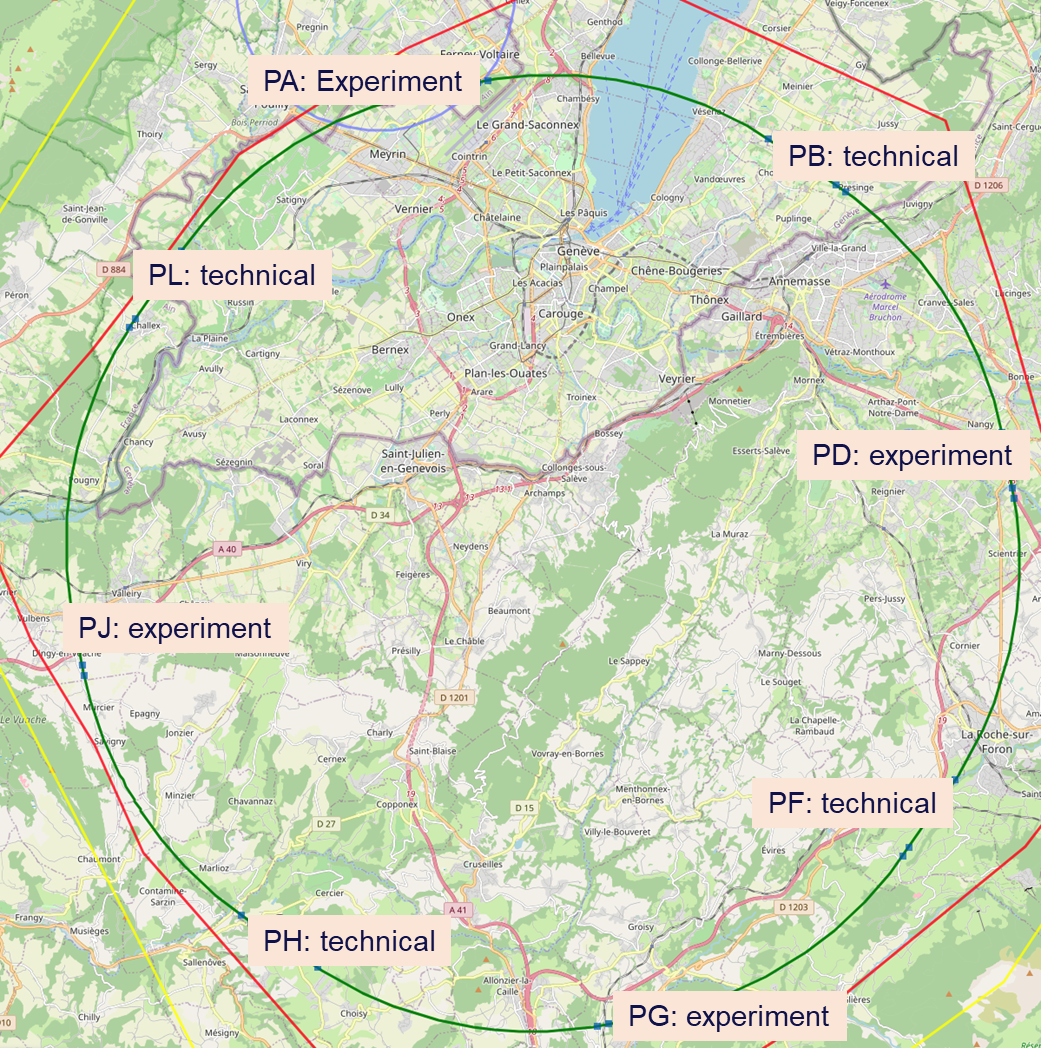}
\vspace*{-3 mm}
\end{center}
\caption{Optimized placement of the FCC in the Lake Geneva basin. }  
\label{optplac}
\end{figure}

\begin{figure}[htb]
\begin{center}
\includegraphics[width=0.98\linewidth]{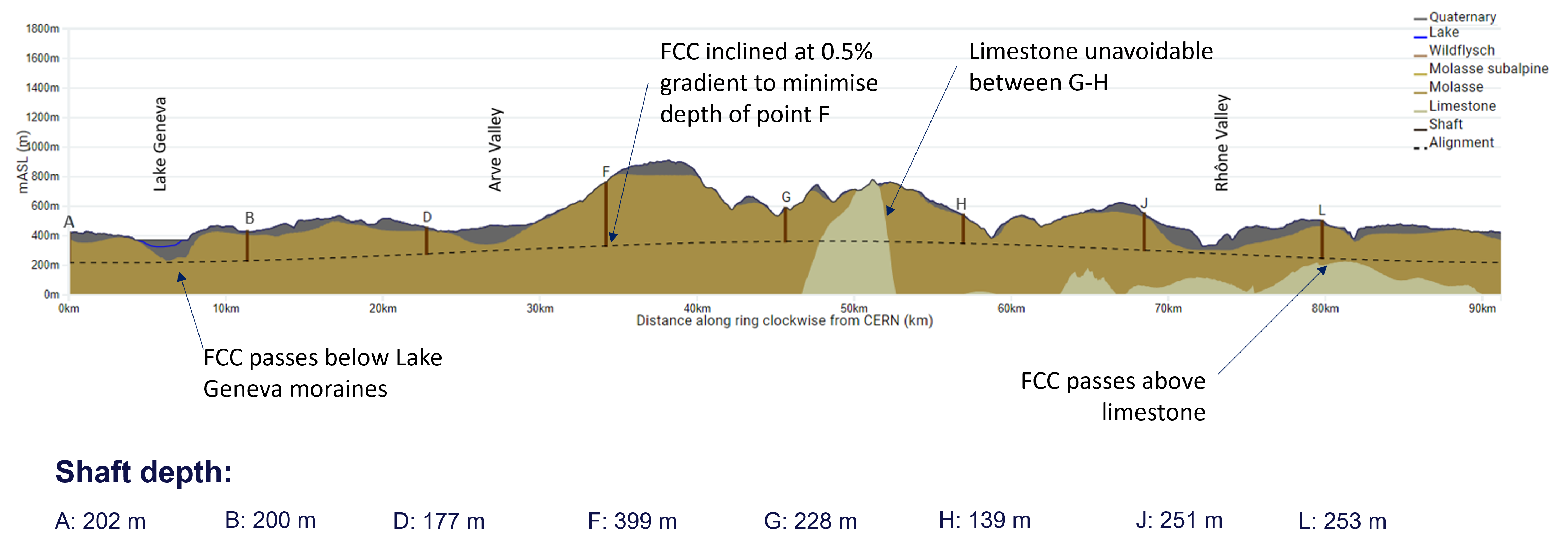}
\vspace*{-3 mm}
\end{center}
\caption{Long section of the optimally placed FCC. }  
\label{longsec}
\end{figure}

\paragraph{Civil Engineering and Site}
In 2021, the placement and layout of the FCC (FCC-ee and FCC-hh)
was optimized, taking into account numerous constraints and considerations,
including geological conditions, depth of access shafts, vicinity of 
access roads, railway connections, etc.,
while avoiding surface sites in water protection
zones, densely urbanized areas, and high mountains.
The number of surfaces sites was reduced from 12 in the Conceptual Design Report to 8, which facilitates the placement and decreases the required surface area from 62 ha to less than 40 ha. 
In addition, the 8 surface sites and new layout 
are arranged with a perfect 4-fold superperiodicity, which allows
for either two or four collision points and experiments. 
This should ensure the best possible beam-dynamics
performance for both lepton and hadron collider.  
The resulting optimized placement is illustrated in Fig.~\ref{optplac}, and the corresponding long section showing the geological situation and the depths of access shafts in Fig.~\ref{longsec}. 
All proposed surface sites are close to road infrastructures, so that in total less than 5 
km of new road constructions are required 
for all sites together. 
Several sites are located in the vicinity of 400 kV electricity grid lines.
Finally, the good road connection of Points PD, PF, PG, PH 
suggest a second operation pole around Annecy (CNRS LAPP) 
in the South.
Detailed site investigations are planned for the period 
2024--2025, with about 40 to 50 drillings and some  
100 km of seismic lines.

 \paragraph{Sustainability}
According to the conceptual design, the FCC-ee 
is the most sustainable of all the advanced Higgs and electroweak 
factory proposals, in that it implies by far the lowest energy consumption
for a given value of total integrated luminosity, 
over the collision energy range from 90 to 365 GeV \cite{naturephysics}. 
This means, for example, that the 
energy needed to produce one Higgs boson at the FCC-ee is much smaller 
than at other realistic Higgs factories under consideration 
(with the possible exception for the CEPC, the design of which is quite 
similar), because of the much 
larger instantaneous luminosity for comparable energy consumption
and the number of interaction points (2 or 4). 

The electrical power
consumption depends on the centre-of-mass energy. 
An estimation of the upper limit of the power 
drawn by the various FCC-ee systems for each mode of operation 
was first presented in 
\cite{Aull:2156972} and updated recently \cite{burnet}.  
Depending on the collision energy the total facility power
extends from about 238 MW at the Z to 
388 MW at the ${\rm t}\bar{\rm t}$ energy.  
These values are comparable in order of magnitude 
with CERN's present  power consumption of about 200 MW,  
when LHC is operating, or with a total 
CERN power consumption of up to $\sim$240 MW  at the time of the previous LEP collider.  
The numbers  include the power required for cooling and ventilation,
for general services, for 
two experiments, for data centres, 
and  for the injector complex.
Although the FCC-ee is three to four times larger than LEP, 
and achieves about $10^{5}$ times the LEP luminosity,
the design concept leads to an overall electrical peak power
of only about 2.5 times the one of LEP, which alone 
consumed  $\sim$120 MW. 
Adding to FCC-ee operation also the powering required
for the present CERN site 
running various lower-energy hadron accelerators,
and for a parallel fixed target proton programme at the existing
CERN SPS North Area,
the total annual energy consumption is expected to 
range from about 1.8 TWh 
at the Z to 2.5 TWh at the ${\rm t}\bar{\rm t}$ \cite{burnet}.  
Additional technology advancements and design optimisation, such as the introduction of  HTS magnets in the collider rings or of permanent magnets in the damping ring, 
will further reduce  the FCC-ee energy consumption. 

The FCC-ee will be powered by a mixture
of renewable and other carbon-free sources.
Today, the electricity produced and consumed in France and Switzerland is already 
more than 90\% carbon-free, an order of magnitude better 
than in most other countries \cite{carbonbrief}. 
Combining this with the high total luminosity per unit electrical power, 
the electricity 
carbon footprint of each Higgs boson produced by the FCC-ee 
is up to two orders of magnitude smaller than for many of the 
alternative Higgs factory proposals \cite{janotgreen2022}.   
% By 2045, the electricity in France and Switzerland 
% is expected to be 100\% carbon free.

The FCC-ee power consumption can be rapidly
adjusted to the power available on the European electricity grid,  simply by 
varying the number of bunches in the collider.

Beside the electrical energy consumed during operation, the tunnel construction itself is also linked to sustainability.  
The optimum use of the FCC-ee tunnel excavation material is being studied through the international competition ``Mining the Future\textsuperscript{\textregistered}'' \cite{miningf}.

\subsubsection{Proposals for upgrades and extensions}

\paragraph{Wide Physics Programme and Multiple Experiments}
The FCC-ee is not only a Higgs, but also a Z and W factory (``TeraZ''). The upgrade to ${\rm t}\bar{\rm t}$ running 
is foreseen, at a cost of about 1 BCHF for additional systems.

Four different FCC-ee detectors placed at the four symmetric 
collision points could be optimized, respectively, for
the {Higgs factory} programme, for {ultraprecise electroweak and QCD} physics, for {Heavy Flavour} physics, and for searching {feebly coupled particles (LLPs)} \cite{dam}.
For the FCC-hh, two high-luminosity general-purpose experiments and two specialized experiments are foreseen \cite{FCC:2018vvp}, 
similar to the present LHC detectors. 

% If warranted by a physics case, also far detectors could be installed at the FCC-ee.

\paragraph{Monochromatized collisions for direct $s$-channel Higgs production}
In addition to the 4 baseline running modes
listed in Table \ref{tab:fccee}, another optional 
operation mode, presently under investigation 
for FCC-ee, is the direct $s$-channel Higgs production, ${\rm e}^{+}{\rm e}^{-}\rightarrow {\rm H}$,
at a centre-of-mass energy of 125 GeV,
which would allow a direct measurement of the electron Yukawa coupling.  
Here, a monochromatization scheme 
should reduce the effective collision energy spread in order for the latter to become 
comparable to the width of the Higgs \cite{afausepj}.  

\paragraph{Hadron Collider FCC-hh}
The FCC integrated program foresees as a second stage, following the FCC-ee, a 100 TeV hadron collider, FCC-hh, which will be installed in the same tunnel and which shares/re-uses much of the technical infrastructure (electric distribution, cooling and ventilation, RF, cryogenics, experimental caverns,...).

\paragraph{Other Extensions}
Numerous other possible extensions
are firmly considered or under study, 
such as heavy-ion collisions, lepton-proton and lepton-hadron collisions (FCC-eh) \cite{FCC:2018vvp},
LHC and FCC based Gamma factories \cite{krasnyGF},
ERL-based upgrades like CERC (discussed in a separate section),   
and a Lemma-type 100 TeV muon collider, FCC-$\mu \mu$ \cite{FCCmumu1,FCCmumu2}, 
which could be based on key elements of the FCC-ee and FCC-hh accelerators. 

% \subsubsection{FCC Schedule}   
% The technical schedule of the FCC integrated project 
% foresees the start of FCC tunnel construction around
% the year 2030 --- or three years 
% after a possible project approval ---, the first 
% e$^+$e$^-$ collisions at the FCC-ee in the mid or late 
% 2040s, and the first FCC-hh hadron 
% collisions by 2065--70 --- see Fig.~\ref{fccschedule}. 
% The FCC integrated project would allow 
% for a seamless continuation of 
% High Energy Physics (HEP) after the
% completion of the High Luminosity LHC (HL-LHC) physics programme. 

% \begin{figure}[htb]
% \begin{center}
% \includegraphics[width=0.95\linewidth]{Accelerator/AF03/figures/FCC-schedule.png}
% \vspace*{-3 mm}
% \end{center}
% \caption{Technical schedule of the FCC integrated project
% with year 1 equal to 2021 (a similar schedule was presented 
% in Ref.~\protect\cite{naturefcc}).}   
% \label{fccschedule}
% \end{figure}

\subsubsection{State of Proposal and R$\&$D plans}
The FCC proposal emerged and evolved during the period 2010--13 (FCC-hh) and in 2011--12 (FCC-ee), 
respectively, originally under the names VHE-LHC and TLEP. 
The 2013 European Strategy Update (ESU) requested a 
Conceptual Design, the four-volume report of which was delivered in 2019
 \cite{Abada:2019zxq,FCC:2018vvp,FCC:2018byv}, describing the 
physics cases, the design of the lepton and hadron colliders,
and the underpinning technologies and infrastructures.
Following the 2020 ESU, the FCC Feasibility Study (FS) 
has been launched by CERN Council in 2021, with 
a Feasibility Study Report (FSR) expected by 2025.
The FSR will address not only the technical design, 
but also numerous other feasibility aspects,
including exact siting, construction, financing, and environment. 

The FCC FS is organized as an international collaboration with, presently, about 150 participating institutes from around the world. 
The FCC FS and a future project will profit from 
CERN's decade-long experience with successful large international accelerator projects, e.g., the LHC and HL-LHC, and the associated global  
experiments, such as ATLAS and CMS.
% to all which the US has made essential contributions. 
% The US participation in CERN based accelerators and experiments during the past decades has been of great mutual benefit.

A comprehensive R\&D program and implementation preparation is presently being carried out in the frameworks of the 
FCC FS, the EU co-financed FCC Innovation Study,
the Swiss CHART program, and the CERN High-Field Magnet Programme.

The first stage of FCC could be approved within a few years 
after the 2027 Strategy Update, if the latter is supportive.
The tunnel construction could then start in the early 2030s and
the FCC-ee physics program begin in the second half of the 2040s, a few years after the completion of the HL-LHC physics runs expected by 2041.

The second stage of the FCC integrated program is the 100 TeV hadron collider FCC-hh, which might start operation around the year 2070.

% \subsubsection{State of Technical Design Report} 

% with the goal to provide input
% to the next Update of European Strategy for Particle Physics, expected in 2026/27.

% The FCC infrastructure will  
% support a century, or more, of frontier particle physics.

%%%%%%%%%%%%%%%%%%%%%%%%%%%%%%%%%%%%%%%%%%%%%%%%%%%%%%%
\newpage

\subsection{Circular Electron Positron Collider (CEPC), \cite{CEPC} }

\subsubsection{Design outline}
\begin{figure}[h]
\centering
\includegraphics[width =
0.75\textwidth]{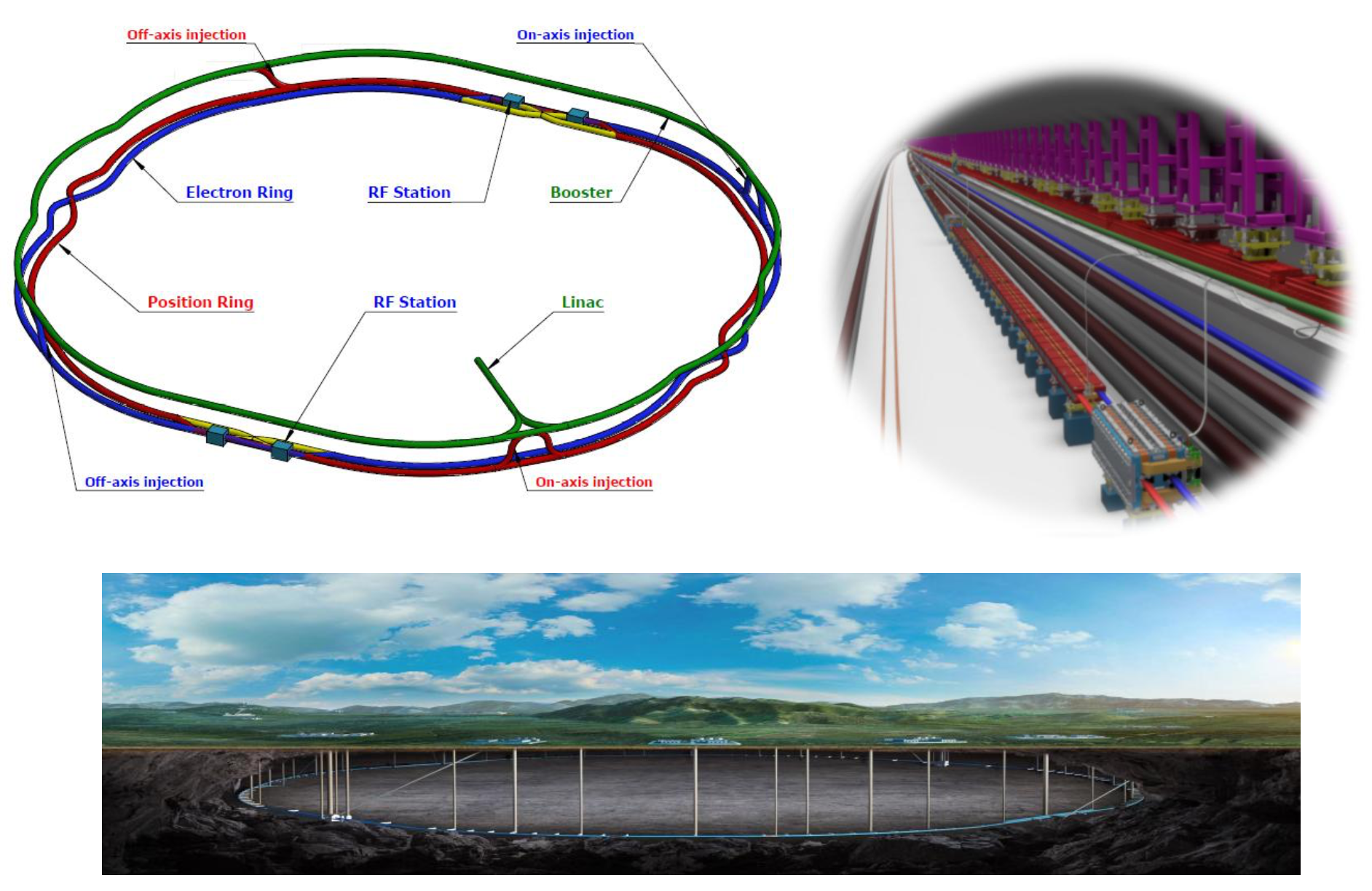}
\caption{CEPC layout: Linac injector, booster and collider rings.}
\label{CEPC_sequence}
\end{figure}

The Circular Electron Positron Collider (CEPC) is a new collider proposed in September 2012, by Chinese scientists  with 240~GeV to serve two large detectors for Higgs studies and other topics (see Fig.\ref{CEPC_sequence}). The $\sim$100~km tunnel for such a machine could also host a Super proton-proton Collider (SppC) to reach energies beyond the LHC.

\paragraph{High efficiency klystron and RF source}

The CEPC two beam synchrotron radiation power is more than 60~MW, it needs high efficiency RF source to minimize CEPC AC power consumption. Considering the klystron operation lifetime and power redundancy, a single 650-MHz 800-kW klystron amplifier will drive two of the collider ring SC cavities through a magic tee and two rated circulators and loads. The CEPC high efficiency 650-MHz klystron design goals are to set the efficiency to be above 80\% and successful industrialization. 

IHEP is developing 650-MHz klystron with 800-kW CW output power and 80\% efficiency. To achieve this goal, a couple of klystron prototypes are being manufactured presently. The first prototype has been completely developed in 2020 with traditional bunching method with the efficiency reaching up to 62\%. The high efficiency klystron prototype has also been developed at the end of 2021 with the output power and efficiency to be 800-kW and 75\% respectively. This prototype is being high power tested on the site of Platform of Advanced Proton Source (PAPS). Moreover, a multi-beam klystron prototype is being developed with designed efficacy of higher than 80\%. The design schemes of high efficiency klystron with other methods are also in progress.

\paragraph{SRF technology}

The CEPC SRF technical challenges that require R\&D include: achieving the cavity gradient and high-quality factor in the real cryomodule environment, robust and variable high power input couplers that are design compatible with cavity clean assembly and low heat load, efficient and economical damping of the HOM power with minimum dynamic cryogenic heat load, and fast RF ramp and control of the Booster.

Impressive test results are obtained on CEPC key SRF components and 650-MHz prototype cryomodule, taking advantages of the new large SRF infrastructure (PAPS). CEPC design goal with world-leading high Q and high gradient 650-MHz and 1.3-GHz cavities have been recently achieved at IHEP with novel recipes. In synergy with CEPC SRF R\&D, large CW XFEL projects in China, such as SHINE (Shanghai HIgh repetition rate XFEL aNd Extreme light facility) in Shanghai etc., will need a total of one thousand high Q 1.3-GHz 9-cell TESLA cavities and their cryomodules in next five years, while IHEP is playing a leading role in the key technology development in this national SRF context. In parallel with design and key R\&D, extensive development of SRF personnel, infrastructure and industrialization is essential for the successful realization of CEPC. Meanwhile, IHEP will maintain and extend CEPC SRF collaborations with international laboratories with strong expertise.

\paragraph{High field superconducting magnets for SppC}

All the superconducting magnets used in existing accelerators are based on NbTi technology. These magnets work at significantly lower field than the SppC 12$\sim$24~T required by SppC.  The upcoming 11~T dipole magnets for HL-LHC project are state-of-the-art superconducting magnets for accelerators.

SppC demands advanced or new type of superconducting materials with low cost and capable of applying in the high fields. Since 2008, iron-based superconductors (IBS) have been discovered and attracted wide interest for both basic research and practical applications. It has high upper critical field beyond 100~T, strong current carrying capacity and lower anisotropy. In 2016, the Institute of Electrical Engineering, Chinese Academy of Sciences (IEE-CAS) manufactured the world's 1st 100-m long 7-filamentary Sr122 IBS tape with critical current of $1\times10^4~{\rm A/cm}^2$ at 10~T successfully, which makes the possibility of fabricating real IBS coils. In 2018, IHEP and IEE fabricated the IBS solenoid coil and tested at 24~T successfully. In 2018 and 2019, IHEP fabricated the IBS racetrack coils wound with 100-m long IBS tapes produced by IEE. The quench current of the IBS coil at 10~T reached 81.25\% of its quench current at self-field. In 2021, the IBS solenoid coil developed by IHEP reached 67~A at 30~T background field. The work verified the IBS conductor could be a promising candidate for the application in high field superconducting magnets.

R\&D of high field model dipole is ongoing at IHEP, and in collaboration with related institutes working on fundamental sciences of superconductivity and the advanced HTS superconductors. A NbTi+Nb$_3$Sn twin-aperture magnet reached 12.47~T at 4.2~K in 2021. After that, Nb$_3$Sn+HTS (IBS or ReBCO) magnet with two $\Phi$ 45 mm apertures will be developed, aiming to reach 16+ T in 5 years, and 20$\sim$24~T in 10 years. The R\&D will focus on the following key issues related with the high field superconducting magnet technology: 
\begin{enumerate}
	\item Explore new methods and related mechanism for HTS materials with superior comprehensive performance for applications. Reveal key factors in current-carrying capacities through studying microstructures and vortex dynamics. Develop advanced technologies of HTS wires for high field applications with high critical current density (J$_{\rm c}$) and high mechanical strength.
	\item Development of novel high-current-density HTS superconducting cables, and significant reduction of their costs. Exploration of novel structures and fabrication process of high field superconducting magnets, based on advanced superconducting materials and helium-free cooling method.
	\item Exploration of novel stress management and quench protection methods for high field superconducting magnets, especially for high field insert coils with HTS conductors. Complete the prototype development with high field and $10^{-4}$ field quality, lay the foundation for the applications of advanced HTS technology in high-energy particle accelerators.
\end{enumerate}

On top of the R\&D for above subsystems, intensive R\&D are operated on the cryogenic system, the collider and booster magnets, the superconducting magnets in the interaction region, the vacuum system, the electrostatic-magnetic separator, the mechanical supports, the remote vacuum connector and collimators, the beam dumps and machine protection, the control system, the beam instrumentation, the septum, the kicker and power supply, the alignment and installation preparations, and the plasma injector as an alternative option to the linac injector, see~\cite{WP_JGao}.
Considering the progress and prototype results of the key technology R\&D, it is concluded that the CEPC accelerator technologies will be ready for construction around 2026. 

\paragraph{Accelerator design}

According to the CEPC TDR baseline physics goals at the Higgs and Z-pole energies, the CEPC should provide $e^+e^-$ collisions at the center-of-mass energy of 240~GeV and deliver a peak luminosity of $5.0 \times 10^{34}~{\rm cm}^{-2}{\rm s}^{-1}$ at each interaction point (IP). The CEPC has two IPs for $e^+e^-$ collisions and is compatible with four energy modes ($t \bar t$, Higgs, W, and Z-pole). At the Z-pole, the luminosity is required to be larger than $1 \times 10^{36}~{\rm cm}^{-2}{\rm s}^{-1}$ per IP. The operation at $t \bar t$ energy is an upgrade at the last stage of CEPC. 

The CEPC TDR design is a 100-km long double ring scheme based on crab-waist collision and 30 MW radiation power per beam at four energy modes, with the shared RF system for Higgs/$t \bar t$ energies and the independent RF system for W/Z energies.

The main parameters of the CEPC for TDR are listed in Table~\ref{tab:cepcParameters}. The luminosity at the Higgs energy is $5 \times 10^{34}~{\rm cm}^{-2}{\rm s}^{-1}$. At the Z-pole, the luminosity is $1.15 \times 10^{36}~{\rm cm}^{-2}{\rm s}^{-1}$ for the 2~T detector solenoid. The limit of the bunch number at the Z-pole comes from the electron cloud instability of the positron beam. A fast transverse feedback system is designed to control the multi-bunch instability induced by the impedance at the Z-pole.

The crab-waist scheme increases the luminosity by suppressing vertical blow up, which is a must to achieve high luminosity. Beamstrahlung is synchrotron radiation excited by the beam-beam force, which is a new phenomenon in a storage ring based collider, especially in the high energy region. It will increase the energy spread, lengthen the bunch, and may reduce the beam lifetime due to the long tail of the photon spectrum. The beam-beam limit at the W/Z is mainly determined by the coherent x-z instability instead of the beamstrahlung lifetime as in the $t \bar t$/Higgs mode. A smaller phase advance of the FODO cell (60$^{\circ}$/60$^{\circ}$) for the collider ring optics is chosen at the W/Z mode to suppress the beam-beam instability when we consider the beam-beam effect and longitudinal impedance consistently. The CEPC TDR design goals have been evaluated and checked from the point view of beam-beam interaction, which is feasible and achievable.
The CEPC TDR upgrade parameters of 50 MW SR power at Higgs, W, Z, and $t \bar t$\ energy operations and luminosities are listed in Table~\ref{tab:cepcParameters}.

\begin{table}[htbp]
\caption{CEPC main parameters in TDR. }
\begin{center}
\begin{tabular}{lcccccc}
Quantity & Symbol & Unit & Higgs & W & Z & $t \bar t$ \\
\hline
Centre of mass energy & $\sqrt{s}$ & ${\mathrm{GeV}}$ & 120 & 80 & 45.5 & 180 \\
Number of IPs & $N_{IP}$ & ~ & 2 & 2  & 2 & 2 \\
SR power per beam & $P_{SR}$ & MW & 30/50&30/50&30/50&30/50 \\
Half crossing angle at IP&$\theta_{IP}$ & mrad & 16.5 &16.5&16.5&16.5 \\
Bending radius & $\rho$ & km & 10.7 & 10.7 &10.7 &10.7 \\
Luminosity per IP & $L_{IP}$ & $10^{34}$cm$^{-2}$s$^{-1}$ & 5.0/8.3 & 16.0/26.7 & 115.0/192 & 0.5/0.8 \\
Energy loss per turn & $U_0$ &GeV & 1.8 & 0.357 & 0.037 & 9.1 \\
Piwinski angle &$\theta_P$& ~ & 5.94 & 6.08 & 24.68 & 1.21 \\
Circumference & C & km & 100&100&100&100 \\ \hline
Bunch number &$N_b$& ~ & 249 & 1297 & 11951 & 35 \\
Bunch spacing & $L_s$&ns & 636 & 257 & 23 (10\% gap) & 4524 \\
Bunch population &$n_e $&$10^{10}$ & 14 & 13.5 & 14 & 20 \\
Beam current & $I$ & mA & 16.7 & 84.1 & 803.5 & 3.3 \\ \hline
Momentum compaction &$ \alpha_p$ & $10^{-5}$ & 0.71 & 1.43 & 1.43 & 0.71 \\
Phase advance of arc FODOs &$ \phi$ & degree & 90 & 60 & 60 & 90 \\
Beta functions at IP &$\beta_x/\beta_y$& m/mm& 0.33/1 & 0.21/1 & 0.13/0.9 & 1.04/2.7 \\
Emittance &$\varepsilon_x/\varepsilon_y$&nm/pm & 0.64/1.3 & 0.87/1.7 & 0.27/1.4 & 1.4/4.7 \\
Beam size at IP & $\sigma_x/\sigma_y$& $\mu$m/nm & 15/36 & 13/42 & 6/35 & 39/113 \\
Bunch length (SR/total) & $\sigma_z$ &mm& 2.3/3.9 & 2.5/4.9 & 2.5/8.7 & 2.2/2.9 \\
Energy spread (SR/total)&$\sigma_e$& \% & 0.10/0.17 & 0.07/0.14 & 0.04/0.13 & 0.15/0.20 \\
Beam-beam parameters &$\xi_x/\xi_y$& ~  & 0.015/0.11 & 0.012/0.113 & 0.004/0.127 & 0.071/0.1 \\ \hline
RF voltage & $V_{RF}$& GV & 2.2 (2cell) & 0.7 (2cell) & 0.12 (1cell) & 10 (5cell) \\
RF frequency & $f_{RF}$ & MHz & 650&650&650&650 \\
Energy acceptance (DA/RF) & ~ &\% & 1.7/2.2 & 1.2/2.5 & 1.3/1.7 & 2.3/2.6 \\
Beam lifetime & $\tau$ &min& 20 & 55 & 80 & 18 \\ \hline
\end{tabular}
\label{tab:cepcParameters}
\end{center}
\end{table}

The CEPC lattice optics is designed with requirements and constraints mainly from top-level parameters, geometry, minimization of cost, compatibility of Higgs, W, Z, and $t \bar t$\ modes, and compatibility with SppC~\cite{3_CDR_Vol1, 5_Wang:2018avj, 6_Wang:2021hyy}.

The interaction region is designed to provide strong focusing and crab-waist collision~\cite{7_Zobov:2010zza}. To obtain a robust lattice, the length from IP to the final strong focusing quadrupole is chosen to be 1.9~m. A large full crossing angle of 33~mrad is chosen to provide large Piwinski angle with constraints from the machine-detector interface. A local chromaticity correction scheme is adopted to achieve a large momentum acceptance. An asymmetric lattice is used to allow softer synchrotron-radiation photons from the upstream part of the IP~\cite{8_Oide:2016mkm}.

In the arc region, twin-aperture dipoles and quadrupoles~\cite{3_CDR_Vol1, 9_Milanese:2016bsr} are used to reduce power. The two beams are separated by 35~cm. The FODO cell structure is chosen to provide a large filling factor of dipoles. The non-interleaved-sextupole scheme is selected due to aberration cancellation. For the Higgs and $t \bar t$\ energy modes, the 90/90-degrees phase advances are chosen to balance the aberration cancellation and the number of magnets. For the W and Z energy modes, 60/60-degrees phase advances are chosen to suppress the microwave and transverse mode coupling instability and to increase the stable tune area considering the beam-beam effect and longitudinal impedance~\cite{10_Zhang:2020jqb}.

\begin{figure}[h]
\centering
\includegraphics[width =
0.75\textwidth]{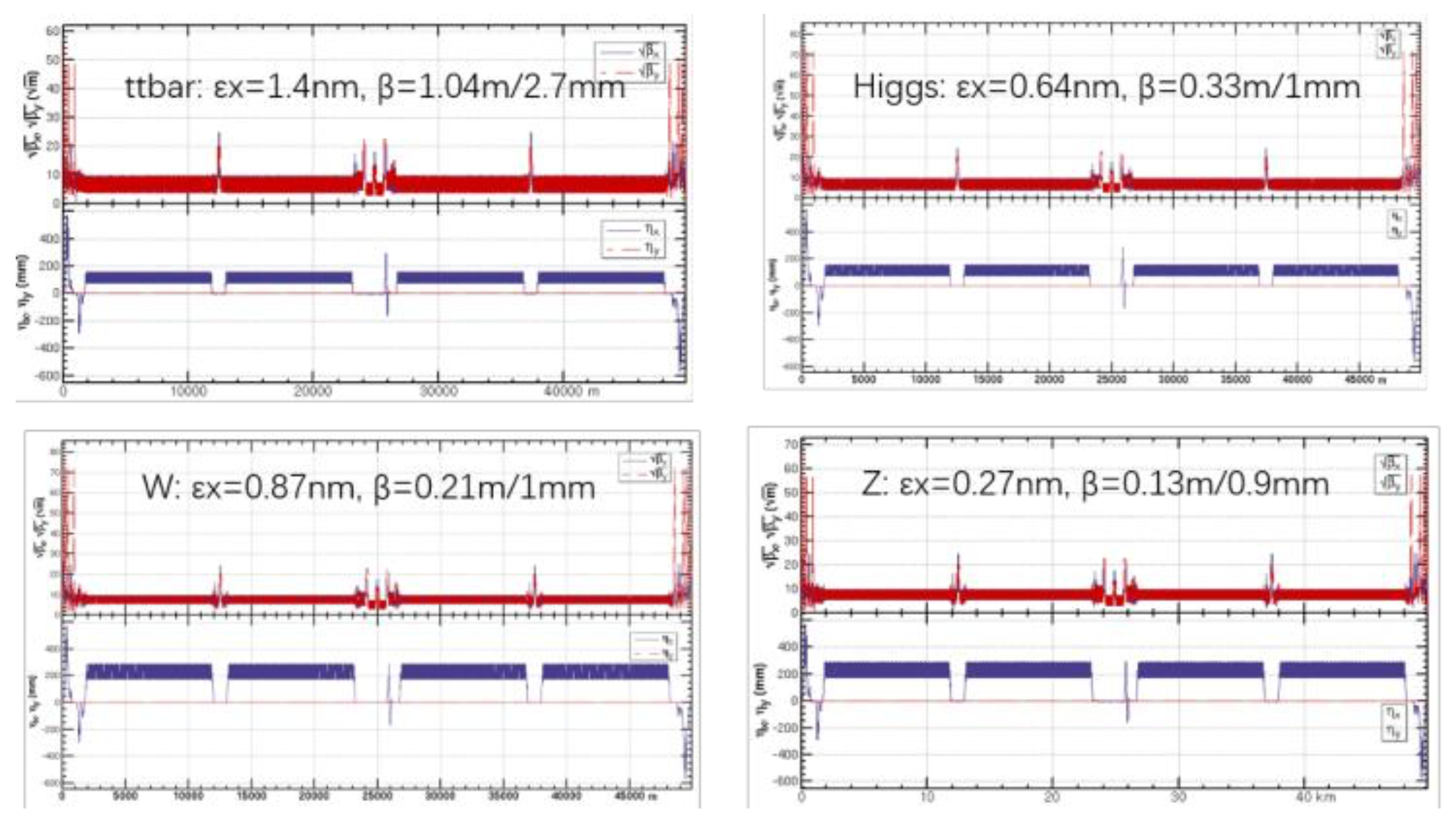}
\caption{Beam optics for the four energies of CEPC collider ring.}
\label{Beam_optics}
\end{figure}

In the RF region, the RF cavities are shared by the two rings. Each RF station is divided into two sections to bypass half of the cavities when operated in W or Z mode~\cite{5_Wang:2018avj, 11_Zhai:2020}. An electrostatic separator combined with a dipole magnet prevents bending of the incoming beam~\cite{8_Oide:2016mkm}. The sawtooth effect is expected to be curable by tapering the magnet strength to account for the beam energy at each magnet. The vertical emittance due to the solenoid field coupling is limited and acceptable. The beam optics of the four energy modes are shown in Fig.~\ref{Beam_optics}.

The requirements of dynamic aperture (DA) are derived from the injection and beam-beam effects to achieve efficient injection and adequate beam lifetime. An optimization code based on the differential evolution algorithm has been developed for CEPC, which is a multi-objective code called MODE~\cite{12_Wu:2020ynh}. The SAD code is used to do the optics calculation and dynamic aperture tracking. Strong synchrotron radiation causes strong radiation damping, which helps enlarge the dynamic aperture to some extent. Quantum fluctuations in the synchrotron radiation are considered in SAD, where the random diffusion due to synchrotron radiation in the particle tracking is implemented in each magnet. Totally, 256 arc sextupole families, 8 IR sextupole families, 4 IR multipoles, and 8 phase advance tuning knobs between different sections can be used to optimize the DA. The error effects were studied with misalignment and main field error for the magnets. 100~$\mu$m are used for the transverse misalignment. Closed orbit correction, dispersion free steering, and beta-beating correction are made to cure the error effects~\cite{13_Wang:2021jwf}. After the corrections, the vertical emittance growth and dynamic aperture are promising.The transverse DA with errors at the Higgs energy satisfies the design requirement.

The CEPC study group also performs design and study on the CEPC MDI, booster, linac, SRF system, beam collective instabilities, injection and extraction transfer lines, timing and bunching patterns, polarization options, and environment effects, see~\cite{WP_JGao}. These studies demonstrate a coherent global design supported by detailed subsystem designs. If properly designed, the CEPC collider could provide the objective luminosities at different physics operations and is compatible with future upgrades. 

\paragraph{Civil Engineering and site}

\begin{figure}[h]
\centering
\includegraphics[width =
0.75\textwidth]{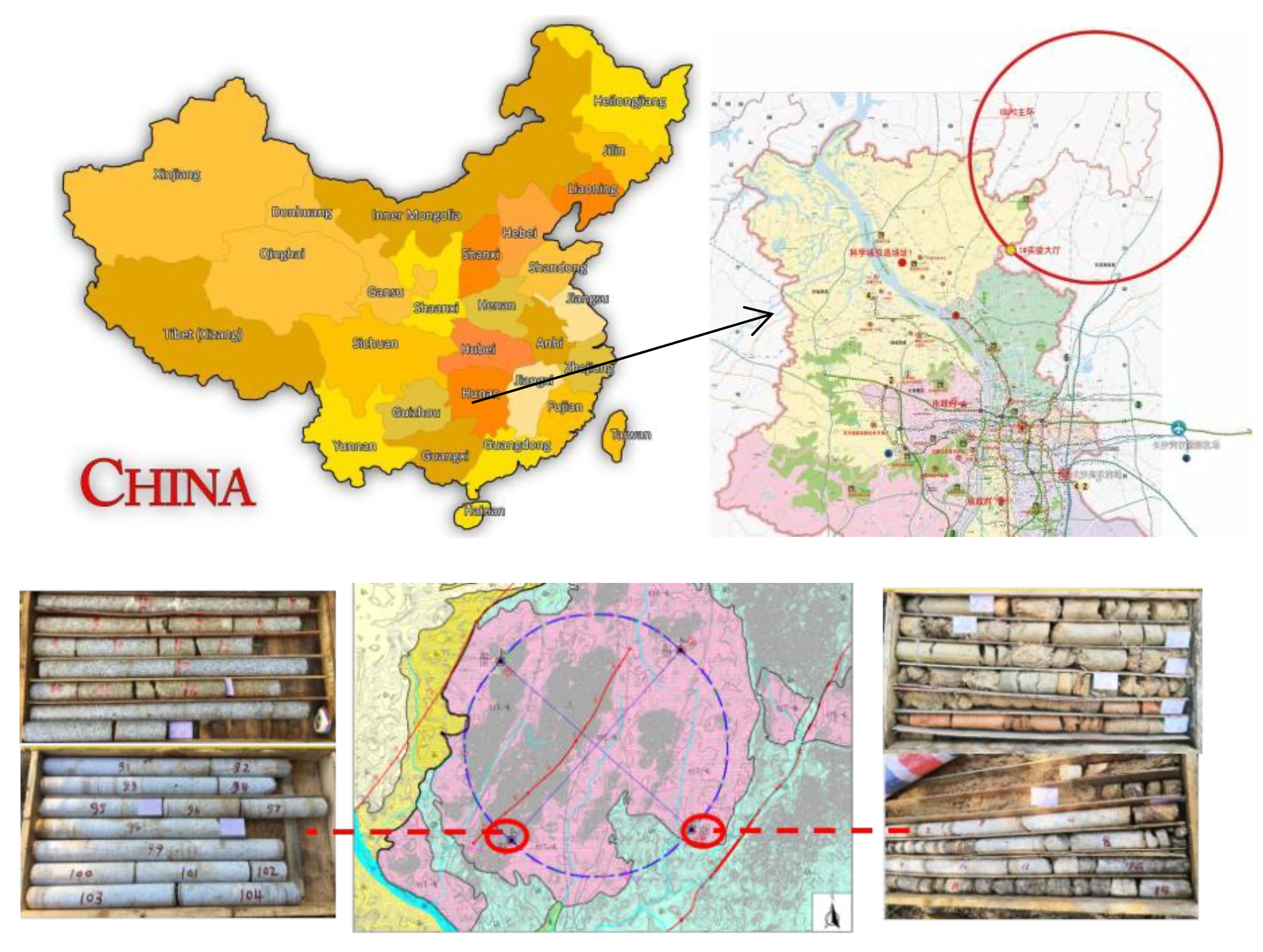}
\caption{CEPC Changsha site, Hunan province and geological condition investigation (one of the site example).}
\label{CEPC_Sites}
\end{figure}

As for CEPC site selection, the technical criteria are roughly quantified as follows: earthquake intensity less than 7 on the Richter scale; earthquake acceleration less than 10\% of gravitational acceleration; ground surface-vibration amplitude less than 20~nm at 1$\sim$100~Hz; granite bedrock around 50$\sim$100~m deep, etc. The site selection process began in February 2015. Preliminary studies of geological conditions for the potential site locations of CEPC have been carried out in Qinhuangdao and Xiongan in Hebei Province, Huangling County in Shanxi Province, Huzhou in Zhejiang Province, Changchun in Jilin Province, and Changsha in Hunan Province, and all of these sites satisfy the CEPC construction requirements. An example site location and geological condition are shown in Fig.~\ref{CEPC_Sites}~\cite{WP_JGao}.

\begin{figure}[h]
\centering
\includegraphics[width =
0.75\textwidth]{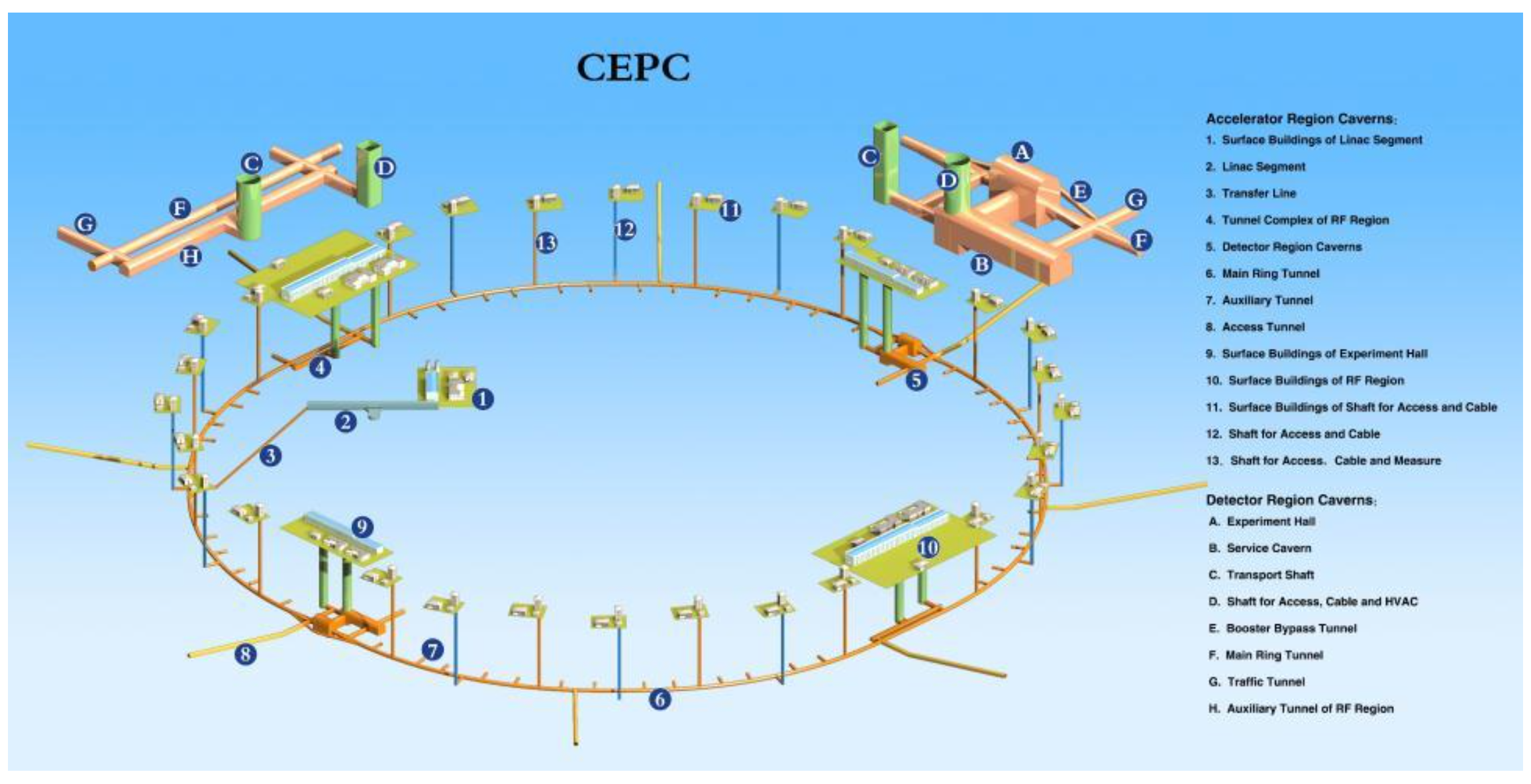}
\caption{CEPC tunnel layout.}
\label{CEPC_Tunnel_Layout}
\end{figure}

According to Chinese civil construction companies involved in the site selection process, it will take less than five years to construct a 100~km tunnel using drill-and-blast methods, followed by the installation of the accelerator and detectors. The total CEPC tunnel civil construction time is 54 months, including 8 months for construction preparation, 43 months for construction of main structures and 3 months for completion. The CEPC tunnel layout is shown in Fig.~\ref{CEPC_Tunnel_Layout}.

\paragraph{Sustainability}

The SppC (Super proton-proton Collider), as an integral part of the CEPC-SppC project, aims to make discoveries at the energy-frontier, which will be a long-term upgrade after the CEPC. It is necessary to investigate the critical physics and technology issues, for example, the iron-based high-field superconducting magnet of at least 20~T to allow proton-proton collisions at a center-of-mass energy of 125~TeV and a luminosity of $4.3 \times 10^{34}~{\rm cm}^{-2}{\rm s}^{-1}$~\cite{17_Snowmass2021_SPPC}.
As for the timeline of SppC, from now to 2035 is the CDR and R\&D period, from 2035 to 2045 is the TDR and EDR periods, from 2045 to 2050 is the construction period, and SppC is expected to be operated after 2050.

The CEPC-SppC, as a Chinese initiated international large science project, will be participated, contributed, and managed internationally, in all aspects and all processes from CDR, TDR, EDR, construction, and operation.

\subsubsection{Proposals for upgrades and extensions}

\paragraph{Luminosity upgrades}

CEPC luminosites at all energies could be upgraded by increasing the SR power per beam from 30MW at the Higgs, W, Z-pole and $ t \bar t$ energies with luminosities $5 \times 10^{34}~ {\rm cm}^{-2}{\rm s}^{-1}$, $16\times 10^{34}~ {\rm cm}^{-2}{\rm s}^{-1}$, $115\times10^{34}~ {\rm cm}^{-2}{\rm s}^{-1}$ and $0.5\times10^{34}~ {\rm cm}^{-2}{\rm s}^{-1}$ per interaction point, respectively, to 50MW SR power/beam at the Higgs, W, Z-pole and ttbar energies with luminosities of $8.3\times10^{34}~{\rm cm}^{-2}{\rm s}^{-1}$, $27\times10^{34}~{\rm cm}^{-2}{\rm s}^{-1}$, $192\times10^{34}~{\rm cm}^{-2}{\rm s}^{-1}$ and $0.8\times10^{34}~{\rm cm}^{-2}{\rm s}^{-1}$ per interaction point, respectively, where the energy upgrade potential to $t \bar t$ energy of 180GeV has been considered with luminosities of $0.5\times10^{34}~{\rm cm}^{-2}{\rm s}^{-1}$ and $0.8\times10^{34}~{\rm cm}^{-2}{\rm s}^{-1}$ corresponding to 30MW and 50MW beam SR power, respectively. 

\paragraph{Energy extension and upgrades }
CEPC is a Higgs factory, and the first priority is to operate the machine at Higgs energy, followed by Z-pole and W energy runs. The norminal SR power per beam is 30MW for all energies. 
CEPC could be upgraded to $t \bar t$ energy of 360 GeV (center of mass), by increasing the SRF cavities and cryogenic system to increase VR voltage from 2.2 GeV to 10 GeV, and the magnets in booster and collider rings have reserved margins to operate at 180 GeV.  
By constructing a Super proton proton Collider (SppC) in the CEPC tunnel, collision energy in center of mass could read as high as 125 TeV.

\subsubsection{Stageability to future experiments}
As experimental system staging from CEPC, a Super proton proton Collider (SppC) could be installed in the same tunnel of CEPC without removing CEPC, and iron-based high-field superconducting magnets of at least 20 T will be used to allow proton–proton collisions at a center-of-mass energy of 125 TeV at a luminosity level of $4.3\times10^{34}~{\rm cm}^{–2}{\rm s}^{–1}$. 
Electron proton collisions can also be realized in the CEPC-SPPC complex by bringing one beam from each of two colliders together and converting two pp collision IR for e-p collisions. The CM energy of e-p collision could reach 6.7 TeV (by 62.5 TeV p  180 GeV e). For 62.5 TeVp  120 GeVe mode, the luminosity is $3.7\times10^{33}~{\rm cm}^{-2}{\rm s}^{-1}$ at one collision point.

\subsubsection{State of Technical Design Report}
CEPC is now in the TDR stage. The TDR will be ended at the end of 2022 as scheduled. After that, CEPC will get into the phase of EDR (2023-2025), which will be the engineering design of CEPC accelerator systems and components towards fabrication in an industrial way. In the meantime, site selection converging to one or two sites with detailed feasibility studies (tunnel and infrastructures, environment), and site dependent civil engineering design will be implemented and prepared.
The EDR document is expected to be completed for government’s approval of starting construction around 2026 (the starting of the “15th five-year plan”of China).

\subsubsection{State of Proposal and R$\&$D plans}
In September 2012, Chinese scientists proposed a 240 GeV Circular Electron Positron Collider (CEPC) as a Higgs factory having two detectors. The 100 km circumference tunnel for such a machine could also host a Super Proton Proton Collider (SPPC) to reach energies around 100 TeV for energy frontier explorations. The CEPC accelerator complex consists of a linear accelerator injector of 30 GeV with a positron damping ring, a full energy injection booster and a double ring collider, both booster and collider rings are located in the same tunnel. CEPC will operate at 240 GeV, W and Z-pole energies with synchrotron radiation power per beam of 30MW (upgradable to 50MW) and has the potential to operate also at $t \bar t$ energy as an upgrade possibility.
In November 2018, CEPC Conceptual Design Report (CDR) has been released formally. In CEPC Technical Design Report (TDR) phase, a full spectrum R\&D activities on the key technologies and prototypes have been conducted, such as 650 MHz 800 kW high efficiency klystrons, 650 MHz and 1.3 GHz SRF cavities and cryomodules, booster low field dipole magnets, dual aperture dipole and quadrupoles for collider rings, final focus SC quadrupole in MDI region, electro-magnetic separators, vacuum chambers with NEG coating technology, instrumentation electronics, high gradient S-band accelerating structures, positron source, high efficiency RF pulse compressor, high precision and high efficiency alignment instrument, etc. In synergy with the 6 GeV High Energy Photon Source (HEPS) under construction by IHEP, many common technologies have been studied and demonstrated, such as various injection/ extraction kickers, high precision magnet power supplies, advanced control system, etc. Many common technology experimental facilities have been established and put to operation,
such as a 4500 $m^2$ SRF laboratory (PAPS), magnets measurement/assembly and vacuum chamber NEG coating laboratories, etc. In addition to CEPC related technologies, the high field SC magnet R\&D for SppC has been launched and made very important progresses. A NbTi+Nb$_3$Sn twin-aperture magnet has reached 12.47 T with $\Phi$ 14 mm at 4.2 K in 2021, and as the next steps, Nb$_3$Sn+HTS (IBS or ReBCO) magnet with two $\Phi$ 45 mm apertures will be developed, aiming to reach $>$16 T in 5 years, and 20-24 T in 10 years. 
In the CDR and TDR phases, CEPC team has established a close collaboration with industries. CEPC Industrial Promotion Consortium (CIPC) of more than 70 members has been established in 2017, and many common development efforts have been put forwards, such as high efficient klystrons, SC cavities, high power cryogenic plant, geological studies in site selections and civil engineering designs, etc. 
CEPC as an international collaboration project has progressed and developed openly and internationally. There are more than 20 collaboration MoUs have been signed and more international collaborators and international industries are welcome to join CEPC and CIPC.
At the end of 2022, CEPC accelerator TDR will be completed and a pre-construction Engineering Deisgn Report (EDR) phase will start from 2023. 
During the CEPC EDR phase, the following activities will be conducted based on TDR, such as engineering design of CEPC accelerator systems and components towards fabrication in an industrial way; CEPC site studies converging to one or two with detailed feasibility studies (tunnel and infrastructures, site dependent civil engineering design and implementation preparation, environmental and social impacts, etc.); EDR document completed for government’s approval of starting construction around 2026 (the starting year of the“15th five year plan”of China). According to the plan, CEPC will be put to operation around 2035.
In 2022, the CEPC team has submitted a formal proposal to a special committee appointed by the Chinese Academy of Sciences, charged to review and develop large particle physics apparatus for the next decades, to begin construction of the CEPC during China's 15th 5-year plan (started from 2026). The committee will conduct reviews involving experts in the field, both domestic and international, and generate formal recommendation to CAS, who may subsequently make recommendation to the central government for the plan of “China Initiated International Large Science and Large Projects”.

As a Higgs factory, the CEPC provides one of the future colliders for the global high energy particle physics community. It was first proposed by Chinese scientists in September 2012, just after the discovery of the Higgs boson at CERN, with strong international and industrial participation. The CEPC CDR was completed in November 2018, and the accelerator TDR will be completed at the end of 2022. The key technologies of CEPC in collider/booster rings and linac injector have been intensively investigated. CEPC will enter the EDR phase in 2023, and the EDR is expected to be delivered at the end of 2025. The CEPC team will work closely with the Chinese central government, international/industrial collaborations, and the local host government during the EDR phase, with the goal of starting CEPC construction around 2026 (within China's 15th Five-Year Plan) and beginning operation around 2035.

%%%%%%%%%%%%%%%%%%%%%%%%%%%%%%%%%%%%%%%%%%%%%%%%%%%%%%%
\newpage
\subsection{e$^{+}$e$^{-}$ circular collider at Fermilab  (EPCCF) \cite{colliders_US} }

EPCCF is a Higgs factory with a circumference of 17 km, which would fit into the FNAL site
\cite{colliders_US,Higgs}; see Fig.~\ref{fig:FNAL-SF}.
The EPCFF design is inspired by the LEP-3 proposal \cite{lep3-0,lep3} (see below). 
EPCFF is expected to 
achieve a luminosity  around $10^{34}$~cm$^{-2}$s$^{-1}$ at a centre-of-mass energy of 
240 GeV at a single collision point, 
assuming two separate collider rings, and a full energy booster for top-up injection.  
Table~\ref{fig:tab-lep3} 
compares EPCCF parameters with those of LEP3, and the FCC-ee CDR  parameters.

\begin{figure}[h]
\centering
\includegraphics[width =
0.75\textwidth]{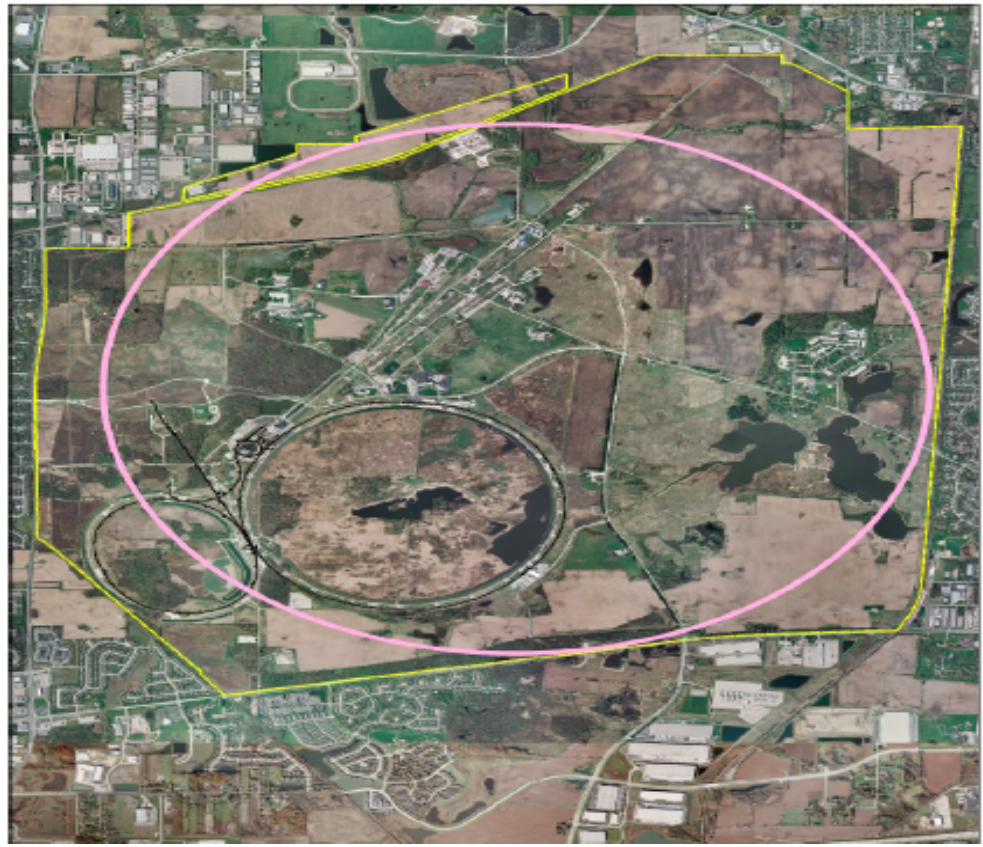}
\caption{Sketch of the FNAL Site Filler, EPCCF.}
\label{fig:FNAL-SF}
\end{figure}

\begin{table}[htbp]
\caption{Parameters of FNAL Site Filler (EPCCF) compared with LEP3 and FCC-ee.}
\label{fig:tab-lep3}
\begin{center}
\begin{tabular}{lccccc}
 Quantity & Symbol & Unit & Site Filler & LEP3 \protect\cite{lep3} & FCC-ee \protect\cite{Abada:2019zxq} \\
\hline
Centre of mass energy & $\sqrt{s}$ & ${\mathrm{GeV}}$ & 120 & 120 & 120 \\
Number of IPs & $N_{IP}$ & & 1 & 2 & 2 \\
Luminosity per IP & $L/{IP}$ & $10^{34}$/cm$^2$s$^{-1}$ & 1.0 & 1.1 & 8.5 \\
Circumference & C & km & 16 & 26.7 & 98 \\ \hline
Bunch number &$N_b$ & & 2 & 4 & 328 \\
Bunch population &$n_e $&$10^{11}$ & 8.3 & 10 & 1.8 \\
Beam current & $I$ & mA & 5 & 7.2 & 29 \\ \hline
Beta functions at IP &$\beta_x/\beta_y$& m/mm & 0.2/1 & 0.2/1 & 0.3/1 \\
Emittance & $\varepsilon_x/\varepsilon_y$ & nm/nm & 21/0.05 & 25/0.1 & 0.63/0.001 \\
Bunch length (SR) & $\sigma_z$ & mm &  2.9 & 2.3 & 3.2 \\
Beam-beam parameters &$\xi_x/\xi_y$ & &  0.075/0.11 & 0.09/0.08 & 0.012/0.12 \\ \hline
RF voltage & $V_{RF}$& GV &  12 & 12 & 2 \\
RF frequency & $f_{RF}$ & MHz & 650 & 1300 & 400 \\
Energy acceptance (DA/RF) &  & $\%$ & $\pm 3$ (RF) & $\pm 4$ (RF) & $\pm 1.7$ (DA) \\
Beam lifetime beamstr. & $\tau_{BS}$ & min & 9--36 & $>17$ & 18 \\
Beam lifetime rad.~Bhabha  &  $\tau_{BB}$ & min & 9 & 18 & 38 \\ \hline
\end{tabular}
\end{center}
\end{table}

The 240 GeV mode of operation requires a total RF voltage of at least 12 GV for each collider ring
and for the booster. EPCCF could also operate at the Z pole and at the WW threshold, with higher luminosity than LEP. 
The EPCCF design has not been worked out in detail.
Due to the small size, even smaller than the LEP/LHC tunnel,
EPCCF cannot be upgraded to a higher collision energy of 365--380 GeV,
and, thus, it cannot probe the WW fusion Higgs production or the 
Higgs self coupling.

%%%%%%%%%%%%%%%%%%%%%%%%%%%%%%%%%%%%%%%%%%%%%%%%%%%%%%%
\newpage
\subsection{Large Electron Positron collider $\#$3 (LEP3)}

LEP3 is a Higgs factory in the LHC tunnel first proposed in 2011 \cite{lep3-0}, 
and submitted as a proposal to
to the 2012/13 European Strategy Process \cite{lep3}. 
With a synchrotron radiation power of 100 MW, LEP3 was predicted to 
achieve a luminosity  around $10^{34}$~cm$^{-2}$s$^{-1}$ at a centre-of-mass energy of 
240 GeV, assuming two separate collider rings, and a full energy booster for top-up injection.  
This mode of operation requires a total RF voltage of 12 GV for each collider ring
and for the booster.
Accommodating the required RF systems inside the existing LHC tunnel is challenging and may require higher-gradient SRF cavities. Also the transverse space constraints need to be considered:  The LHC tunnel has a diameter of 3.8 m, which is significantly smaller than the 5.5 or 6.0 m required for RF installations at the proposed future FCC-ee.   
The construction of LEP3 requires the prior dismantling of the LHC accelerator \cite{lhec}.
So the earliest start of physics operation would be 5--10 years after the end of the
HL-LHC physics program presently scheduled for 2042.
LEP3 cannot be upgraded to a higher collision energy of 365--380 GeV,
and, thus, it cannot probe the WW fusion Higgs production or the 
Higgs self coupling.

%%%%%%%%%%%%%%%%%%%%%%%%%%%%%%%%%%%%%%%%%%%%%%%%%%%%%%%
\newpage

\subsection{Circular e$^{+}$e$^{-}$ Collider using Energy Recovery Linac (CERC), \cite{CERC}}

\subsubsection{Design outline}

\begin{figure}[h]
\centering
\includegraphics[width =
0.50\textwidth]{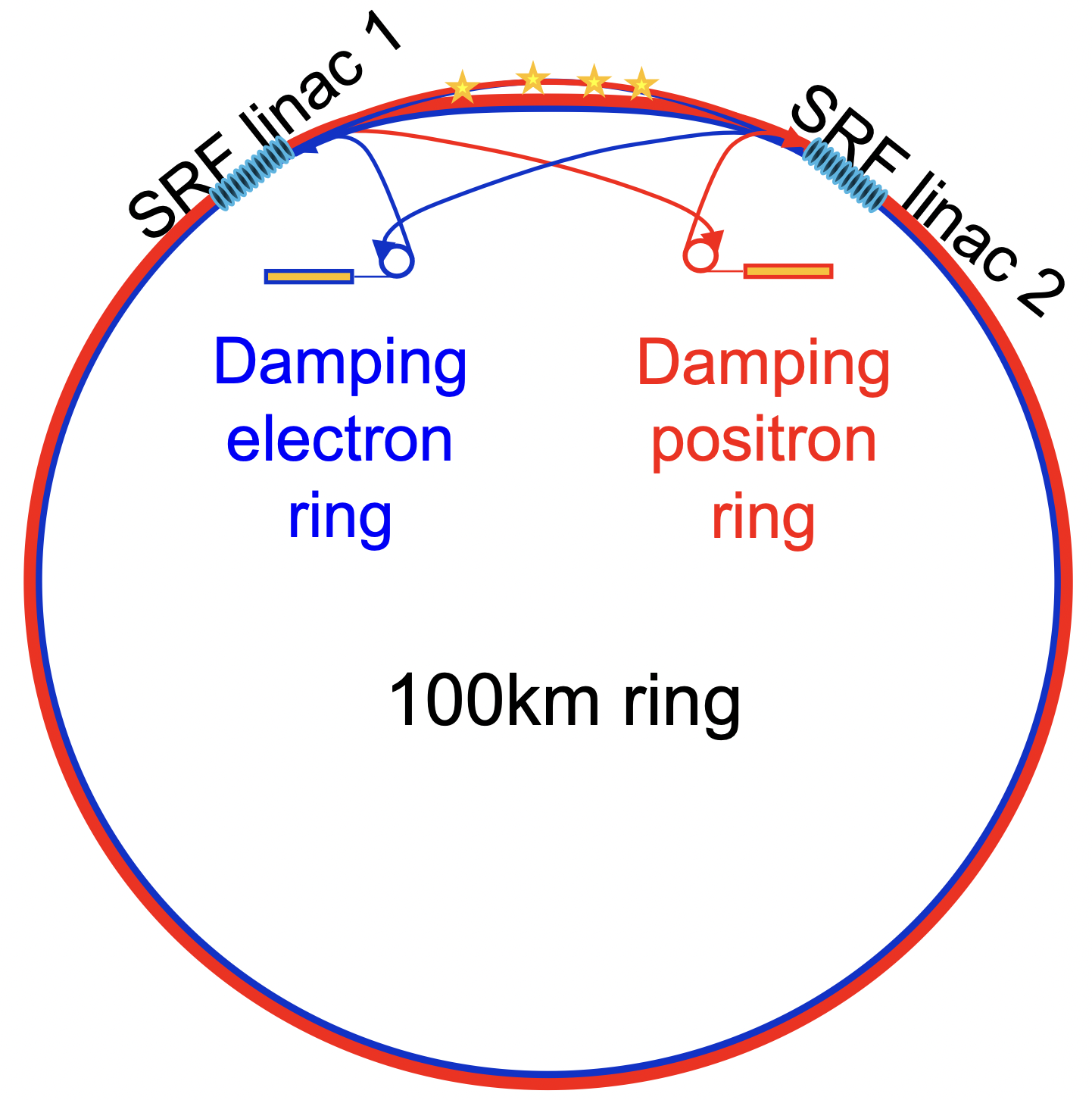}
\caption{CERC layout containing a ring of FCC dimensions, two straight sections for ERL linads, and two damping rings for recovering, damping, and polarizing recaptured electrons and positrons.}
\label{fg:CERClayout}
\end{figure}

A novel approach for a high-energy high-luminosity electron-positron collider was presented in \cite{CERC}. It is based on a ring akin to the FCC-ee that is filled by an Energy Recovery Linac as shown in Fig.~\ref{fg:CERClayout}. This addresses the shortcoming that ring-based collider like the FCC-ee have very high electric power consumption to compensate for the beam energy losses from around 100 MW of synchrotron radiation power. An ERL located in the same-size 100 km tunnel mitigates this drawback. An ERL allows large reduction of the beam energy losses while providing higher luminosity in the full center-of-mass energy that has been evaluated. It allows for extending the CM energy up to 600 GeV, which would enable double-Higgs production, and ttH production.

The CERC e$^{+}$/e$^{-}$ collider is based on ERLs for each beam and on two damping rings that are  used for particle recycling, similar to the ReLiC design. The IRs use a linear collider approach: flat low emittance beams with large vertical disruption parameters. The ERL recycling as much beam energy as possible. Because of self polarization in the damping rings, the CERC can provide collisions of highly polarized electron and position beams. Beam losses are being topped of in the damping ring

This approach combines the advantages of  colliders with those of linacs: Storage ring colliders: recycling beam energy and particles, and 
linear provide efficient collisions
using a large disruption parameter.

\begin{figure}[h]
\centering
\includegraphics[width =
0.75\textwidth]{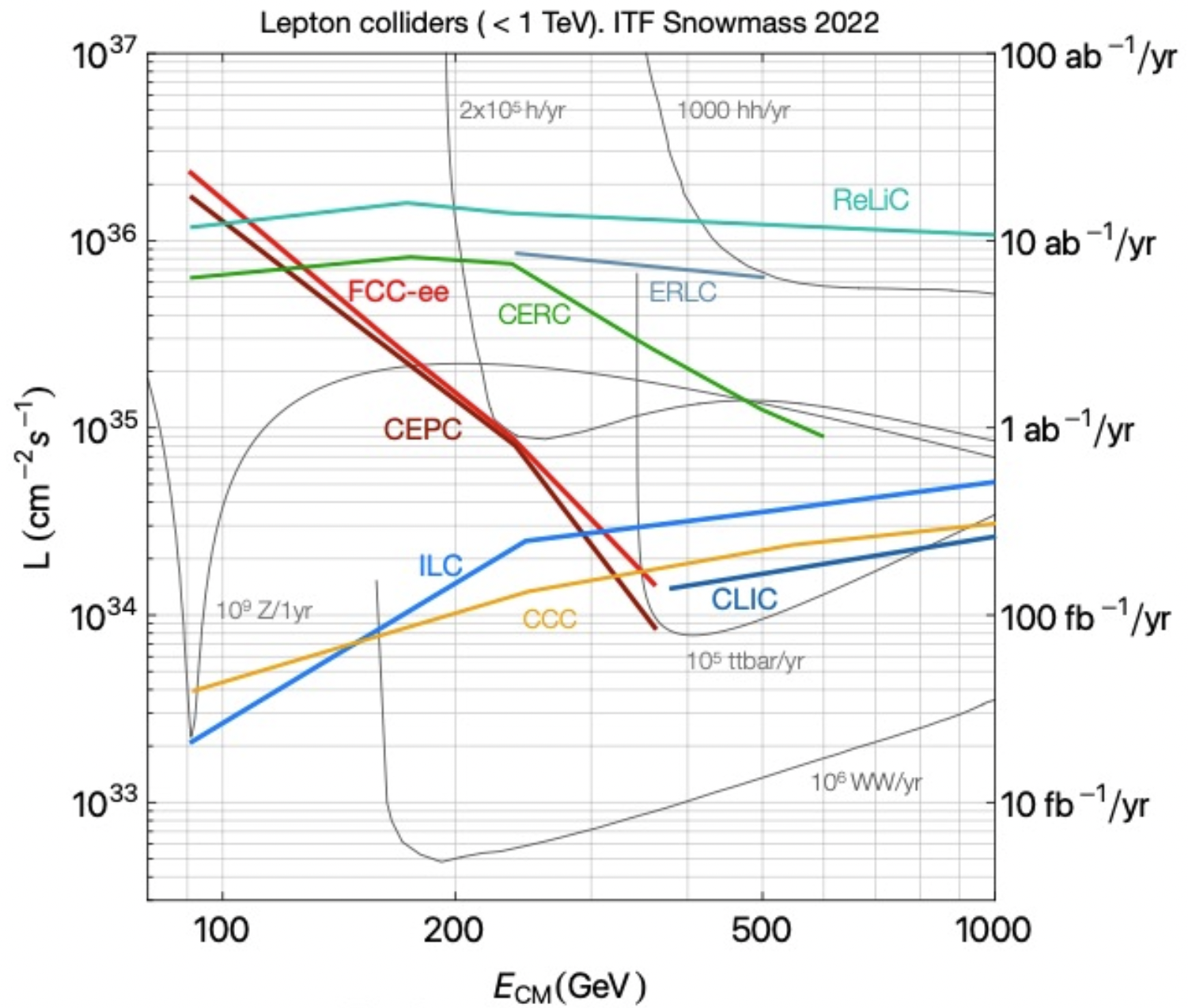}
\caption{The CERC's luminosity potential as compared to other lepton collider projects.}
\label{fg:CERCluminosity1}
\end{figure}

This approach grew out of ideas developed for an
ERL-based electron-ion collider at Brookhaven National Laboratory
where a 20 GeV electron beam collides with a 275 GeV proton
beam. The luminosity potential is shown in Fig.~\ref{fg:CERCluminosity1} as compared to other lepton collider projects.

The key technologies are:
\begin{itemize}
\item Superconducting RF Linacs optimized for ERLs, i.e. with high Q0, low input power, and strong HOM extraction.
\item  An efficient 1.5~K cryoplant, the white paper specifies a~703 MHz 5-cell cavity of the BNL-3 design, operating around 20~MV/m. (d) A 16 m long cryostat housing 10 five-cell cavities, (e) Low emittance damping rings for electron and positron recapture. (f) kickers to extract/inject bunches into the damping rings with 0.1~MHz repetition rate.
\end{itemize}

The white paper evaluates a 4-turn ERL. Different from other ERLs, the accelerating and decelerating beams will have separate beamlines, because their energy will not be the same due to synchrotron radiation. 
 
 \paragraph{Accelerator Design}
 
The main parameters from the white paper design are compiled in Table~\ref{tab:CERCparam}. 
These parameters rely on developing solutions 
for the following main challenges:
\begin{itemize}
\item a novel optics design with 6250 FODO cells with combined function (dipole, quadrupole and sextupole) magnets and zero chromaticity to achieve small emittances, 
\item Polarized beams with less than 0.1\% depolarization, 
\item SRF with Q of $1\cdot 10^{11}$ 
\item Beamline transport preserving the small vertical emittance, 
\item Damping rings with very flat beams 
$\varepsilon_{h}/\varepsilon_{v}=1,000$,
and large energy acceptance of about 5\%,
\item (De-)compressing of electron/positron bunches to match the energy acceptance of the damping rings,  
\item Use of small gap magnets to reduce power consumption and cost of the multiple 100km beamlines, 
\item absolute beam energy measuring systems with accuracy $10^{-5}$ at IRs as pioneered at CEBAF, 
\item High repetition rate extraction and injection kickers for the damping rings.
\end{itemize}

 \begin{table}[htbp]
 \caption{Parameters of the CERC.}
\begin{center}
\begin{tabular}{lcccccccc}
Quantity & Symbol & Unit & Z & W & ZH & $t \bar t$ & HHZ & H$t \bar t$ \\ \hline
Centre of mass energy & $\sqrt{s}$ & ${\mathrm{GeV}}$ & 91.2 & 160 & 240 & 365 & 500 & 600 \\
SR power per beam & $P_{SR}$ & MW & 15 & 14.9 & 14.9 & 15.0 & 16.8 & 16.9 \\
Luminosity per IP & $L_{IP}$ & $10^{35}$cm$^{-2}$s$^{-1}$ &  6.7 & 8.7 & 7.8 & 2.8 & 1.3 & 0.9 \\
Energy loss per turn & $U_0$ &GeV & 4.0 & 4.4 & 6.0 & 17 & 48 & 109 \\
Bends filling factor & & & 0.9 & 0.9 & 0.9 & 0.9 & 0.9 & 0.9 \\
Circumference & C & km & 100 & 100 & 100 & 100 & 100 & 100 \\ \hline
Bunch population &$n_e $&$10^{11}$ & 0.78 & 0.78 & 1.6 & 1.4 & 1.2 & 1.2 \\
Beam current & $I$ & mA  & 3.71 & 3.37 & 2.47 & 0.90 & 0.31 & 0.16 \\
Bunch frequency & & kHz & 297 & 270 & 99 & 40 & 16 & 9 \\\hline
Bunch length  & $\sigma_z$ &mm& 2 & 3 & 3 & 5 & 7.5 & 10 \\
Beta functions at IP &$\beta_x/\beta_y$& m/mm& 0.5/02 & 0.6/0.3 & 1.75/0.3 & 2.0/0.5 & 2.5/0.75 & 3.0/1 \\
Emittances &$\varepsilon_x/\varepsilon_y$&$\mu$m/nm & 3.9/7.8 & 3.9/7.8 & 6.0/7.8 & 7.8/7.8 & 7.8/7.8 & 7.8/7.8 \\
Hor/Vert.~disruption  & $D_x$/$D_{y}$ & & 2.2/503 & 1.9/584 & 0.8/544 & 0.5/505 & 0.3/459 & 0.3/492 \\  \hline
Total ERL voltage & $V_{RF}$& GV & 10.9 & 19.6 & 29.8 & 46.5 & 67.4 & 89 \\ \hline
\end{tabular}
\label{tab:CERCparam}
\end{center}
\end{table}

 \paragraph{Sustainability}
 
 The Energy Recovery Linac principle leads to large energy savings per luminosity. However, because the linac is on continuously, cooling needs become very large. Advances in SRF $Q$ values and in the efficiency of cryoplants therefore have to be found in order to reduce the energy consumption to the order of 100--200 MW, similar to that of other projects, but providing much higher luminosity. The energy consumption for different CERC scenarios is summarized in Table~\ref{tab:CERCpower}.

 \begin{table}[htbp]
 \caption{Energy consumption of the sub-components of the CERC at different energies.}
\label{tab:CERCpower}
 \begin{center}
    \begin{tabular}{lccccccc}
Quantity & Unit & Z & W & ZH & $t \bar t$ & HHZ & H$t \bar t$ \\ \hline
Centre of mass energy & ${\mathrm{GeV}}$ & 45.6 & 80 & 120 & 182.5 & 250 & 300\\ \hline 
SR power &  MW & 30 & 30 & 30 & 30 & 30 & 30 \\
Microphonics  & MW & 1.6 & 2.9 & 4.5 & 7.0 & 10.1 & 13.4 \\
HOM power  & MW & 0.1 & 0.2 & 0.3 & 0.2 & 0.1 & 0.0 \\
Total RF power  & MW &  31.7 & 33.1 & 34.8 & 37.2 & 40.2 & 43.4 \\
Magnets  & MW & 2.0 & 6.2 & 13.9 & 32.0 & 60.1 & 86.6 \\
1.8 K cryo load   & kW & 5 & 01 & 15 & 23 & 34 & 45 \\
Cryoplant AC power   & MW & 6.25 & 12.5 & 18.75 & 28.75 & 42.5 & 56.25 \\ \hline
Total AC power  & MW & 60 & 74 & 90 & 123 & 169 & 215 \\ \hline 
     \end{tabular}
 \end{center}
 \end{table}
 
\subsubsection{Proposals for upgrades and extensions and their stagability}

The potential of the CEPC has been worked out for a Higgs factory as well as for higher energy options, up to 600~GeV, as shown above. They all assume the same radius, that of the FCC. However, the tunnel needs two straight sections that each accommodate half the linac. The length of these straights limits the top energy from the linac and has to be chosen long enough for all desired upgrades.

CERC upgrades in luminosity are also possible, which will increase the SR power linearly. CERC can also be used for hadron-electron and hadron-positron colliders in conjunction with the FCC-hh.
 
\subsubsection{State of Technical Design Report}

A conceptual evaluation has been published in Phys. Rev., a full design report has however not been developed yet. A small group, mostly at BNL continues to develop the design. 

Preliminary simulations have studied the beam-beam effects, the lattice, transport along the lattice, and the optimization of accelerator parameters.
However, full 3D simulations are still required to study collisions with flat beams and high disruption parameters, and optics for the bunch de-compression schemes for the damping rings have to be found. The cost strongly depends on progress on the following R\&D items: (a) High Q0 SRF, (b) verification of Multi-turn, high current, high energy ERLs, (c) High rep-rate kickers, (d) High-efficient cryo plants.

\subsubsection{State of Proposal and R\&D plans}

Publication of the CERC concept idea points out the large advantage of up to a factor of 200 in luminosity, however this realise on several assumption of improved SRF, on improved cryoplants, and on very low losses during energy recovery. R\&D on these fronts is essential to realize the large luminosity potential.

%%%%%%%%%%%%%%%%%%%%%%%%%%%%%%%%%%%%%%%%%%%%%%%%%%%%%%%
\newpage
\section{ Conclusions and General Comments}

This report describes, discusses,  and compares the goals, the designs, the technical state of readiness, and the critical R$\&$D needs of the accelerators that are currently under discussion as Electroweak Higgs factories. We have also addressed their staging options towards future energy-frontier colliders. 

Different linear and circular colliders Higgs factories, e.g., based on conventional SRF or NCRF (including cold-NCRF), or ones including energy recovery linacs or not, have been considered, but only a small number of the proposals are ``shovel-ready'' 
or close to a construction phase, while 
most of the proposals are still in the conceptional or pre-conceptional design stage. 
For colliders of the first category, the  R$\&$D issues have been properly identified and planned, for the latter a focused  R$\&$D  will be needed to identify the possible showstoppers and to move forward to a global project with a self-consistent performance parameter table integrating all the subsystems.

The ERL-based Higgs factories offer the prospect of a longer-term evolution in energy and luminosity that could be achieved through upgrades of either the circular or linear collider infrastructures. However, a strong R$\&$D program would be needed to possibly make this happen. 

Many challenges remain before a muon collider could be realized, implying a relatively long R$\&$D period. 

Finally, we like to issue a few general comments: 
\begin{itemize}

\item Transfer of know-how, experience and expertise to the young generation is  crucial. The proposed facilities will be the colliders for the next generation of accelerator physicists. Our projects need to be attractive and motivating to them. Attractiveness could be increased by granting more 
co-ownership and responsibilities, and by better career perspectives for students, postdocs, and young staff.

\item  Test facilities at ANL, BNL, Cornell, FNAL, LBNL and SLAC in USA, STF, KEK in Japan or CERN in Europe, oﬀer a wide range of opportunities for R$\&$D on new technologies and methods for accelerators. These facilities will promote between others, new RF techniques, development and test of new NCRF - SRF technology, Damping Ring issues related, nanobeams tunability and handling and also advances in plasma and dielectric structure wakeﬁeld acceleration. Their operation fosters the training of accelerator scientists over a broad range of topics, giving the basis for undertaking major new projects. The choice of scale for a demonstrator needs to balance the desirability of having a machine which is useful scientiﬁcally with the desire to invest as minimally as possible in this R\&D phase. Furthermore complementary and synergies between facilities in different part of the world is advisable to optimizing the global scientiﬁc payoﬀ.

\item The few colliders worldwide which are either under construction or in operation 
offer rare and excellent training and learning opportunities, in preparation 
for the ambitious future facilities, and, in particular, for the proposed 
Higgs and electroweak factories.  
In the U.S., the Electron-Ion Collider (EIC), expected to start operation in the early 2030s, 
will become the next such machine.  In Japan, SuperKEKB is already
pushing the frontiers of accelerator physics with (FCC-ee-type!)  ``virtual'' crab-waist collisions, a world-record luminosity of $4.7 \times 10^{34}$~cm$^{-2}$s$^{-1}$, 
and a vertical rms beam spot size of 300 nanometers. The ultimate goal is achieving $6 \times 10^{35}$~cm$^{-2}$s$^{-1}$, and  a beam spot size of 50 nm.
Evidently, SuperKEKB is an important test-bed for all the proposed future Electroweak Higgs  Factories,  
and also a key facility for training the next generation of accelerator physicists.

\item 
The future Higgs factory 
will certainly be unique and most likely be realized as a global enterprise. 
The U.S.~should naturally be one of the major players in this endeavour, 
wherever it is finally realized.

\item Even though Snowmass is a U.S.-centered process, 
the coordination and harmonization with the European EPPSU 2020 and subsequent European Laboratory Directors Group (LDG) efforts on important topics of common interest will be highly beneficial.   
Several novel instruments 
are at the disposal of our communities 
to improve and strengthen intercontinental collaboration. 
One example is the recently approved EC-cofunded EAJADE project (``Europe–America–Japan Accelerator Development and Exchange programme''), which is focused on Higgs Factories, and in which major European (CERN, INFN, CEA, DESY, CNRS, CSIC, UOXF), Japanese (KEK, Tokyo Univ., Tohoku Univ.), and North-American labs (BNL, FNAL, SLAC, JLAB, LBNL, Cornell Univ., VISPA) are participating. 
We also note that the Snowmass process is being organized, or co-organized, by several divisions of the American Physical Society, which aspires to become a ``global hub''.

\item Lastly, the enormous societal impact (medical, industrial, security,...) of collider projects has to be better explained, communicated and exploited. 
Since all modern colliders are expensive projects, we have to convince the society and its decision makers of the need for, and benefits of,  constructing and operating this kind of facilities. The proliferation of accelerator technology from its initial use for basic science to applications more directly benefiting society has been a very visible trend in recent decades; and this represents only the first step in a major evolution for particle accelerators.   
In a parallel effort, careful analyses for a few prominent large collider 
projects have revealed that 
the direct (i.e.,   non-scientific) economic impact more than compensates
the initial investments made by the taxpayers.

\end{itemize}

%%%%%%%%%%%%%%%%%%%%%%%%%%%%%%%%%%%%%%%%%%%%%%%%%%%
%\newpage
%\appendix
%\chapter{TRL level description}
%\begin{figure}[h]
%\centering
%\includegraphics[width = 0.75\textwidth]{Accelerator/AF03/figures/TRL_levels.pdf}
%\caption{Technology Readiness Level (TRL) Measurement.}
%\label{TRLs}
%\end{figure}

%%%%%%%%%%%%%%%%%%%%%%%%%%%%%%%%%%%%%%%%%%

%  If you would like to use BibTEX for the bibliography, please feel free to do so.  It is not required.

%  To use BibTeX,

%    1.  uncomment the following two lines,
%    2.  comment out everything below from  \begin{thebibliography}{99}   to \end{thebibliography).
%    3.  create the file  myreferences.bib in this directory, and process this file in the usual way

%\bibliographystyle{JHEP}
%\bibliography{Accelerator/AF03/myreferences} 

%%%%%%%%%%%%%%%%%%%%%%%%%%%%%%%%%%%%%%%%%

\end{document}